\documentclass[longauth]{aa}

\usepackage{latexsym}		
\usepackage{graphicx}		
\usepackage{rotating}		
\usepackage{natbib}  
\usepackage{savesym}
\usepackage{amssymb}
\usepackage{amsmath}
\usepackage{morefloats}
\usepackage{nicefrac}
\savesymbol{doublespace}
\usepackage{xspace}
\usepackage{color}
\usepackage{mdframed}
\usepackage{url}
\usepackage{grffile}
\usepackage{import}
\usepackage[utf8]{inputenc}
\usepackage{longtable}
\usepackage{array}
\usepackage{booktabs}
\usepackage{fancyhdr}
\usepackage{subfig}

\usepackage[hang,flushmargin]{footmisc}
\usepackage{ifpdf}

\newcommand{\msun}{\ensuremath{M_{\odot}}\xspace}			

\newcommand{\formaldehyde}{\ensuremath{\textrm{H}_2\textrm{CO}}\xspace}

\newcommand{\threeohthree}{\ensuremath{3_{0,3}-2_{0,2}}\xspace}

\newcommand{\kms}{\textrm{km~s}\ensuremath{^{-1}}\xspace}	

\def\eqref#1{Equation \ref{#1}}

\newenvironment{rotatepage}
{}{}

\newcommand{\bsens}{\texttt{bsens}\xspace}
\newcommand{\cleanest}{\texttt{cleanest}\xspace}
\newcommand{\paperone}{Paper I; \cite{Motte2021}\xspace}
\newcommand{\paperthree}{Paper III; Louvet et al (in prep)\xspace}

\def\refreport#1{{#1}}

\begin{document}

\title{ALMA-IMF II - investigating the origin of stellar masses: Continuum Images and Data Processing}
\titlerunning{ALMA-IMF II: Continuum Data}

\author{A. Ginsburg\inst{1}\and 
T. Csengeri\inst{2}\and 
R. Galv\'an-Madrid\inst{3}\and 
N. Cunningham\inst{4}\and 
R. H. Álvarez-Gutiérrez\inst{5}\and 
T. Baug\inst{6}\and 
M. Bonfand\inst{2}\and 
S. Bontemps\inst{2}\and 
G. Busquet\inst{7,8,9}\and 
D. J. D\'iaz-Gonz\'alez\inst{3}\and 
M. Fern\'andez-L\'opez\inst{10}\and 
A. Guzmán\inst{11}\and 
F. Herpin\inst{2}\and 
H. Liu\inst{12}\and 
A. L\'opez-Sepulcre\inst{13,14}\and 
F. Louvet\inst{15,16}\and 
L. Maud\inst{17}\and 
F. Motte\inst{4}\and 
F. Nakamura\inst{18,19,20}\and 
T. Nony\inst{3}\and 
F. A. Olguin\inst{21}\and 
Y. Pouteau\inst{4}\and 
P. Sanhueza\inst{11,22}\and 
A. M.\ Stutz\inst{5,23}\and 
A. P. M. Towner\inst{1}\and 
\\
\textbf{The ALMA-IMF Consortium:}
M. Armante\inst{24}\and 
C. Battersby\inst{25}\and 
L. Bronfman\inst{26}\and 
J. Braine\inst{2}\and 
N. Brouillet\inst{2}\and 
E. Chapillon\inst{2,13}\and 
J. Di Francesco\inst{27}\and 
A. Gusdorf\inst{24}\and 
N. Izumi\inst{11,28,29}\and 
I. Joncour\inst{7}\and 
X. (Walker) Lu\inst{11}\and 
A. Men'shchikov\inst{30}\and 
K. M. Menten\inst{31}\and 
E. Moraux\inst{4}\and 
J. Molet\inst{2}\and 
L. Mundy\inst{32}\and 
Q. Nguyen Luong\inst{30}\and 
S. D. Reyes-Reyes\inst{5}\and 
J. Robitaille\inst{4}\and 
E. Rosolowsky\inst{33}\and 
N. A. Sandoval-Garrido\inst{5}\and 
B. Svoboda\inst{34}\and 
K. Tatematsu\inst{11}\and 
D. L. Walker\inst{25}\and 
A. Whitworth\inst{35}\and 
B. Wu\inst{36,11}\and 
F. Wyrowski\inst{31}}

\institute{Department of Astronomy, University of Florida, PO Box 112055, USA \and
Laboratoire d'astrophysique de Bordeaux, Univ. Bordeaux, CNRS, B18N, allée Geoffroy Saint-Hilaire, 33615 Pessac, France\and
Instituto de Radioastronom\'ia y Astrof\'isica, Universidad Nacional Aut\'onoma de M\'exico, Morelia, Michoac\'an 58089, M\'exico\and
Univ. Grenoble Alpes, CNRS, IPAG, 38000, Grenoble, France\and
Departamento de Astronom\'{i}a, Universidad de Concepci\'{o}n, Casilla 160-C, 4030000 Concepci\'{o}n, Chile\and
S. N. Bose National Centre for Basic Sciences, Block JD, Sector III, Salt Lake, Kolkata 700106, India\and
Univ. Grenoble Alpes, CNRS, IPAG, 38000, Grenoble, France\and
Institut de Ci\`encies de l'Espai (ICE, CSIC), Can Magrans, s/n, 08193, Cerdanyola del Vall\`es, Catalonia, Spain\and
Institut d'Estudis Espacials de Catalunya (IEEC), 08340, Barcelona, Catalonia, Spain\and
Instituto Argentino de Radioastronom\'\i a (CCT-La Plata, CONICET; CICPBA), C.C. No. 5, 1894, Villa Elisa, Buenos Aires, Argentina\and
National Astronomical Observatory of Japan, National Institutes of Natural Sciences, 2-21-1 Osawa, Mitaka, Tokyo 181-8588, Japan\and
Department of Astronomy, Yunnan University, Kunming, 650091, PR China\and
Institut de Radioastronomie Millim\'etrique (IRAM), 300 rue de la Piscine, 38406 Saint-Martin-D'H\`eres, France\and
Univ. Grenoble Alpes, CNRS, IPAG, 38000, Grenoble, France\and
AIM, CEA, CNRS, Universit{\'e} Paris-Saclay, Universit{\'e} Paris Diderot, Sorbonne Paris Cit{\'e}, F-91191 Gif-sur-Yvette, France\and
LERMA (UMR CNRS 8112), \'Ecole Normale Sup\'erieure, 75231 Paris Cedex, France\and
ESO Headquarters, Karl-Schwarzchild-Str 2 D-85748 Garching\and
National Astronomical Observatory of Japan, 2-21-1 Osawa, Mitaka, Tokyo 181-8588, Japan\and
Department of Astronomy, The University of Tokyo, 7-3-1 Hongo, Bunkyo, Tokyo 113-0033, Japan\and
The Graduate University for Advanced Studies (SOKENDAI), 2-21-1 Osawa, Mitaka, Tokyo 181-0015, Japan\and
Institute of Astronomy, National Tsing Hua University, Hsinchu 30013, Taiwan\and
Department of Astronomical Science, SOKENDAI (The Graduate University for Advanced Studies), 2-21-1 Osawa, Mitaka, Tokyo 181-8588, Japan\and
Max-Planck-Institute for Astronomy, K\"{o}nigstuhl 17, 69117 Heidelberg, Germany\and
Laboratoire de Physique de l\'{E}cole Normale Sup\'{e}rieure, ENS, Universit\'{e} PSL, CNRS, Sorbonne Universit\'{e}, Universit\'{e} de Paris, 75005, Paris, France\and
University of Connecticut, Department of Physics, 196A Auditorium Road, Unit 3046, Storrs, CT 06269 USA\and
Departamento de Astronomía, Universidad de Chile, Casilla 36-D, Santiago, Chile\and
Herzberg Astronomy and Astrophysics Research Centre, National Research Council of Canada, 5071 West Saanich Road, Victoria, BC V9E 2E7, Canada\and
College of Science, Ibaraki University, 2-1-1 Bunkyo, Mito, Ibaraki 310-8512, Japan\and
Institute of Astronomy and Astrophysics, Academia Sinica, No. 1, Section 4, Roosevelt Road, Taipei 10617, Taiwan\and
AIM, IRFU, CEA, CNRS, Universit{\'e} Paris-Saclay, Universit{\'e} Paris Diderot, Sorbonne Paris Cit{\'e}, F-91191 Gif-sur-Yvette, France\and
Max Planck Institute for Radio Astronomy, Auf dem H\"{u}gel 69, 53121 Bonn,  Germany\and
Department of Astronomy, University of Maryland, College Park, MD 20742, USA\and
4-183 CCIS, University of Alberta, Edmonton, Alberta, Canada\and
Jansky fellow of the National Radio Astronomy Observatory, Socorro, NM 87801 USA\and
School of Physics and Astronomy, Cardiff University, Cardiff, UK\and
NVIDIA Research, 2788 San Tomas Expy, Santa Clara, CA 95051, USA}\authorrunning{Ginsburg et al}

\abstract{
We present the first data release of the ALMA-IMF Large Program, which covers
the 12m-array continuum calibration and imaging.  The ALMA-IMF Large Program is
a survey of fifteen dense molecular cloud regions spanning a range of
evolutionary stages that aims to measure the core mass function (CMF).  We
describe the data acquisition and calibration done by the Atacama Large Millimeter/submillimeter Array (ALMA) observatory and
the subsequent calibration and imaging we performed.  The image products are
combinations of multiple 12m array configurations created from a selection of
the observed bandwidth using multi-term, multi-frequency synthesis imaging and
deconvolution.  The data products are self-calibrated and exhibit substantial
noise improvements over the images produced from the delivered data.  We
compare different choices of continuum selection, calibration parameters, and
image weighting parameters, demonstrating the utility and necessity of our
additional processing work.  Two variants of continuum selection are used and
will be distributed: the ``best-sensitivity'' (\bsens) data, which include the
full bandwidth, including bright emission lines that contaminate the continuum,
and ``cleanest'' (\cleanest), which select portions of the spectrum that are
unaffected by line emission.  We present a preliminary analysis of the spectral
indices of the continuum data, showing that the ALMA products are able to
clearly distinguish free-free emission from dust emission, and that in some
cases we are able to identify optically thick emission sources.  The data
products are made public with this release.
}

\maketitle

\section{Introduction}
\label{sec:intro}

In our Galaxy, stars form out of dense, dust-rich gas that has its peak
emission in the far-infrared and is bright at millimeter wavelengths.
Observations of the thermal continuum emission from dust grains have become the most important
tool for determining the mass of the pre-stellar material that collapses under self-gravity
to form stars \citep[e.g.,][]{Motte1998,Enoch2008}.
While star formation within the local kiloparsec is well-observed with
single-dish instruments
and small interferometers, the Atacama Large Millimeter/submillimeter Array
(ALMA) has opened new opportunities to study star formation at
solar system scale resolution throughout the Galaxy
\citep[e.g.,][]{Ginsburg2017,Motte2018,Csengeri2018,Sanhueza2019}.

We have therefore undertaken a large observing program to take advantage of
these new capabilities.  ALMA-IMF is an ALMA Large Program\footnote{Program ID
2017.1.01355.L;  PIs: Motte, Ginsburg, Louvet, Sanhueza,
\url{https://www.almaimf.com}} to
survey fifteen high-mass star-forming regions in the Galactic plane.  The
survey overview is given in \paperone. 

The primary goal of ALMA-IMF is to measure the gas-phase precursor to the
stellar initial mass function (IMF), the core mass function (CMF).  This
distribution function has previously been observed, in local clouds, to share a
shape with the IMF \citep[e.g.,][]{Motte1998,Alves2007,Konyves2015}, leading to the suggestion that
the origin of stellar masses is in this gas phase, though other interpretations
of this similarity are possible \citep{Offner2014}.  The local-cloud
observations were limited both in the upper mass limit
($M_{core,max}\lesssim10\msun$) and in the range of physical conditions probed,
especially in terms of feedback from high-mass stars and protostars.  The
precursor works that motivated ALMA-IMF \citep{Ginsburg2017,Motte2018,Sanhueza2019} have shown that a range of CMF shapes
exist in high-mass star-forming regions \citep[e.g.,][]{Beuther2004,Zhang2015,Ohashi2016,Lu2020}, driving
the need to observe a larger sample.

The ALMA-IMF sample has been selected to probe the full range of evolutionary
stages and a wide range of Galactic environmental conditions.
The selection, described in the companion overview paper (\paperone), is based
on the ATLASGAL survey \citep{Schuller2009,Csengeri2014} and ancillary multi-wavelength data.
It consists of 15 regions in the process of forming star clusters at
different evolutionary stages: young regions, with no signs of high-mass stars
having ignited HII regions; intermediate, with only ultracompact or
hypercompact HII regions present and feedback effects confined to a small
region, and evolved, in which HII regions coexist with ongoing star
formation.

In this paper, we present the data reduction, imaging,
and characterization to obtain continuum maps in ALMA's
Band 3 centered at 99.66~GHz, and Band 6 centered at 230.6~GHz.
\paperone describes the sample selection and early results.
\paperthree describes the core catalog extracted from the
data presented here.

Section \ref{sec:observations} describes the observations and data acquisition.
Section \ref{sec:data} describes the processing performed to produce
the delivered data products.
Section \ref{sec:dataproducts} describes the data products.
Section \ref{sec:analysis} demonstrates some preliminary science applications
of the data, focusing on the spectral index measurements.
We summarize the result in Section \ref{sec:conclusions}.

There are several appendices discussing self-calibration comparison (Appendix \ref{appendix:selfcalcompare}),
\refreport{self-calibration parameter details (Appendix \ref{sec:selfcaldetails}),}
data processing and handling (Appendix \ref{sec:datahandling}),  
listing central frequencies (Appendix \ref{sec:centralfreq}),
describing different data releases (Appendix \ref{sec:datareleases}),
describing the W43-MM1 B6 archival data (Appendix \ref{sec:w43mm1b6}),
describing additional data products produced excluding CO and N$_2$H+ (Appendix \ref{appendix:bsens_noco}),
and listing the supplemental figure sets
and additional overview
figures 
(Appendix \ref{appendix:suppfigures}).

The data are released on Zenodo at \url{https://zenodo.org/record/5702966}.

\section{Observations}
\label{sec:observations}
We report a summary of the observations taken by ALMA and a brief description of the target selection. 
Table \ref{tab:observations} lists the details and the observing setup for the targeted fields.

The observing strategy for the ALMA-IMF program was to take a homogeneous
approach to imaging 15 of the most extreme Galactic massive clumps covering a
distance range between 2 and 5.5~kpc (Fig. \ref{fig:overview_multicolor_B3} and
\ref{fig:overview_multicolor_B6}). The mosaics in ALMA's band~3 (B3; 91-106
GHz) and band~6 (B6; 216-234 GHz) were set up to map a $\gtrsim$1$\times$1~pc area
covering the  highest column density region of each protocluster as determined
from ATLASGAL and Hi-GAL imaging \citep{Csengeri2018,Molinari2010}. The angular
resolution for each individual protocluster was chosen to achieve a physical
resolution $\lesssim2000$~au for all regions.
All target fields were observed with two 12 m array configurations in band~3 to
achieve both high spatial resolution and high dynamic range.
In band~6, the more distant regions ($d>3.9$~kpc), W43, W51, G338.93, G337.92, G333.60, and G010.62
required two 12 m configurations, while the more nearby used only one.
The long- and short-baseline observations are denoted TM1 and TM2, respectively, in the ALMA-delivered data products.
Full details of the array configurations are given in Table \ref{tab:observations}.
The resulting angular resolution is between 0.3\arcsec and 1.5\arcsec using a
robust weighting of 0 (see more details in Sect.\,\ref{sec:imaging}).

The mosaics have varying fields of view (FOVs) to accommodate different clouds.
Generally, the Band 3 FOV is larger
than that of Band 6 because of the intrinsically larger primary beam at Band 3.
The fields of view are shown overlaid on Spitzer GLIMPSE
\citep{Benjamin2003,Churchwell2009} images in Appendix
\ref{appendix:suppfigures}, Figure \ref{fig:overview}.

All fields also included 7m array and total power observations in the same spectral setup.  The total power
observations cannot be used to create images of the continuum and therefore are not discussed
here.
Although the data products presented here make no use of the 7m array data, the properties of the short spacing information are discussed in Sect.\,\ref{sec:7m12m}.

The program data were originally retrieved from the ALMA archive shortly after passing the quality assessment by the observatory, and were further inspected by our data reduction team.
We examined the pipeline-produced calibration web logs in detail, noting any clear problems in the data. In several cases, 
this process enabled reports back to the observatory that data quality failed to meet standards
and triggered additional observations.
Weblog examination and initial tests were distributed over the whole data reduction team.

The data presented in this paper were later retrieved
from the ALMA archive using \texttt{astroquery} \citep{Ginsburg2019} 
between June 2019 and June 2020. 
These data were restored to measurement sets using the \texttt{scriptForPI.py} files
provided by the ALMA archive, and further batch processed with the custom scripts and
imaging parameters determined from the individual tests discussed below. 

All of these measurement sets have been subsequently taken back to the ALMA
observatory for QA3 reprocessing, and therefore their latest archival versions
may show differences compared to the version used for this work.  Members of
the FAUST Large Program (Project code: 2018.1.01205.L) reported that the system
calibration temperature approach adopted by ALMA sometimes results in
artificial suppression of bright lines\footnote{ALMA ticket:
\url{https://help.almascience.org/kb/articles/607},
\url{https://almascience.nao.ac.jp/news/public-announcement-of-casa-imaging-issues-affecting-some-alma-products}}.
The issues amount to a combination of spectral normalisation and system
temperature calibration problems.  The ALMA-IMF data were affected by these
issues and returned to the Joint ALMA Observatory for further QA3 processing in
November 2020.  Reprocessing was completed in March 2021.  Because continuum data are minimally
affected (the expected effect is proportional to the affected bandwidth, and
the bright lines affected are generally excluded in this work), the data
presented here did not undergo this QA3 reprocessing.  However,
we will also release the reprocessed data; see Appendix \ref{sec:datareleases}.

The W43-MM1 B6 data were taken as part of the pilot program, 2013.1.01365.S
\citep[][]{Motte2018}.  These data were also reprocessed following the
same QA3 procedure as the 2017.1.01355.L data.

\section{Data}
\label{sec:data}

We present the data obtained from the ALMA-IMF Large Program (2017.1.01355.L,
plus W43-MM1 data from 2013.1.01365.S) and discuss the data reduction process
followed to obtain images of the continuum emission.

\subsection{ALMA-IMF data pipeline}
\label{sec:almaimfpipeline}

We describe the ALMA-IMF data pipeline and the subsequent data quality
assessment steps we performed in this section. The pipeline can be found on the
github repository\footnote{\url{https://github.com/ALMA-IMF/reduction}}. 

Our custom pipeline is used to perform several essential steps on the continuum data:
\begin{enumerate}
    \item Combination of different array configurations (the ALMA-IMF data include up to two 12m array
    configurations for each field).
    \item Masked deep cleaning of the images.
    \item Self-calibration of the mosaic data.
\end{enumerate}
The main advantages of our processing are the masked deep cleaning with
parameters optimized for each field and the  self-calibration that greatly (by
up to a factor of 5) increases the signal-to-noise ratio in several fields.

The ALMA-IMF data pipeline starts from the ALMA pipeline-calibrated data 
and restores the archival data products to measurement sets using the standard ALMA pipeline
procedures.  We verified the observatory's quality assessment analysis by
examining the weblogs.  While several issues of potential concern were noted,
such as high phase variations in the calibrators in some execution blocks, all
pipeline products were good enough for initial imaging, and we determined that
further correction via self-calibration was the best approach for improving the
images.

To enable continuum selection, faster cleaning, and self-calibration, the
science target data were split out from the original pipeline-processed data sets.
The continuum selection process is described in \ref{sec:continuumselection},
and the subsequent spectral averaging is described in \ref{sec:splitting}.

\subsubsection{Implementation details}
\label{sec:pipelineimplementation}
The ALMA-IMF data pipeline is designed to run in the CASA \citep{McMullin2007}
environment, and is implemented as a suite of python scripts.
The workflow is as follows:

\begin{enumerate}
    \item Retrieve and extract the data from the ALMA archive. 
    \item Run  \texttt{scriptForPI.py} to restore the measurement sets. 
    \item Run the pipeline script \texttt{split\_windows.py} to create
        the separate continuum and line measurement sets. 
    \item Run the \texttt{continuum\_imaging\_selfcal.py} script to perform
        the imaging and self calibration. 
\end{enumerate}

The pipeline relies on \texttt{astroquery} \citep{Ginsburg2019} to retrieve
the data.
Several of the analysis routines use \texttt{astropy}
\citep{AstropyCollaboration2013,AstropyCollaboration2018},
\texttt{spectral-cube}\footnote{\url{https://spectral-cube.readthedocs.io/en/latest/}},
and \texttt{radio-beam}\footnote{\url{https://radio-beam.readthedocs.io/}}.
The usual suite of python numerical tools, numpy, scipy, and matplotlib, serve
as the base of these other packages \citep{Hunter2007,Harris2020,vanderWalt2011,Virtanen2020}.

The pipeline contains many other support files included beyond those described above.
Most important is the \texttt{imaging\_parameters.py} file,
which contains the complete listing of the user-specified parameters used both
for imaging and self-calibration.

\subsubsection{Processing and data Storage}

The data processing was done in several stages.
In the first, distributed stage, each member of the data reduction team
downloaded a small number of target fields (one to four) and processed
them locally.  They delivered processed products and the corresponding 
imaging parameters to a central repository.

In a second stage, all of the data were collected on one machine,
the University of Florida's \texttt{hipergator} supercomputer, and the pipeline
was re-run following all steps in \ref{sec:pipelineimplementation}.
Each complete run of the continuum pipeline takes up to about a week,
though the majority of fields complete processing, self-calibration,
and final imaging in less than a day.  The largest fields, W43-MM2, W51-IRS2, and W51-E B3, take much longer because they are $>4000$ pixels on a side, and both the minor and major clean cycles are slow.

The data products during pipeline running can require up to 250 TB of storage space.
The raw data are $\sim30$ TB, but they are duplicated many times over when creating additional measurement sets and line cubes.  
The continuum image release is much smaller, totaling $<100$ GB. 
Further details of the computing setup and data processing are given in Appendix \ref{sec:datahandling}.

\begin{figure*}
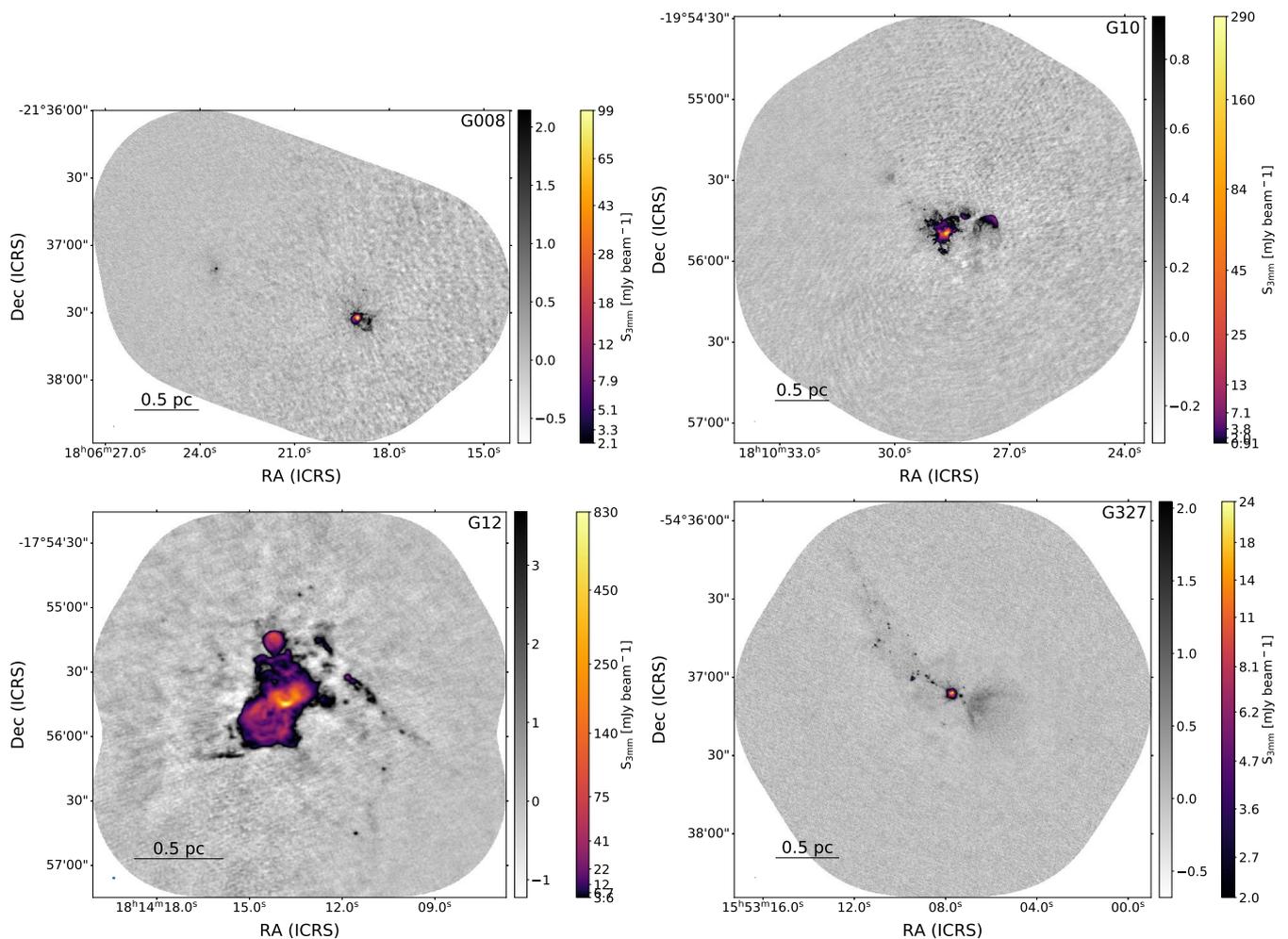

    \centering
    \includegraphics[width=0.48\textwidth]{f1}
    \includegraphics[width=0.48\textwidth]{f2}
    \includegraphics[width=0.48\textwidth]{f3}
    \includegraphics[width=0.48\textwidth]{f4}
    \caption{Overview plot showing B3
    continuum emission maps.
    The plots are shown with two colorbars, the first (grayscale) showing -5 to
    +15 times the noise on a linear scale, then a second (matplotlib's inferno
    colorscheme) showing the range +15 times the noise to the peak intensity of
    the image in an arcsinh stretch.
    The field names are labeled in the top-right corner.
    A scalebar in the bottom-left shows the size of 0.1 pc at the target's distance.
    The synthesized beam is shown as a blue ellipse in the bottom-left corner;
    it may be too small to print properly.
    Only the first four fields in B3 are shown here.  The remaining figures are 
    shown in Appendix \ref{appendix:suppfigures}.
    }
    \label{fig:overview_multicolor_B3}
\end{figure*}

\subsubsection{Continuum selection process}
\label{sec:continuumselection}

\begin{table*}[htbp!]
\centering
\small
\caption{Spectral setup of the ALMA-IMF Large Program}
\label{tab:lines}
\begin{tabular}{lcccccl}
\hline
\hline \noalign {\smallskip}
ALMA  & Spectral  & Frequency & Bandwidth & \multicolumn{2}{c}{Resolution}  & Bright contaminant lines \\
band & window & [GHz] & [MHz] & [kHz] & [\kms] &  \\
\hline \noalign {\smallskip}
Band 6	& SPW0 & 216.200  & 234 & 244 & 0.34 &  \\ 
    	& SPW1 & 217.150  & 234 & 282 & 0.39  & SiO(5-4)  \\ 
    	& SPW2 & 219.945  & 117 & 282 & 0.38  & SO(6-5)  \\ 
    	& SPW3 & 218.230  & 234 & 244 & 0.33  & H$_{2}$CO(3-2), O$^{13}$CS(18-17), HC$_{3}$N(24-23) \\
    	& SPW4 & 219.560  & 117 & 244 & 0.33  & C$^{18}$O(2-1) \\ 
    	& SPW5 & 230.530  & 469 & 969 & 1.3  & CO(2-1),  CH$_{3}$OH \\
    	& SPW6 & 231.280  & 469 & 488 & 0.63 & $^{13}$CS(5-4), N$_{2}$D$^{+}$(3-2), OCS(19-18), CH$_{3}$OH  \\ 
    	& SPW7 & 232.450  & 1875   & 1130 & 1.5 & H30$\alpha$ (RRL), CH$_{3}$OH  \\
    	\hline
Band 3	& SPW0 & 93.1734 & 117 & 71 & 0.23 & N$_{2}$H$^{+}$(1-0) \\ 
    	& SPW1 & 92.2000 & 938 & 564 & 1.8 & CH$_3$CN(5-4), H41$\alpha$ (RRL)\\
    	& SPW2 & 102.600 & 938 & 564 & 1.6 & CH$_3$CCH(6-5), CH$_3$OH, H$_2$CS(3-2)  \\
    	& SPW3 & 105.000 & 938 & 564 & 1.6 & CH$_{3}$OH \\
\hline \noalign {\smallskip}

\end{tabular}
\end{table*}

The continuum channels were selected from subsections of the observed bandpass.
We lay out the spectral coverage in Table \ref{tab:lines}.
The total bandwidth covered in B3 is 2.93 GHz and B6 is 3.75 GHz.
We created two different groups of measurement sets for continuum imaging:

\begin{enumerate}
    \item In the default (labeled \cleanest throughout this text, to indicate
        that it is the less line-contaminated of the two), we used
        the ALMA pipeline \texttt{find\_continuum} tool developed by Todd
        Hunter to reject line-contaminated channels.
        \texttt{find\_continuum} was run independently on each of the array configurations (7M-only, 12M-long, 12M-short), resulting in three \texttt{cont.dat} files that describe which parts
        of the spectrum are contaminated by lines and which are continuum-only.
    \begin{enumerate}
        \item We merged these continuum selections by union, counting a spectral
            region as continuum if it was identified as continuum in either of the
            12m configuration observations.
        \item We plotted the continuum selection over a variety of spectra
            extracted from the measurement set (the $uv$-averaged spectrum) and from
            an early version of the imaged full cubes (spatially averaged spectra).
        \item Based on the resulting plots, we removed several spectral regions that clearly contained line emission but were identified as continuum by the original script.
    \end{enumerate}
    \item In a second approach to averaging, we used all bandwidth  
    whether or not it was line-contaminated (labeled \bsens, short for ``best sensitivity'';
    Section \ref{sec:bsens}). 
    This data product should give the best continuum sensitivity in regions without line emission.
    These images are optimized for detection of faint sources.  
\end{enumerate}

\refreport{We explain in more detail the \cleanest approach.}
In the ALMA pipeline approach, described in Section 10.28 of the ALMA pipeline users guide
for CASA 5.6.1\footnote{\url{https://almascience.nrao.edu/documents-and-tools/alma-science-pipeline-users-guide-casa-5-6.1}}, 
a dirty cube is created, then the brightest region from the peak intensity map is spatially
selected.  \refreport{The region selection is done by applying a threshold to the moment-0 (integrated intensity) and another threshold on the peak intensity maps.  
The thresholds are determined based on automatic noise determination and a preselected set of heuristics.
The two masks are combined by union.
A more detailed description is expected in a forthcoming paper led by Todd Hunter.}
That region is averaged over to create a representative spectrum;
this spectrum is dominated by the emission of the brightest regions, which in our
data typically correspond to hot cores.  The line-containing regions are
then automatically identified based on a threshold and excluded.
\refreport{There are several additional steps in the contaminant-rejection process,
including handling of spectral edges and atmospheric emission features,
but we leave a full description of this process to the paper on this topic.}

\begin{figure}[htp]
    \includegraphics[width=0.5\textwidth]{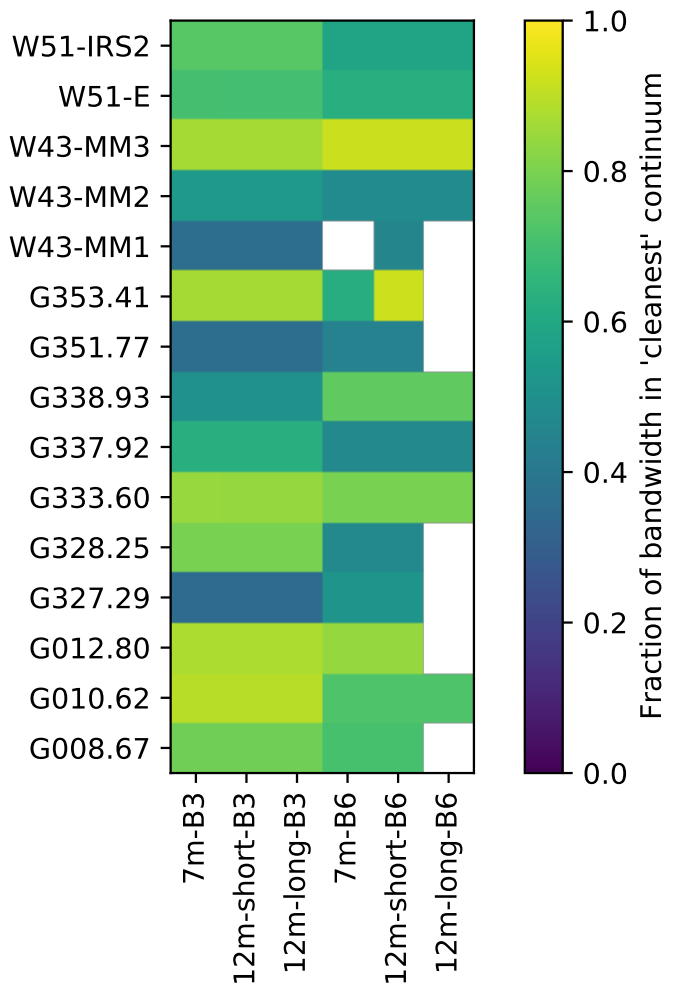}
    \caption{Graphical illustration of the fraction of the bandwidth included
    in each \cleanest continuum band for each field.  Empty (white) squares are
    those where no data were taken in ALMA-IMF, including fields that have only
    one 12m configuration and W43-MM1 B6, for which data were taken separately
    in the pilot program \citep[2013.1.01365.S;][]{Motte2018}.}
    \label{fig:contselfrac}
\end{figure}

\begin{figure*}[htp!]
    \includegraphics[width=0.9\textwidth]{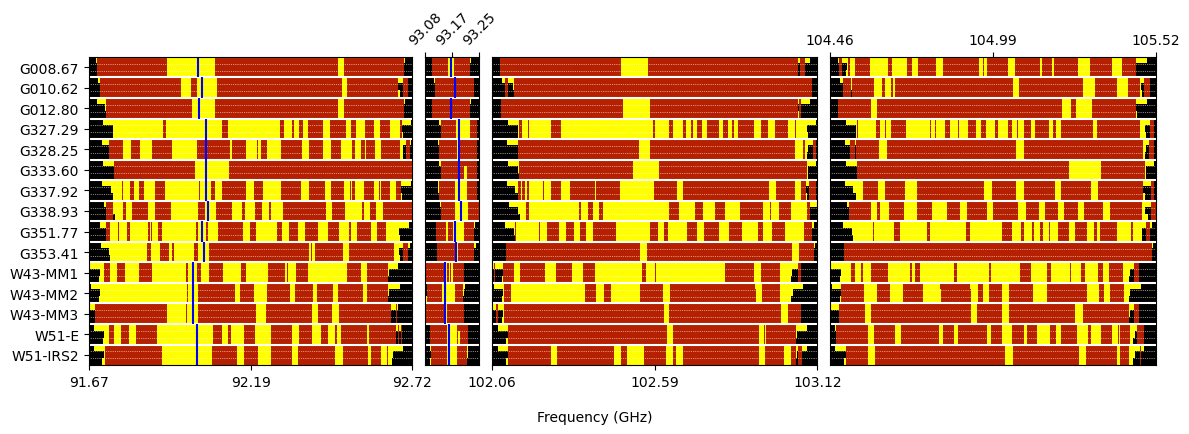}
    \includegraphics[width=0.9\textwidth]{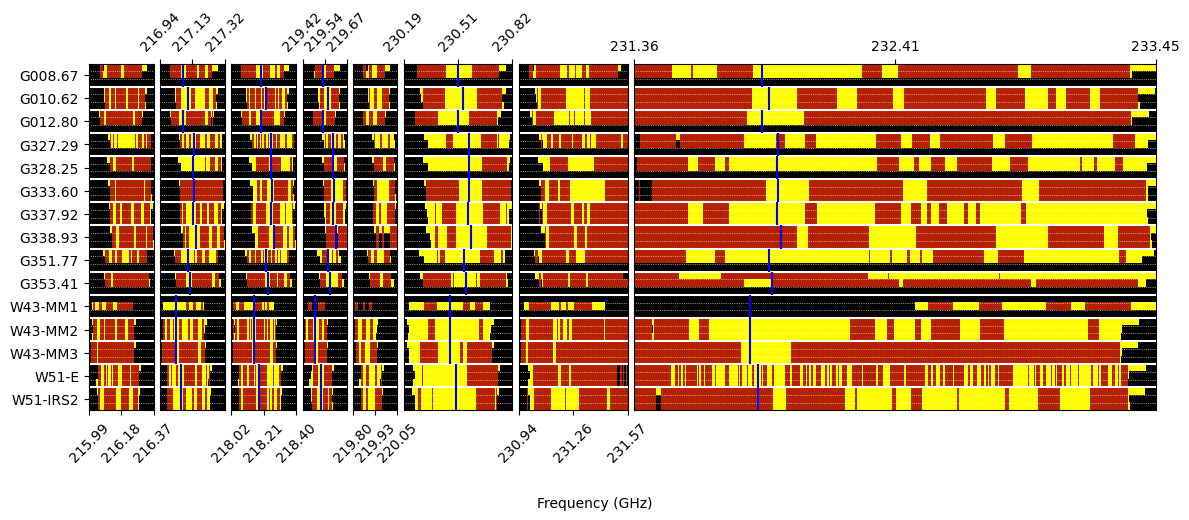}
    \caption{Continuum selection figures for band 3 (top) and band 6 (bottom).  For each
    field, there are up to three rows: the first is 7m, the second is the short-baseline configuration
    of the 12m array, and the third - when present - is the long-baseline configuration of the 12m array.
    Red shows data included in the continuum, yellow shows data excluded from the continuum, and
    black shows where no data were taken (for several fields, only one 12m configuration was used).
    The blue vertical lines show selected bright emission lines doppler shifted to the target velocity
    in these fields.  Selected lines are N$_2$H$^+$ 1-0, SiO 5-4, \formaldehyde \threeohthree, $^{12}$CO 2-1,
    H30$\alpha$, H41$\alpha$, and C$^{18}$O 2-1.  The X-axis shows frequency in the kinematic
    local standard of rest (LSRK) frame.
    W43-MM1 B6 was observed with a slightly different frequency setup; its spectral coverage continues beyond the right edge of the plot.
        }
    \label{fig:contsel}
\end{figure*}

While this approach is generally as good as can be done in a reasonably automatic way,
in regions like those targeted that contain line-rich hot cores, it is imperfect \refreport{(though more recent versions of the pipeline apparently perform well on hot cores too; Todd Hunter, private communication)}.
We therefore inspected the spectra created from line cubes at varying levels of reduction
(some were moderately well-cleaned, others were dirty) and modified the
continuum selection based on those cubes where needed.
\refreport{We inspected both the peak intensity spectrum and the mean spectrum (i.e., the spectrum created by taking the maximum value from each channel and the average value of each channel in the \texttt{image} cube, respectively).  We expanded or contracted the continuum regions based on a by-eye assessment of whether there was substantial line contamination.}
Figure \ref{fig:contselfrac} shows the fraction of continuum included in each spectral window
for each field.
Figure \ref{fig:contsel} shows the continuum selection for each field and for each
array configuration.  While there are similarities between each
target field, the continuum selection is not uniform.

Several data packages in the archive do not include the \texttt{findcont} step in the
calibration files because they were re-imaged by the archive to account for a mosaicing bug.
We have restored their \texttt{cont.dat} files from the original weblogs and included them in the data
reduction repository.

The effective central frequencies for a range of assumed spectral indices $\alpha$, where $I_\nu \propto \nu^\alpha$, are given in detail in Appendix \ref{sec:centralfreq}.
In brief, $\nu_{3mm}\approx 100$ GHz and $\nu_{1mm}\approx 228$ GHz, with variations up to $\sim2$ GHz.
The central frequencies are calculated as the intensity-weighted average frequency
\begin{equation}
\nu_{eff} = \frac{\int \nu I_\nu d\nu}{\int I_\nu d\nu} = \frac{\int \nu^{\alpha + 1} d\nu}{\int \nu^{\alpha} d\nu},
\end{equation} where the bounds of integration and $d\nu$ are computed for each band included in the continuum image.
We assume the sensitivity per unit frequency is constant across each spectral window.
The \cleanest images have different spectral coverage than the \bsens images and therefore have different
central frequency as a function of $\alpha$.

\subsubsection{Largest angular scale}
Interferometers are not sensitive to all angular scales on the sky.
Like single-dish, filled-aperture telescopes, they are limited in the smallest
measurable size scale by the aperture diameter or longest baseline length.
We reported the smallest measurable angular size scale, the synthesized beam,
in Table \ref{tab:selfcal}.  The reported sizes correspond to a 
two-dimensional Gaussian beam.

The largest angular scale recoverable in an image is similarly limited by sampling
in the Fourier ($uv$) domain.  However, unlike the conventional beam size, there
is no agreed upon standard for describing the largest recovered angular scale.
The \texttt{CLEAN} algorithm, by adding spatial model components with power at
all angular scales into the final images, breaks the simple assumption that there
is a trivial largest-angular-scale cutoff above which no flux is recovered.
Instead, the final images contain flux on large angular scales, including a net
``direct current'' (DC) component, even though the interferometer did not directly
measure flux on these scales.  On the largest scales that are measured, though,
different weighting schemes can dramatically change how much flux is present;
the Briggs weighting adopted in this work, with robust=0, down-weights the largest
angular scales in favor of producing a smaller resolution element.  Additionally,
the observations were performed as mosaics,
which can recover more flux on large angular scales than single-pointing interferometric
images.

\begin{figure*}
    \centering
    \includegraphics[width=\textwidth]{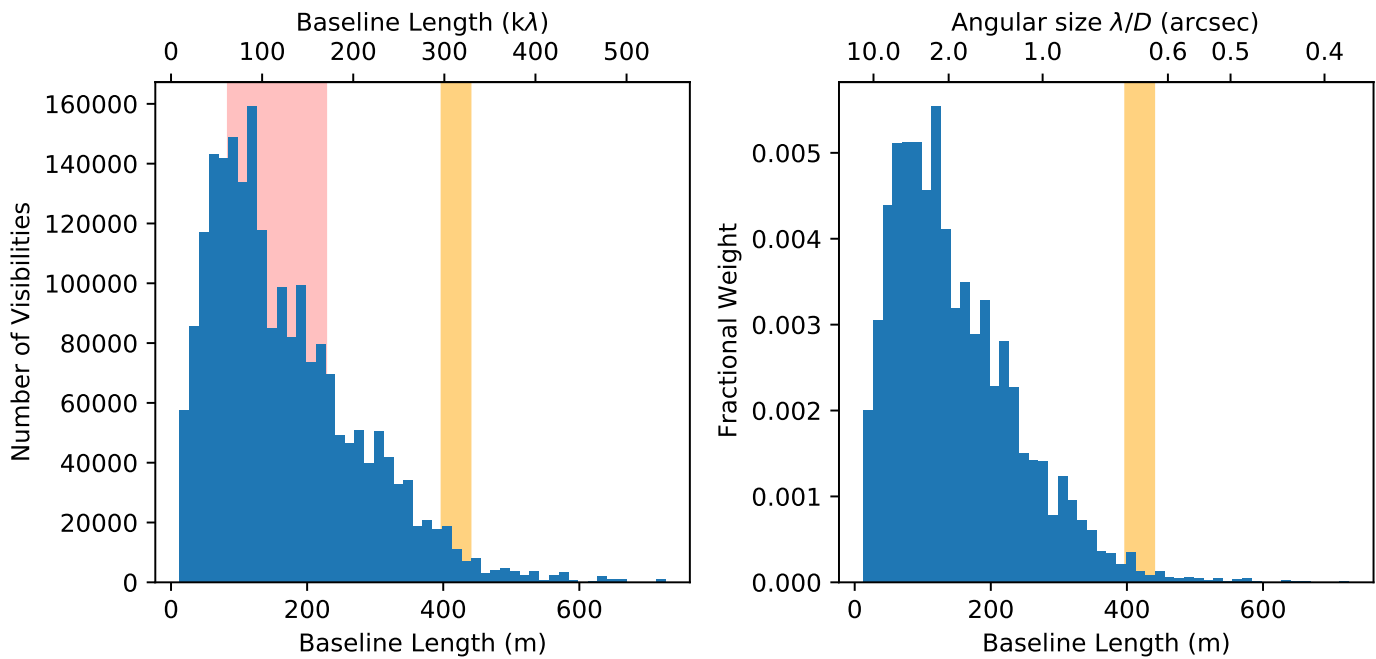}
    \caption{Histograms of the $uv$ sampling of the observations of G327.29 in
    Band 6.  The \emph{left} histogram shows the number of visibilities as a function of baseline length.
    In this panel, the top axis indicates the baseline length in units of kilolambda, that is, thousands
    of wavelengths.
    The red highlighted region shows the 25th-75th percentile of baseline
    lengths: half of the data are in this range, illustrating that scales
    $\lesssim5\arcsec$ are well-covered in this data set (the peak of the histogram
    is near $\sim5\arcsec$).
    The \emph{right} histogram shows the fractional weight in the visibilities as a function
    of baseline length; the similarity of the left and right panels shows that
    the visibility weighting does not substantially deviate from uniformity.
    In this panel, the top axis indicates the corresponding angular size scale inferred
    from the equation $\theta = \lambda / B$, where $\lambda$ is the observed wavelength
    and $B$ is the baseline length.
    Note that the weights are the per-visibility weights derived from 
    the measurement calibration process; the final weights used for gridding are modified by
    the \texttt{CLEAN} algorithm gridding.
    In both panels, the orange highlighted region covers the range from the beam major
    to minor axis.
    }
    \label{fig:g327b6_uvhistogram}
\end{figure*}

\begin{figure*}
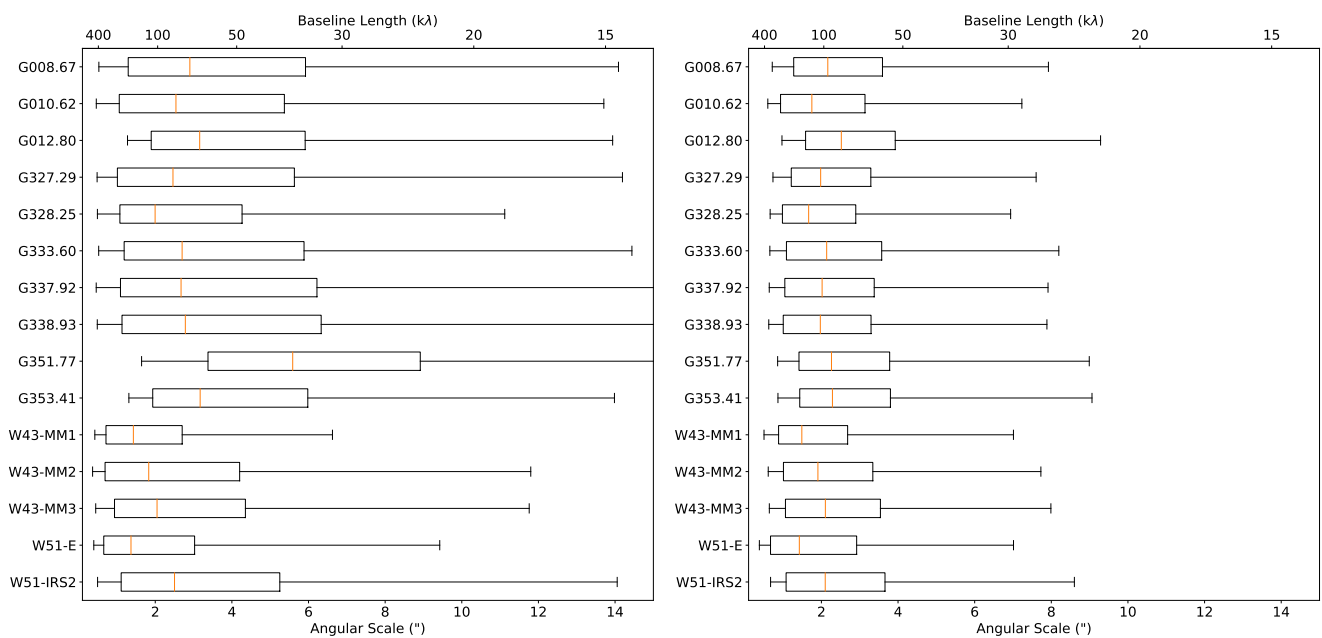

    \centering
    \includegraphics[width=0.47\textwidth]{f11}
    \includegraphics[width=0.47\textwidth]{f12}
    \caption{Summary of observed, but un-weighted, $uv$ spacing in all data sets. B3 is left, B6 is right.  Each row shows a box plot in which the ``whiskers'' at either end show the 5th and 95th percentile, the box ends show the 25th and 75th percentile, and the orange line shows the median of baseline length or angular scale.
    More detailed histograms are given in Figure \ref{fig:g327b6_uvhistogram}, demonstrating that the weighted and unweighted baseline lengths are similar.
    A similar overview figure scaled to the distance of the individual sources is shown in Figure \ref{fig:uvdistribution_phys}.
    }
    \label{fig:uvdistribution}
\end{figure*}

\begin{figure*}
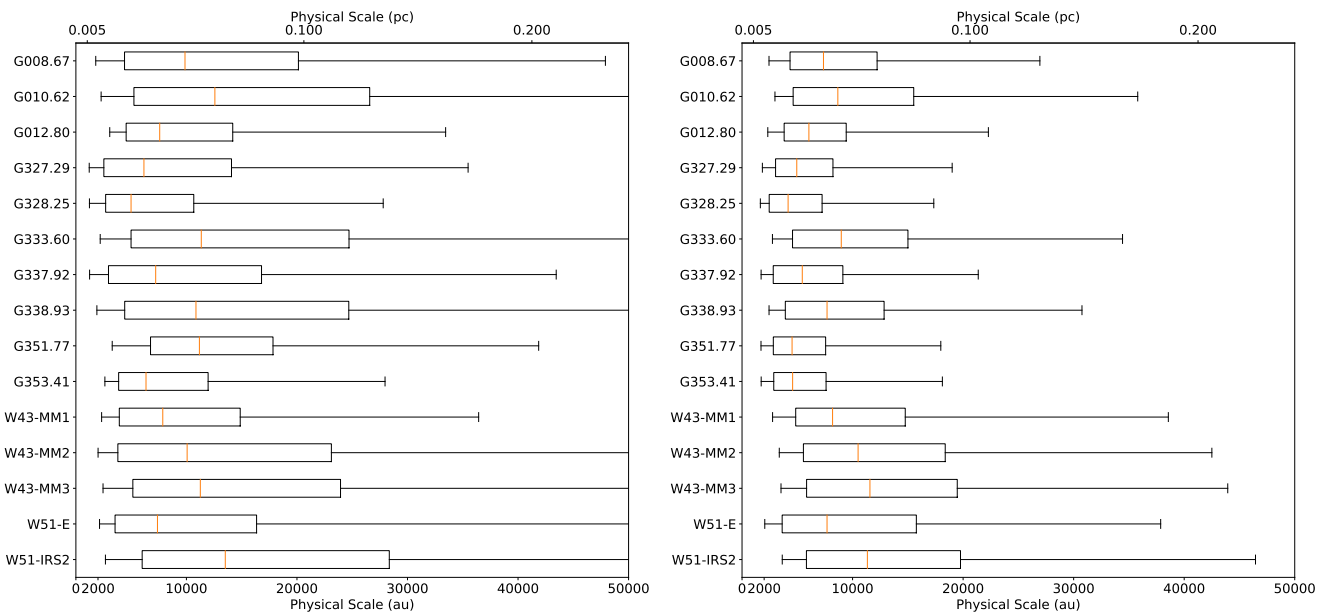

    \centering
    \includegraphics[width=0.47\textwidth]{f13}
    \includegraphics[width=0.47\textwidth]{f14}
    \caption{Summary of observed, but un-weighted, $uv$ spacing in all data sets
    scaled to the source distance.  B3 is left, B6 is right.  Unlike Figure
    \ref{fig:uvdistribution}, the scales are shown as physical scales.
    }
    \label{fig:uvdistribution_phys}
\end{figure*}

We therefore do not report a single largest angular scale.  Instead, we provide
histograms of the baseline length.  Figure \ref{fig:g327b6_uvhistogram} shows
G327.29 as an example; the remainder are provided as a digital supplement\footnote{The file \texttt{combined\_uvhistograms.pdf}.}.
The histogram illustrates that most baselines are relatively short, densely packed
around 100-200m; this pattern holds for all observations.
Both the number of visibilities and the histogram of the visibility weights are shown
to demonstrate that the weighting prior to imaging does not affect the $uv$ coverage.
The synthesized beam size generally corresponds to the $\gtrsim95$ percentile
of baseline lengths.  
An overview of all of the baseline lengths is given in Figure \ref{fig:uvdistribution}.
While the angular resolution in the B3 and B6 data sets of the same region are
generally the same, the largest angular scale recovery may be substantially
different.
The baseline lengths are transformed to physical scale in Figure \ref{fig:uvdistribution_phys},
emphasizing that the smallest scale probed is similar between regions, but the largest
may vary substantially.

\subsubsection{Splitting}
\label{sec:splitting}

To create the measurement sets to be used for continuum imaging and
self-calibration, we identified the line channels (see Section
\ref{sec:continuumselection}), flagged them out, then ran the
\texttt{split} CASA command to average the data spectrally.  The spectral averaging
widths were selected to keep bandwidth smearing to $<2\%$ based on VLA
guidelines\footnote{\url{https://science.nrao.edu/facilities/vla/docs/manuals/oss2016A/performance/fov/bw-smearing}}.
In band 6, this requirement is a channel width $\Delta\nu<0.5$ GHz, while at
band 3, it requires $\Delta\nu<76$ MHz, for a beam size of $0.3\arcsec$.  In
some cases, this would have allowed us to average down the entire spectral
window into a single channel, but instead we opted to have a minimum of two
channels per spectral window.

After each individual scheduling block (SB) had
its continuum split out, the flags were restored to their original state.  The
split continuum data were then concatenated into single merged measurement
sets for further processing\footnote{In principle, the split and concatenate
process is simply a matter of bookkeeping that should have no effect on the eventual data,
but in practice, the internal handling of concatenated and non-concatenated data sets
within CASA can have substantial effect.  For example, we found that, if one 
attempts to image any data selected from a concatenated data set that includes
7m antennae, the primary beam will be based on the 7m antenna, even if the
selection includes only 12m antennae. }. 
The \texttt{bsens} data were split in the same way as the \cleanest,
but no flagging beyond the original ALMA calibration pipeline's flagging was performed.

\subsubsection{Cleaning and Imaging}
\label{sec:imaging}
We jointly image the multiple 12-m configurations using the \texttt{tclean} task in CASA.  This process is a
straightforward joint cleaning of multiple measurement sets (MSes) that have
been concatenated.

The ALMA-IMF pipeline uses a simple set of heuristics to identify the mosaic
center and pixel scale. The mosaic center (\texttt{phasecenter} in
\texttt{tclean}) is set to be the mean position of all individual pointings.
The pixel scale is set to $dx=\nicefrac{1.22 \lambda}{4 B_{max}}$, where
$B_{max}$ is the longest baseline in the MS\footnote{This approach can result
in unnecessarily small pixels when there is an unusually
distant baseline included in the MS, but this was never a severe issue in the
ALMA-IMF data.  The choice of 1.22 is arbitrary; it comes from the Rayleigh
resolution criterion for a circular filled aperture, which does not necessarily
apply to the non-circular synthesized aperture, but this arbitrary scaling of
order unity has only a small effect on the resulting images.},
that is, we chose to sample the expected synthesized beam minor axis FWHM
with four pixels.

The image size is set to cover the full area of the mosaic.
The extrema of the image are found in RA and Dec by identifying the pointing
centers of each of the mosaic pointings, then going out further from the phase
center by one primary beam full width half maximum (FWHM), which provides
padding around the image edge.  The CASA \texttt{synthesisutils} tool
\texttt{getOptimumSize} is then used to round the image size up to a value
that is best suited to FFTs (i.e., a number whose prime factors are 2, 3,
and 5).  The code used to obtain these heuristics was based on Todd Hunter's
\texttt{analysisUtils}
package\footnote{\url{https://safe.nrao.edu/wiki/bin/view/Main/CasaExtensions}}.

\paragraph{Masking:}
We created custom clean masks for each field and each band.  Two types of clean
mask were used: hand-drawn polygonal regions and local-threshold-based regions.

The local-threshold regions are created in the following process:
\begin{enumerate}
    \item A first-pass image is created; in the first pass, this is a dirty image,
        in later passes, it is a cleaned image.
    \item A hand-drawn \texttt{ds9} region, generally a circle, rectangle, or other polygon, is placed on the
    image encompassing a region containing emission that is to be included in the cleaning.
    \item A threshold in Janskys per beam is selected for that region by the user. 
        This threshold is specified in the \texttt{text} attribute of the \texttt{ds9}
        region file.
    \item A boolean mask image is created including only pixels above the
    threshold in the hand-drawn region.
    \item The steps above are repeated for each hand-drawn region.
    \item The individual masks are combined by union; that is, any pixel included in any
    of the masks is included in the final mask.
\end{enumerate}

The hand-drawn polygonal ``clean boxes'' were made simply using CASA CRTF
regions.  The choice of threshold-based or hand-drawn regions was left to the
individual team member performing the data processing.  No differences in the
final product are expected from choosing one approach over the other, as both
approaches are adequate to ensure that clean model components are only added to
regions expected to contain signal during the self-calibration process.

For each target field and each observing band, at least one, but sometimes
several, masks were created in this fashion.  In the multiple-iteration
self-calibration, different masks were needed for each iteration, with
subsequent iterations including a larger area.
The final cleaning is done over a more inclusive area.

The regions used for each field and each iteration of self-calibration are
distributed in the github repository \footnote{\url{https://github.com/ALMA-IMF/reduction/tree/master/reduction/clean_regions}}.

\begin{figure}
    \centering
    \includegraphics[width=0.5\textwidth]{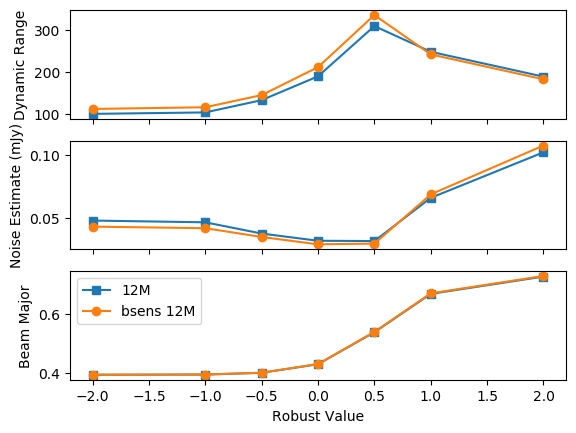}
    \caption{
    (top) Estimated dynamic range (peak signal divided by noise estimate)
    as a function of Briggs robust parameter for the W43-MM3 B3 mosaic.
    (middle) Estimate of the noise as a function of Briggs robust parameter.
    (bottom) Beam major axis FWHM in arcseconds as a function of the Briggs robust parameter.
    In all three figures,
    the lines show the \cleanest (blue squares) and ``best sensitivity'' (\bsens; orange circles) data.
    The noise is estimated in a relatively signal-free region selected from the robust=0 maps; the rise in noise to higher robust values is partly or entirely caused by added large-scale signal in these regions.
    }
    \label{fig:noise_vs_robust}
\end{figure}

\paragraph{Visibility weighting:}
We created test images with Briggs weighting and a range of robust parameters, which control
the relative weighting of long and short baselines from $-$2 to +2.
Smaller (more negative) values of the robust parameter result in smaller synthesized
beams, while larger values result in larger beams but, potentially, greater sensitivity.
However, we found that, for the majority of our fields, there was a minimum in the noise
at robust $\sim$0-0.5 (description of the noise estimation method is given in
Section \ref{sec:noiseestimation}).
While larger robust values should result in lower thermal noise levels, the
greater observed noise is most likely caused by un-modeled, mostly resolved-out,
large-angular-scale structure that contributes to noise on larger scales.
A representative example is shown in Figure \ref{fig:noise_vs_robust}.
Figures showing the noise as a function of robust parameter for each field
are presented as a supplementary product\footnote{File \texttt{combined\_noise\_and\_beams\_vs\_robust.pdf},
see Appendix~\ref{appendix:suppfigures}.}.
Since we found that robust=0 provided the best compromise between resolution
and sensitivity, we adopted it for our continuum images.
All data products presented in this work use robust=0 unless noted otherwise.

\paragraph{Deconvolution:}
We deconvolved the image using the \texttt{tclean} method in CASA
\citep{McMullin2007}.  Since we expect the continuum emission to exhibit a
large spatial dynamic range and some spectral structure, we used the
multi-scale multi-frequency synthesis (MTMFS)  method described by
\citet{Rau2011}.  In all cases, we used two Taylor terms, representing a
constant (\texttt{tt0}) and a term that encodes the spectral index
(\texttt{tt1} = $\alpha$ \texttt{tt0}).  The MTMFS method allows us to recover
a large flux dynamic range in the images that a single-frequency clean would
not be able to achieve \citep{Rau2011}.  Multi-scale cleaning was used,
generally with 3-4 independent scales following a geometric series; the default
choice was scales \texttt{[0, 3, 9, 27]}, corresponding to point sources and
several larger scales.  The resulting images were found to depend only weakly
on the choice of scales, so these defaults were only modified in cases where
the cleaning process failed to converge.

\subsubsection{Self-calibration}
\label{sec:selfcal}
The ALMA-pipeline products delivered by the observatory generally suffer from
dynamic range limitations when bright sources are in the field of view. 
The dynamic range limitations, resulting from bright ($>100$ mJy) sources
and extended structures, create artifacts and excessive negative features and add noise. 
For several fields, we determined that self-calibration was necessary to achieve
our requested sensitivity (see also Sect.\,\ref{sec:selfcalcompare}).

Self-calibration was attempted on all fields for both the \cleanest and \bsens data.
The self-calibration procedure here follows suggestions of \citet{Brogan2018}.
We iteratively image, calibrate, and reimage each field for 2-9 iterations.
Early iterations use conservative clean masks: they select only bright regions
that appear to have been imaged successfully.
\refreport{The first iteration always used \texttt{solint=`inf'}, the maximum solution interval.  Over the course of several iterations, the solution interval was progressively decreased for some fields when adequate solutions were obtained; for some fields, the solution interval was not decreased.  Further details are given in Appendix \ref{sec:selfcaldetails}.}
The self-calibration was applied with \texttt{applymode=`calonly'} and
\texttt{calwt=False} such that no data are thrown out and data are not
re-weighted during self-calibration; this approach was adopted as the most
conservative, since iteratively changing the data could have surprising
results.  In some fields, the total integrated
flux is dominated by compact sources, which are easily selected and included in masks,
while in others, extended emission dominated the recovered flux, requiring a more inclusive
mask to obtain good calibration solutions.
The clean masks were expanded and included more total flux in progressive iterations.
For the majority of fields, we used phase-only self-calibration (but see the amplitude self-calibration paragraph below).

The self-calibration parameters are publicly  available\footnote{\url{https://github.com/ALMA-IMF/reduction/blob/master/reduction/imaging_parameters.py}} 
along with the corresponding imaging parameters.  \refreport{They are also summarized in Table \ref{tab:selfcaldetails}.}
Most self-calibration solutions were obtained by averaging both polarizations
(\texttt{gaintype=`T'}), but in some high S/N cases single-polarization was used
(\texttt{gaintype=`G'}).
The single-polarization self-calibration shows no obvious benefit
over polarization-averaged self-calibration, however.

Because our data were taken as mosaics, some pointings in each target region include no bright sources. 
Within each target region, we selected mosaic
pointings to use for calibration only if they passed two criteria:
\begin{enumerate}
    \item The mean signal-to-noise ratio in the self-calibration solutions was
        at least $\left<SNR\right> > 5$
    \item The standard deviation of the phase solutions was $\sigma_{\theta} <
        \pi / 4$.  This choice of threshold is arbitrary, but means that, assuming the phase solutions are Gaussian distributed, phase wraps - phase solutions with $\Delta\theta>\pi$ - will be 4-$\sigma$ events, happening in $<0.01\%$ of solutions, and therefore adding negligibly to noise.
\end{enumerate}
These criteria exclude pointings within the mosaic that have too little flux to
achieve a high-quality solution.  The phase solutions obtained from the high
signal-to-noise fields, which were generally the central several pointings,
were then applied to all scans and pointings in the observation.  This
field-specific mosaic self-calibration has been applied in \citet{Ginsburg2018} 
and has the advantage of including the signal from many mosaic pointings in the
model creation but excluding solutions that may worsen the overall
calibration\footnote{In one case, G328.25, we were only able to obtain
solutions for the central field by manually selecting it; in this case, the
overall emission is weak anyway,
and the improvement from self-calibration was minimal.}.

\refreport{The adopted approach has two theoretical advantages:
the calibration solutions are obtained closer on the sky and closer in time to the
data.  Separations between the source and the calibrator ranged from 1--14$^\circ$,
while separations between phase calibrator observations were $\sim10$ minutes.
Self-calibrating based on fields in the mosaic always reduced the on-sky separation to
$<2 \arcmin$ and usually reduced the time difference to $<5$ minutes.
For the B3 observations, each mosaic pointing was included at least once in every
cycle between quasar phase-calibrator observations, and so the time interval was always
decreased.  For the B6 observations, however, the larger mosaics were only about half
covered during each inter-phase-calibrator cycle.  Therefore, it was possible to have a
phase solution from self-calibration be further away in time than the phase calibrator
solution.  The most affected fields were G333.60 and G328.25, with about half of the fields
having longer time separations to the calibrator.  Since these fields
still showed improvement after self-calibration, the net effect of self-calibration
was positive.
}

Table \ref{tab:selfcal} summarizes the quantitative improvement produced by
self-calibration and the noise levels achieved.  This comparison is further
discussed in Section \ref{sec:selfcalcompare}.

\begin{table*}[htp]
\caption{Selfcal Summary}
\resizebox{\textwidth}{!}{
\begin{tabular}{lllllllllllllll}
\label{tab:selfcal}
Region & Band & $n_{sc}$ & $\theta_{maj}$ & $\theta_{min}$ & BPA & $\theta_{req}$ & $\Omega_{syn}^{1/2}/\Omega_{req}^{1/2}$ & $S_{peak}$ & $\sigma_{MAD}$ & $\sigma_{req}$ & $\sigma_{MAD}/\sigma_{req}$ & DR$_{pre}$ & DR$_{post}$ & DR$_{post}$/DR$_{pre}$ \\
 &  &  & $\mathrm{{}^{\prime\prime}}$ & $\mathrm{{}^{\prime\prime}}$ &  & $\mathrm{{}^{\prime\prime}}$ &  & $\mathrm{mJy\,beam^{-1}}$ & $\mathrm{mJy\,beam^{-1}}$ & $\mathrm{mJy\,beam^{-1}}$ &  &  &  &  \\
\hline
G008.67 & B3 & 5 & 0.62 & 0.43 & 58 & 0.67 & 0.77 & 99 & 0.14 & 0.09 & 1.6 & 300 & 700 & 2.3 \\
G008.67 & B6 & 5 & 0.73 & 0.60 & -84 & 0.67 & 0.99 & 210 & 0.37 & 0.3 & 1.2 & 440 & 520 & 1.2 \\
G010.62 & B3 & 9a & 0.40 & 0.33 & -78 & 0.37 & 0.98 & 290 & 0.059 & 0.03 & 2.0 & 860 & 4700 & 5.4 \\
G010.62 & B6 & 5 & 0.53 & 0.41 & -78 & 0.37 & 1.3 & 380 & 0.084 & 0.1 & 0.84 & 2600 & 3200 & 1.2 \\
G012.80 & B3 & 7a & 1.4 & 1.2 & 88 & 0.95 & 1.4 & 830 & 0.24 & 0.18 & 1.3 & 1300 & 3400 & 2.6 \\
G012.80 & B6 & 6a & 1.1 & 0.70 & 75 & 0.95 & 0.92 & 400 & 0.35 & 0.6 & 0.58 & 690 & 720 & 1.0 \\
G327.29 & B3 & 2 & 0.43 & 0.37 & 70 & 0.67 & 0.59 & 24 & 0.13 & 0.09 & 1.5 & 170 & 170 & 1.0 \\
G327.29 & B6 & 5 & 0.69 & 0.63 & -41 & 0.67 & 0.99 & 830 & 0.36 & 0.3 & 1.2 & 1200 & 1800 & 1.4 \\
G328.25 & B3 & 4 & 0.60 & 0.43 & -82 & 0.67 & 0.76 & 12 & 0.091 & 0.09 & 1.0 & 110 & 130 & 1.2 \\
G328.25 & B6 & 4 & 0.62 & 0.47 & -11 & 0.67 & 0.81 & 150 & 0.37 & 0.3 & 1.2 & 330 & 400 & 1.2 \\
G333.60 & B3 & 6a & 0.47 & 0.45 & 39 & 0.51 & 0.89 & 220 & 0.090 & 0.06 & 1.5 & 920 & 1700 & 1.8 \\
G333.60 & B6 & 6a & 0.59 & 0.52 & -33 & 0.51 & 1.1 & 240 & 0.11 & 0.2 & 0.57 & 1300 & 1300 & 1.1 \\
G337.92 & B3 & 4 & 0.39 & 0.35 & 75 & 0.51 & 0.73 & 17 & 0.056 & 0.06 & 0.94 & 220 & 240 & 1.1 \\
G337.92 & B6 & 4 & 0.61 & 0.48 & -56 & 0.51 & 1.1 & 280 & 0.22 & 0.2 & 1.1 & 1200 & 1400 & 1.2 \\
G338.93 & B3 & 3 & 0.43 & 0.42 & 17 & 0.51 & 0.83 & 11 & 0.071 & 0.06 & 1.2 & 140 & 150 & 1.0 \\
G338.93 & B6 & 6 & 0.56 & 0.51 & -85 & 0.51 & 1.1 & 150 & 0.17 & 0.2 & 0.87 & 490 & 670 & 1.4 \\
G351.77 & B3 & 4 & 1.5 & 1.3 & 89 & 0.95 & 1.5 & 86 & 0.25 & 0.18 & 1.4 & 330 & 340 & 1.0 \\
G351.77 & B6 & 4 & 0.89 & 0.67 & 87 & 0.95 & 0.81 & 540 & 0.42 & 0.6 & 0.69 & 1000 & 1100 & 1.1 \\
G353.41 & B3 & 6 & 1.3 & 1.1 & 76 & 0.95 & 1.3 & 170 & 0.16 & 0.18 & 0.89 & 860 & 910 & 1.1 \\
G353.41 & B6 & 6 & 0.94 & 0.67 & 85 & 0.95 & 0.83 & 110 & 0.32 & 0.6 & 0.54 & 270 & 280 & 1.0 \\
W43-MM1 & B3 & 4 & 0.53 & 0.31 & -74 & 0.37 & 1.1 & 14 & 0.044 & 0.03 & 1.5 & 220 & 310 & 1.4 \\
W43-MM1 & B6 & 4 & 0.51 & 0.36 & -77 & 0.37 & 1.1 & 360 & 0.10 & 0.1 & 1.0 & 2300 & 2700 & 1.2 \\
W43-MM2 & B3 & 4 & 0.30 & 0.24 & -73 & 0.37 & 0.74 & 3.6 & 0.037 & 0.03 & 1.2 & 110 & 94 & 0.82 \\
W43-MM2 & B6 & 5 & 0.52 & 0.41 & -75 & 0.37 & 1.3 & 150 & 0.12 & 0.1 & 1.2 & 1000 & 1100 & 1.1 \\
W43-MM3 & B3 & 5 & 0.43 & 0.29 & -85 & 0.37 & 0.95 & 6.0 & 0.032 & 0.03 & 1.1 & 180 & 190 & 1.0 \\
W43-MM3 & B6 & 5 & 0.53 & 0.45 & 89 & 0.37 & 1.3 & 56 & 0.072 & 0.1 & 0.72 & 630 & 650 & 1.0 \\
W51-E & B3 & 7 & 0.29 & 0.26 & 70 & 0.37 & 0.74 & 400 & 0.061 & 0.03 & 2.0 & 1200 & 4800 & 4.0 \\
W51-E & B6 & 7 & 0.34 & 0.27 & 26 & 0.37 & 0.82 & 400 & 0.19 & 0.1 & 1.9 & 740 & 1500 & 2.0 \\
W51-IRS2 & B3 & 4 & 0.29 & 0.27 & -60 & 0.37 & 0.75 & 79 & 0.062 & 0.03 & 2.1 & 560 & 1000 & 1.9 \\
W51-IRS2 & B6 & 9a & 0.51 & 0.44 & -26 & 0.37 & 1.3 & 880 & 0.095 & 0.1 & 0.95 & 2800 & 4500 & 1.6 \\
\hline
\end{tabular}
}\par
$n_{sc}$ is the number of self-calibration iterations adopted.  Those with a final iteration of amplitude self-calibration are denoted with the `a' suffix.  $\theta_{maj}, \theta_{min}$, and BPA give the major and minor full-width-half-maxima (FWHM) of the synthesized beams.  $\theta_{req}$ is the requested beam size, and $\Omega_{syn}^{1/2}/\Omega_{req}^{1/2}$ gives the ratio of the synthesized to the requested beam area; larger numbers imply poorer resolution.  $\sigma_{MAD}$ and $\sigma_{req}$ are the measured and requested RMS sensitivity, respectively, and $\sigma_{MAD}/\sigma_{req}$ is the excess noise in the image over that requested.  $\sigma_{MAD}$ is measured on the \texttt{cleanest} images.  $DR_{pre}$ and $DR_{post}$ are the dynamic range, $S_{peak} / \sigma_{MAD}$, for the pre- and post-self-calibration data; $DR_{post}/DR_{pre}$ gives the improvement factor.
\end{table*}

\begin{figure*}
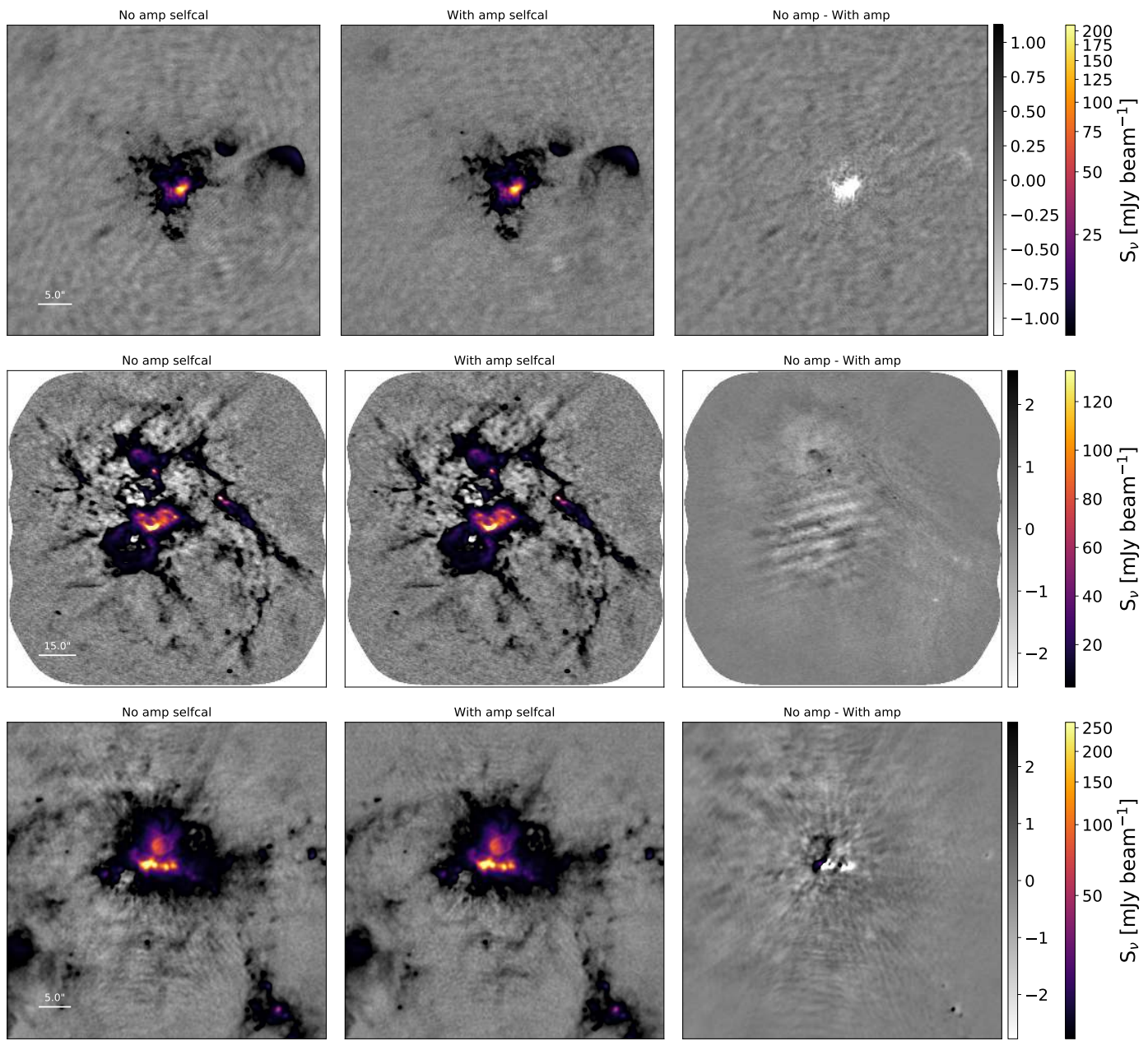

    \centering
    \includegraphics[width=\textwidth]{f16}
    \includegraphics[width=\textwidth]{f17}
    \includegraphics[width=\textwidth]{f18}
    \caption{Figures demonstrating the change before amplitude self-calibration (\emph{left})
    vs after amplitude self-calibration (\emph{middle}), with the difference of no-amplitude minus with-amplitude self calibration
    shown on the \emph{right} to highlight the differences.
    These three images, G010.62 B3 (top), G012.80 B6 (center), and W51-IRS2 B6 (bottom) showed the greatest
    structural difference and greatest noise improvement with amplitude self-calibration included.
    While the noise decreased, and structure moved, the sum of the flux and the peak intensity changed by $<2\%$
    in each case.
    The images use the \cleanest spectral selection.
    }
    \label{fig:ampselfcal_improvement}
\end{figure*}

\paragraph{Amplitude self-calibration}
\label{sec:ampselfcal}
We explored using amplitude self-calibration.
This approach is generally considered higher-risk, since it has the potential
to introduce systematic offsets in the calibration.
We therefore only adopted the amplitude self-calibrated images as the final
products after extensive analysis.
We performed several iterations of phase-only self-calibration, followed by a
deep clean, prior to performing amplitude self-calibration in order to ensure
that systematic errors are not introduced.
\refreport{Solutions were calculated with \texttt{solint=`inf'} to maximize signal-to-noise for amplitude self-calibration.}
There are two regions, G010.62 B3 and G012.80 B3, in which a single iteration of
amplitude self-calibration resulted in a very large noise reduction - 32\%
(G010.62) and 46\% (G012.80) improvement - and little or no change in the flux
in recovered objects.
W51-IRS2 B6 shows a 7\% reduction in noise (15\% for the \bsens images), but a
very substantial reduction in obvious artifacts, so we elect to use amplitude
self-calibration on this source.  
Similarly, G012.80 B6 shows a small (3\%) reduction in noise, but a substantial
qualitative improvement.
We note that the cores appear to brighten by $\sim1-2\%$ from amplitude self
calibration, which is negligible compared to the overall systematic calibration
uncertainties (which are assumed\footnote{\url{https://library.nrao.edu/public/memos/alma/main/memo599.pdf}} to be $\sim10\%$).
Examples of the improvement from amplitude self-calibration are shown in Figure \ref{fig:ampselfcal_improvement}.

\subsection{Best sensitivity images}
\label{sec:bsens}
In order to obtain the best possible continuum sensitivity, in addition
to the images created with line-contaminated channels flagged out,
we also created continuum images using all the available bandwidth.
This approach gives the best achievable sensitivity in those regions where
contamination from molecular lines is not severe.
These images were self-calibrated in the same way as the \cleanest
images, using the same cleaning parameters, masks, and thresholds.

The brightest sources in the field are generally line-rich, and therefore
suffer from substantial (and difficult to disentangle) line contribution.
For the brightest 1 mm continuum source in G351.77, for example, the peak
brightness changes by $\sim$30\% between the line-contaminated and
uncontaminated images, which is larger than the calibration uncertainty.
Most of the lines producing the contaminating emission are relatively
compact and hence confined to the surroundings of the brightest continuum
sources in the field; the complex organic molecules giving rise to this contamination
are discussed further in \paperone and Csengeri et al (in prep).
The exceptions are those regions with bright and broad CO outflows, such as W43-MM1
and G351.77; for such regions, a modified \bsens image, excluding the CO window,
may be more useful, and such products will be made available (see Appendix \ref{appendix:bsens_noco}).
The images with the best possible sensitivity are useful for direct continuum
measurements in the emission-poor regions of the maps, and they can be used as
boosted signal-to-noise ratio maps for source selection.

In general, we expect that the line-contaminated versions should have
higher observed brightness and poorer image quality.  Specifically,
we expect imaging artifacts to manifest as amplitude errors, since
the amplitude of the visibilities deviate from a smooth continuum.

There are intriguing features in the \bsens minus \cleanest images
that show locations with excess line emission.  These are most likely
hot cores, which have excess line emission throughout the spectrum,
HII regions, which have bright and broad recombination line emission,
or outflows, which again are spectrally broad but spatially compact.
Appendix \ref{appendix:selfcalcompare} 
shows comparisons between the \bsens and \cleanest images.

\begin{figure}[htp]
    \includegraphics[width=0.5\textwidth]{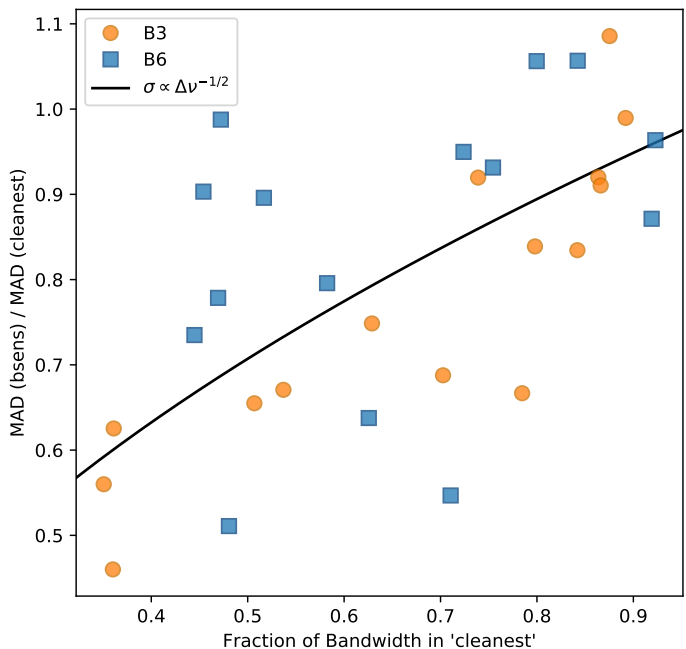}
    \caption{Comparison between the \bsens and \cleanest data sets.  The
    X-axis shows
    the fraction of the bandwidth used in the
    \cleanest continuum.  The Y-axis shows the ratio of the noise
    in \bsens vs that of \cleanest.  The data are the
    MAD-measured standard deviation in the noise measurement regions.
    The black curve shows the theoretical
    expectation that the noise goes down as the square root of the bandwidth,
    $\sigma \propto \Delta \nu^{-1/2}$.  Points above the curve have excess
    noise in the \bsens data, while points below improved more than expected.
    }
    \label{fig:bsens_improvement}
\end{figure}

Figure \ref{fig:bsens_improvement} shows the improvement in noise level
from the \cleanest to the \bsens images.
The improvement in the noise from \cleanest to \bsens is clear, and it is
correlated with the fraction of bandwidth included in measuring
the \cleanest continuum.  However, there is also substantial scatter, some of
which is accounted for by the line contamination added into the \bsens data:
the line forests in hot cores behave as higher-amplitude noise when
they are averaged into continuum visibilities.
We also observe that there are several fields in which the noise improved more
than expected based on the simple expectation that $\sigma \propto \Delta \nu^{1/2}$.
The \bsens self-calibration solutions achieved substantially higher signal-to-noise ratios, which may partly explain this phenomenon.

\begin{table*}[htp]
\caption{Best Sensitivity vs Cleanest Continuum comparison}
\resizebox{\textwidth}{!}{
\begin{tabular}{lllllllllll}
\label{tab:bsens_cleanest}
Region & Band & $\sigma_{MAD}$(bsens) & $\sigma_{MAD}$(cleanest) & $\frac{\sigma_{MAD}(\mathrm{bsens})}{\sigma_{MAD}(\mathrm{cleanest})}$ & $f_{BW,cleanest}$ & $S_{peak}$(bsens) & $S_{peak}$(cleanest) & $\frac{S_{peak}(\mathrm{bsens})}{S_{peak}(\mathrm{cleanest})}$ & Requested $\sigma$ & $\sigma_{\mathrm{bsens}}/\sigma_{\mathrm{req}}$ \\
 &  & $\mathrm{mJy\,beam^{-1}}$ & $\mathrm{mJy\,beam^{-1}}$ &  &  & $\mathrm{mJy\,beam^{-1}}$ & $\mathrm{mJy\,beam^{-1}}$ &  & $\mathrm{mJy\,beam^{-1}}$ &  \\
\hline
G008.67 & B3 & 0.093 & 0.14 & 0.67 & 0.78 & 94 & 99 & 0.95 & 0.090 & 1.0 \\
G008.67 & B6 & 0.20 & 0.37 & 0.55 & 0.71 & 210 & 210 & 1.0 & 0.30 & 0.68 \\
G010.62 & B3 & 0.059 & 0.059 & 0.99 & 0.89 & 290 & 290 & 1.0 & 0.030 & 2.0 \\
G010.62 & B6 & 0.081 & 0.085 & 0.95 & 0.72 & 380 & 380 & 1.0 & 0.10 & 0.81 \\
G012.80 & B3 & 0.24 & 0.22 & 1.1 & 0.88 & 860 & 830 & 1.0 & 0.18 & 1.3 \\
G012.80 & B6 & 0.78 & 0.74 & 1.1 & 0.84 & 400 & 400 & 1.0 & 0.60 & 1.3 \\
G327.29 & B3 & 0.073 & 0.13 & 0.56 & 0.35 & 32 & 24 & 1.4 & 0.090 & 0.82 \\
G327.29 & B6 & 0.32 & 0.36 & 0.90 & 0.52 & 950 & 830 & 1.1 & 0.30 & 1.1 \\
G328.25 & B3 & 0.076 & 0.091 & 0.84 & 0.80 & 15 & 12 & 1.2 & 0.090 & 0.85 \\
G328.25 & B6 & 0.29 & 0.37 & 0.78 & 0.47 & 200 & 150 & 1.3 & 0.30 & 0.97 \\
G333.60 & B3 & 0.075 & 0.090 & 0.83 & 0.84 & 220 & 220 & 1.0 & 0.060 & 1.2 \\
G333.60 & B6 & 0.12 & 0.11 & 1.1 & 0.80 & 230 & 240 & 0.94 & 0.20 & 0.61 \\
G337.92 & B3 & 0.049 & 0.065 & 0.75 & 0.63 & 19 & 17 & 1.1 & 0.060 & 0.81 \\
G337.92 & B6 & 0.22 & 0.22 & 0.99 & 0.47 & 340 & 280 & 1.2 & 0.20 & 1.1 \\
G338.93 & B3 & 0.047 & 0.071 & 0.65 & 0.51 & 13 & 11 & 1.2 & 0.060 & 0.78 \\
G338.93 & B6 & 0.16 & 0.17 & 0.93 & 0.75 & 160 & 150 & 1.1 & 0.20 & 0.81 \\
G351.77 & B3 & 0.12 & 0.25 & 0.46 & 0.36 & 110 & 86 & 1.2 & 0.18 & 0.65 \\
G351.77 & B6 & 0.31 & 0.42 & 0.73 & 0.44 & 660 & 540 & 1.2 & 0.60 & 0.51 \\
G353.41 & B3 & 0.17 & 0.19 & 0.92 & 0.86 & 190 & 170 & 1.1 & 0.18 & 0.96 \\
G353.41 & B6 & 0.40 & 0.42 & 0.96 & 0.92 & 110 & 110 & 1.0 & 0.60 & 0.67 \\
W43-MM1 & B3 & 0.031 & 0.049 & 0.63 & 0.36 & 18 & 14 & 1.3 & 0.030 & 1.0 \\
W43-MM1 & B6 & 0.18 & 0.20 & 0.90 & 0.45 & 370 & 360 & 1.0 & 0.10 & 1.8 \\
W43-MM2 & B3 & 0.026 & 0.038 & 0.67 & 0.54 & 5.0 & 3.6 & 1.4 & 0.030 & 0.85 \\
W43-MM2 & B6 & 0.062 & 0.12 & 0.51 & 0.48 & 200 & 150 & 1.3 & 0.10 & 0.62 \\
W43-MM3 & B3 & 0.029 & 0.032 & 0.91 & 0.87 & 6.1 & 6.0 & 1.0 & 0.030 & 0.96 \\
W43-MM3 & B6 & 0.062 & 0.072 & 0.87 & 0.92 & 57 & 56 & 1.0 & 0.10 & 0.62 \\
W51-E & B3 & 0.042 & 0.061 & 0.69 & 0.70 & 400 & 400 & 1.0 & 0.030 & 1.4 \\
W51-E & B6 & 0.12 & 0.19 & 0.64 & 0.63 & 380 & 400 & 0.93 & 0.10 & 1.2 \\
W51-IRS2 & B3 & 0.057 & 0.062 & 0.92 & 0.74 & 80 & 79 & 1.0 & 0.030 & 1.9 \\
W51-IRS2 & B6 & 0.075 & 0.095 & 0.80 & 0.58 & 910 & 880 & 1.0 & 0.10 & 0.75 \\
\hline
\end{tabular}
}\par
Like Table \ref{tab:selfcal}, but comparing the cleanest and bsens data.  $\sigma_{MAD}$(bsens) and $\sigma_{MAD}$(cleanest) are the standard deviation error estimates computed from a signal-free region in the map using the Median Absolute Deviation as a robust estimator.  Their ratio shows that the broader included bandwidth increases sensitivity; $f_{BW,cleanest}$ specifies the fraction of the total bandwidth that was incorporated into the cleanest images.  $S_{peak}$ is the peak intensity in the images.
\end{table*}

\subsection{Image quality assessment process}
\label{sec:qa}
Both the visibility data and the processed images went through extensive
quality assessment beyond that performed by the ALMA pipeline and data
reduction experts.

\begin{figure*}[htp]
    \includegraphics[width=\textwidth]{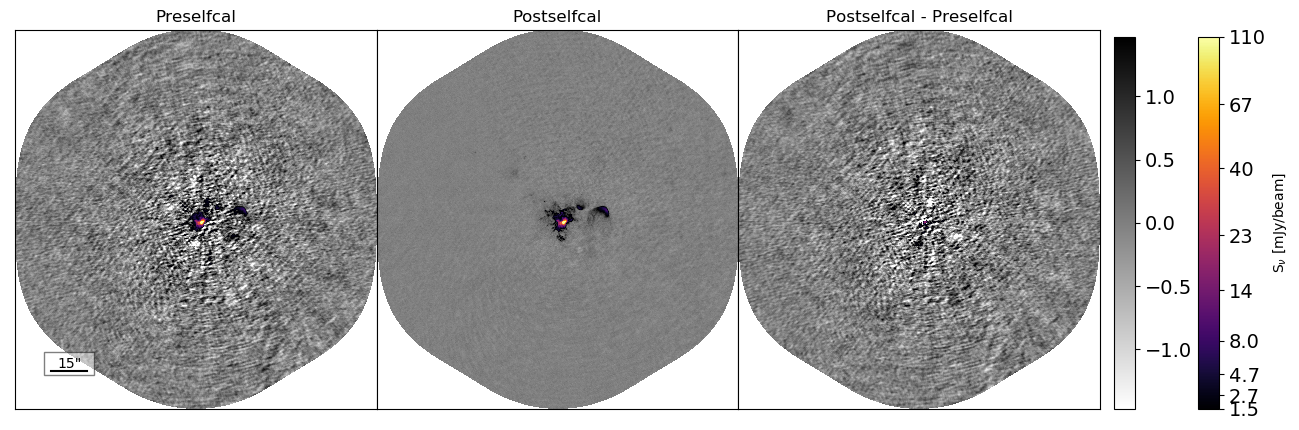}
    \caption{Comparison of the imaging results for the region G010.62 before (left) and after (middle)
    self-calibration.  The right panel shows the difference image,
    self-calibrated minus pre-self-calibrated.   The images being compared
    use the same model, so they are cleaned to the same depth. }
    \label{fig:comparison1}
\end{figure*}

To assess the imaging and self-calibration, we created a set of pre- and
post-self-calibration images and displayed them in a form similar to Figure
\ref{fig:comparison1}.
A web form was created to
display each image and allow feedback on the general image quality.  The web
form data were fed in to a common spreadsheet.  Each delivered image was inspected by
5-10 members of the ALMA-IMF team, noting any data artifacts or clear problems
and reporting them back to the data reduction team for further processing.
These QA comments were passed to the individual responsible for imaging the
data and were corrected if possible.

We present further analysis of the final data products here.  Summary
statistics are given in Tables \ref{tab:selfcal} and \ref{tab:bsens_cleanest}.

\subsubsection{Noise estimation}
\label{sec:noiseestimation}
To measure the noise in each field, we used a median-absolute-deviation (MAD)
estimator of the standard deviation, since the MAD is robust to small numbers
of outliers\footnote{We scaled the MAD by 1.4826 such that the reported value is equivalent to the standard deviation if the underlying data are normally distributed.}.
However, even with this approach, many of the target fields
are dominated by signal, so empirical estimation of the noise level is not
trivial.
We have identified the regions in each of the maps with
little or no signal and estimated the noise from the selected sub-images;
the regions used are available from the reduction
repository\footnote{\url{https://github.com/ALMA-IMF/reduction/tree/master/reduction/noise_estimation_regions}}.
The true noise level in the maps is variable, as it depends on the strength
of the bright sources and the degree to which our cleaning and self-calibration
removed sidelobes in the vicinity of these sources, so we focus our
analysis on the minimum noise level in the maps.
We estimate the noise levels from the cleaned images uncorrected by the primary beam
response pattern; the noise in the primary beam corrected data is
higher and non-uniform throughout the images.

Figure \ref{fig:noisehistogram} shows histograms of the B3 and B6 data for one
field, G351.77, with a Gaussian profile overlaid to illustrate the measured
noise.  This field is a typical case, where the Gaussian captures most
of the histogram, but not all.  As with all fields, there is a substantial positive
tail from the detected signal.
Histograms for the other fields are available in an online
supplement\protect\footnote{File \texttt{combined\_flux\_histograms.pdf}; see Appendix \ref{appendix:suppfigures}.}. 

\begin{figure}
    \centering
    \includegraphics[width=0.5\textwidth]{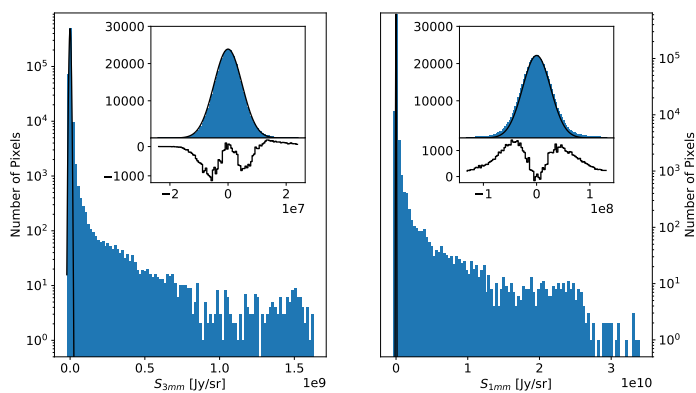}
    \caption{Histograms of the G351.77 \cleanest self-calibrated continuum data,
    which illustrate the noise distribution.  The left panel shows B3, the right B6.
    The inset shows a linearly-scaled zoom-in to the $\pm5\sigma$ region centered around
    zero, with the residual of the histogram minus the noise model shown below.
    The noise model is a Gaussian with width calculated from the median absolute deviation estimated from a signal-free area
    of the field as described in \ref{sec:noiseestimation}.
    }
    \label{fig:noisehistogram}
\end{figure}

\subsubsection{Pre-selfcal to post-selfcal comparison}
\label{sec:selfcalcompare}
We compare the cleaned images before and after self-calibration to determine
how much self-calibration improved the data.

To ensure a fair comparison between the images, they needed to have the same degree
of cleaning.  The final self-calibrated images were restored with deeply-cleaned
models that represent our best estimate of the sky brightness.
We used these final models to create images with the un-self-calibrated visibility
data, which are distributed as files with the suffix \texttt{preselfcal\_finalmodel}.
We use these most highly-cleaned images to measure the noise level before and
after self-calibration.  We show the difference between the self-calibrated
image and the un-self-calibrated image in Figure \ref{fig:comparison1} and
similar figures are shown in Appendix \ref{appendix:selfcalcompare}.

Table \ref{tab:selfcal} summarizes the results of self-calibration.
It provides information about the self calibration (i.e., the number of
self-calibration iterations and the dynamic range improvement
obtained through self-calibration) and on the output images (beam
size, peak intensity).  Two columns, $\theta_{req}/\theta_{maj}$
and $\sigma_{req}/\sigma_{MAD}$, compare the requested
to the observed beam size and noise level, respectively.
We report noise levels calculated from signal-free parts of the
non-primary-beam-corrected maps (see \S \ref{sec:noiseestimation}); the primary-beam correction is essential for
actual source property calculation, but is less useful for determining the
noise floor of the data.

For several fields, even when self-calibration solutions were obtained,
no improvement was seen.  Particularly for the lowest dynamic range images,
those with peak signal-to-noise ratio $\sim100$, the improvement was negligible:
G327.29, G337.92, G338.93, G351.77, W43-MM2, W43-MM3 in B3 and W43-MM3 and G353.41 in B6.
For most of these fields, the achieved sensitivity is close to the requested,
and the peak intensity in the field is quite faint, so little improvement
was theoretically possible.
For W43-MM2 B3, which has the faintest peak intensity of all fields at only 4 mJy/beam,
the dynamic range decreased, suggesting the gain solutions were harmful
to the image.  For this field only, we therefore recommend using the un-self-calibrated
data.

\subsubsection{Noise target}
\label{sec:noisegoals}
\label{sec:calerrors}

ALMA-IMF was planned to reach a uniform gas mass sensitivity of $\sim0.2$
\msun (3-$\sigma$) assuming optically thin dust at a temperature of $20$ K. 
This requirement led to a range
of flux sensitivity requests.  While the delivered data generally met or
exceeded the requested sensitivity within the requested beam size, there are
several large outliers (Figure \ref{fig:noise_excess}).
\refreport{Figure \ref{fig:noise_excess} shows the measured noise scaled to the requested beam divided by the requested noise level.
The scaling is done assuming that the variance $\sigma^2\propto\Omega$, which is valid over a narrow range of angular scales around the beam size; this scaling means that if our synthesized beam is larger than the requested beam, we would expect a correspondingly lower noise by $(\Omega_{synthesized}/\Omega_{requested})^{1/2}$.
The noise ratio is plotted as a function of both integrated and peak intensity to determine whether bright emission is responsible for driving the noise.}
There is no clear
correlation between total or peak flux and excess noise, which would be
expected for dynamic-range-limited data when the limitation is driven by broad
extended emission or very bright, barely resolved sources, respectively. 

One of the most notable problem cases is W51-E B3, which has a recovered noise level
$\sim2\times$ greater than requested.  This high noise level persisted despite
extensive phase self-calibration resulting in a $\sim3\times$ noise reduction
(in the un-self-calibrated data, the noise level is $\sim6.4\times$ higher than
requested). 
While W51-E contains perhaps the most egregiously complicated
spectrum in B6, its B3 spectrum is relatively tame (there are few emission
lines in B3), so poor continuum selection is not a good explanation for the
excess.  We therefore attempted to check the data for variability among the
seven observing blocks of 12-m data.  We imaged each scheduling block
independently, then convolved the images to common resolution and measured
their difference.  No significant variability was observed, ruling out
variability as the explanation for the high noise.  We conclude that the most
likely explanation for the noise excess is multi-scale emission, including
resolved-out and poorly-$uv$-sampled structure, combined with some residual line contamination, but we
acknowledge that this is not a completely satisfactory outcome.
Similar problems are likely the explanation for the noise excess in
G010.62 B3 and W51-IRS2 B3.  For these three fields, the noise excess
remains in the \bsens data, while in W43-MM1 B3, which has a similar excess
in the \cleanest data, the target noise level is achieved when using the full
bandwidth.

\subsubsection{PSF properties}
\label{sec:psfprops}

The observations were designed to achieve beam sizes $\theta_{FWHM}\sim2000$ au at
the distances of each of the targets (see Table \ref{tab:selfcal}).
There is substantial
variation around the requested beam sizes.  This variation is mostly within the
ALMA QA2 boundaries of $\sim30\%$, with exceptions in G012.80, G353.41, and
G351.77, in which the beam area was $\gtrsim50\%$ greater than requested.  In
these cases, both of the individual contributing scheduling blocks (the
``short'' and ``long''  baseline 12-m array configurations) independently passed
the $\sim30\%$ criterion, but when combined, because of the weighting given to
the ``short'' baseline configuration, had a beam that would not have passed
under the ALMA QA2 requirements if the TM1 and TM2 measurement sets had been assessed together.

Figures \ref{fig:b3psfs} and \ref{fig:b6psfs} show the PSFs from each field.
We computed elliptical radial profiles of the square of the PSF and used a
simple peak-finding tool (\texttt{scipy.ndimage.find\_peaks}) to locate the
first minimum in the radial profile (blue dashes in Figure \ref{fig:b3psfs} and
\ref{fig:b6psfs}).  We locate the first peak in the PSF beyond the radius of
the first minimum, and we call this peak the first sidelobe (green dotted line
in Figure \ref{fig:b3psfs} and \ref{fig:b6psfs}).  These features are
highlighted in the PSF figures.  There are substantial variations in the shapes
of the PSFs that should be noted when examining the images.
Note that the peaks seen surrounding the red contours in Figures \ref{fig:b3psfs}
and \ref{fig:b6psfs} but within the blue contours are part of the main dirty
beam, occur within the first null and are therefore considered part of the peak
rather than sidelobes.
Figure \ref{fig:beamsize} summarizes the relation between the requested and
achieved beam sizes and the requested and achieved noise in the data.

\subsection{Combination between 7m and 12m data}
\label{sec:7m12m}

We attempted to combine the 7m and 12m array data sets for each of our
fields.  In principle, the 7m/ACA data should recover spatial scales up to
$\sim70\arcsec$ (3 mm / B3) and $\sim25\arcsec$ (1 mm / B6).
For a proper combination, the noise on the overlapping baselines between the 7m and 12m array observations needs to be similar. We find, however, that for most of the fields 
the 7m array observations were noisier and, therefore, the combination added substantial noise on the angular scales
covered by these baselines.

While our ALMA-IMF data pipeline is capable of combining the 7m array, and the two 12m array configurations and perform a joint deconvolution, we find that it provides a satisfactory result only for a fraction of the targets. One of such examples is shown in Fig.\,\ref{fig:G328_7m12m}, which is the G328.25 clump in ALMA's band 3. This particular region has a synthesized beam size of $0.62\times0.47$\arcsec\ in the 12m only data, and a synthesized beam size of $0.72\times0.62$\arcsec\ in the 7m and 12m combined dataset using the same parameters as before (where robust = 0).
Taking the geometric mean of the beam major and minor axes, we find that it degrades from 0.54\arcsec\ to 0.67\arcsec\ corresponding to a 20\% larger synthesized beam FWHM (50\% larger beam area) in the combined data.
The $rms$ noise is about 0.39~mJy/beam in the 12m-only dataset (Table\,\ref{tab:selfcal}), while in the 12m and 7m array combined dataset we measure a noise of 0.50~mJy/beam in the same region on images prior to primary beam correction. Considering the different beam sizes, the $rms$ noise in the 7m and 12m combined dataset would translate to a noise of  0.77~mJy in a 0.54\arcsec\ beam, corresponding to a factor of two worse $rms$ noise. However, as shown in Fig.\,\ref{fig:G328_7m12m} the combined image suggests a complex structure of emission from extended structures that are not visible in the 12m only image.

Because in some cases the noise in the fields degraded significantly by including the 7m array data, 
we only present here the 12m array observations alone, and defer a homogeneous discussion of the images including the short spacing information to a future work.

\section{Data product summary}
\label{sec:dataproducts}
Data processing was described in Section \ref{sec:data}.
The data are released at
\url{https://zenodo.org/record/5702966}.
Links to the data are hosted at the ALMA-IMF webpage, \url{http://almaimf.com/}.

The delivery includes a subset of the products output from \texttt{tclean}.
We deliver the \texttt{tt0} and \texttt{tt1} images of the model, residual,
image, and psf, where \texttt{tt0} and \texttt{tt1} correspond to the first and
second  term of the multi-frequency synthesis. The approximate monochromatic
flux is given by the \texttt{tt0} data product. We also provide the  masks used
in the different steps for the data reduction.  The \texttt{image.tt0} and
primary-beam-corrected \texttt{image.tt0.pbcor} images are provided as FITS files.
Each of the above file types is produced for both the \cleanest and \bsens data.
We provide only the final, self-calibrated images.

The \texttt{image.tt0} files contain in their headers a list of the parameters
used to create them in \texttt{tclean}.
All of these parameters are listed as key-value pairs in the \texttt{HISTORY} header entries.
They also include the version
number of the pipeline encoded as a git commit tag; the images were produced with
different versions of the pipeline by necessity, so the commit tag should be used
to track down the exact code used to produce the images.

\section{Analysis}
\label{sec:analysis}

\subsection{Spectral indices and HII regions}
\label{sec:spectralindex}

Since we used the multi-scale, multi-frequency synthesis method with two Taylor terms, we have produced images of the spectral index $\alpha$ (\texttt{tt1} = $\alpha$ \texttt{tt0}).
While most of the images we obtain are
well-represented by a constant value with respect to frequency (i.e., there is little significant signal in the
\texttt{tt1} image), the brighter sources, and especially the bright extended
objects, contain enough emission in \texttt{tt1} to recover the intra-band spectral index
$\alpha$.

Several examples of high-signal regions where the spectral index $\alpha$ could
be accurately measured are shown in Figures \ref{fig:W51e_zooms} and
\ref{fig:w51spindx} (W51-E), \ref{fig:g327contour_freefree} (G327), and Figures
\ref{fig:W51irs2_zooms} \ref{fig:w51irs2spindx} (W51-IRS2).  These images
highlight several salient features: first, while the $\alpha$ images clearly
contain signal, they are noisy and, in general, not trivial to evaluate.
Measured $\alpha$ values frequently have uncertainties that cover the entire
physically plausible range.  Second, there are clear differences in the
spectral indices of known HII regions (detected at lower frequencies with the
VLA, for example) and in evidently dust-dominated sources.  This information
can be used, with appropriate caution, to infer the emission properties of
individual sources.

We specifically explore the brightest sources in the W51-E field in Figure
\ref{fig:w51spindx} because these sources proved to be some of the most
surprisingly problematic for deconvolution.  While the deconvolution of
extended structures throughout these mosaics was expected to be difficult,
point-like sources should not pose a problem for deconvolution and
self-calibration.  In W51-E, however, substantial residual PSF-like artifacts
remained after self-calibration and deep cleaning despite an overall very good
improvement in the noise level and dynamic range.  In Section
\ref{sec:calerrors}, we explored and ruled out the possibility that one of the
central sources was varying.  By examining the spectral index, we see that the
continuum in these sources is structured and complex; there is modest evidence
for a change in spectral index from B3 to B6 (93-100 to 217-233 GHz).  The pair
of sources, seen in the two middle panels in Figure \ref{fig:w51spindx},
separated by only $\lesssim0.5$\arcsec, have dramatically different  spectral
indices in B3, and have much shallower indices than the surrounding material in
B6, highlighting the importance of the multi-term modeling approach.  There are
hints of spectral structure detected within B3 toward e2w, but we were
unable to obtain a reliable determination of $\alpha$ in the low ($\sim92.5$
GHz) and high ($\sim103.8$ GHz) subbands independently, so we cannot provide
detailed estimates of the spectral curvature within B3.

In stark contrast to the complicated W51 e2 region, W51 IRS2 has clean,
self-consistent spectral shape across B3 and between B3 and B6 (Figure
\ref{fig:w51irs2spindx}).  The figures show substantial noise on the spectral
index where physically none is expected, suggesting caution in interpretation
of variations of the spectral index, but qualitative interpretation of $\alpha$
maps should be useful for distinguishing physical emission processes.  These
two fields are adjacent on the sky and therefore have similar $uv$ coverage, so
they are a fair comparison for assessing image quality properties.

While the in-band spectral indices highlight the quality of the ALMA data and
the performance of our data reduction pipeline, the inter-band spectral indices have a much
greater frequency lever arm and therefore much greater signal to noise.
The bottom row of Figure \ref{fig:w51irs2spindx} highlights this improvement,
showing that the IRS2 region splits into a free-free dominated ($\alpha\sim0$)
extended area and a dust-dominated ridge much larger than can be seen in the
single-window $\alpha$ maps.  Interpretation of the spectral indices is further
discussed in \paperone.

\subsection{Hot cores and outflows}
\label{sec:hotcoreoutflow}
The difference images between the \bsens and \cleanest data products contain,
in many cases, substantial structure.  These structures come from excess emission
in the line data that are averaged into the continuum created by \bsens.
The \bsens - \cleanest difference images therefore represent integrals of the
total line intensity in the resulting images.
Most fields show a net excess
of emission.

The emission comes from two primary origins: \emph{hot cores} and \emph{outflows}.
Detailed analysis and cataloging of these objects is deferred to a later
paper, but we highlight some example cases.
In G351.77 (Figure \ref{fig:bsenscompG351_B6}), the excesses surrounding the central hot core
come primarily from broad linewidth emission features that track the bow shocks of material
flows from the central region.
In W51-IRS2 (Figure \ref{fig:absorption_bsens_diff}), excess emission is visible from hot cores
toward the center.
However, a deficit of emission is also seen toward the HII region because of molecular absorption
against the bright continuum.

The excess features in the \bsens-\cleanest difference images highlight the wide variety of spectral
features we anticipate mapping with the ALMA-IMF data.

\section{Conclusions}
\label{sec:conclusions}
We present the ALMA-IMF continuum image mosaics in Band~3 and Band~6, produced
with a custom data reduction pipeline.  This pipeline, with input parameters
fine-tuned by the ALMA-IMF data team for each field, produced self-calibrated
continuum images from multi-configuration ALMA data.  The data underwent
several stages of quality assessment.

The final products exhibit noise levels within a factor of two of those
requested from ALMA, and synthesized beam linear sizes within 40\% of the
expected range, except for one field.  The self-calibration process improved
the dynamic range by up to a factor of five for most of the fields.  Only
those fields with the weakest continuum sources show small improvement by the
self-calibration.

We performed a preliminary analysis of the spectral indices of the mosaics
calculated both in-band and between bands.  This analysis serves both as a
demonstration of the data quality and as a preliminary science demonstration.
The spectral index maps directly identify regions of interest: HII regions
stand out as low-$\alpha$ regions ($\alpha\sim0$), and dust-dominated areas
have high index ($\alpha>2$).

These data will serve as the basis of several ongoing and planned studies on
the development of the stellar initial mass function via the core mass
function as outlined in Paper I.

\begin{acknowledgements}
A.G. acknowledges support from the National Science Foundation under grant AST-2008101.
T.Cs. has received financial support from the French State in the framework of the IdEx Universit\'e de Bordeaux Investments for the future Program. 
R.G.-M. acknowledges support from UNAM-PAPIIT project IN104319 and from CONACyT Ciencia de Frontera project ID: 86372. 
RA gratefully acknowledges support from ANID Beca Doctorado Nacional 21200897.
TB acknowledges the support from S. N. Bose National Centre for Basic Sciences under the Department of Science and Technology, Govt. of India. 
M.B. has received financial support from the French State in the framework of the IdEx Universit\'e de Bordeaux Investments for the future Program. 
S.B. acknowledges support by the French Agence Nationale de la Recherche (ANR) through the project \textit{GENESIS} (ANR-16-CE92-0035-01).
ALS, YP, and BL acknowledge funding from the European Research Council (ERC) under the European Union’s Horizon 2020 research and innovation programme, for the Project “The Dawn of Organic Chemistry ” (DOC), grant agreement No 741002.
FL acknowledges the support of the Marie Curie Action of the European Union (project \textit{MagiKStar}, Grant agreement number 841276).
This project has received funding from the ERC under the European Union’s Horizon 2020 research and innovation programme (ECOGAL, grant agreement no. 855130). FM acknowledges the support of the French Agence Nationale de la Recherche (ANR) under reference ANR-20-CE31-0009, of the Programme National de Physique Stellaire and Physique et Chimie du Milieu Interstellaire (PNPS and PCMI) of CNRS/INSU (with INC/INP/IN2P3).
P.S. was supported by a Grant-in-Aid for Scientific Research (KAKENHI Number 18H01259) of the Japan Society for the Promotion of Science (JSPS). P.S. and H.-L.L. gratefully acknowledge the support from the NAOJ Visiting Fellow Program to visit the National Astronomical Observatory of Japan in 2019, February.
AS gratefully acknowledges funding support through Fondecyt Regular (project code 1180350) and from the Chilean Centro de Excelencia en Astrofísica y Tecnologías Afines (CATA) BASAL grant AFB-170002.
CB and DW gratefully acknowledges support from the National Science Foundation under Award No. 1816715.
LB acknowledges support from ANID BASAL grant AFB-170002.
ER acknowledges the support of the Natural Sciences and Engineering Research Council of Canada (NSERC), funding reference number RGPIN-2017-03987.
BW was supported by a Grant-in-Aid for Scientific Research (KAKENHI Number 18H01259) of Japan Society for the Promotion of Science (JSPS)
We thank the referee for a helpful and constructive report.
The authors acknowledge University of Florida Research Computing for providing computational resources and support that have contributed to the research results reported in this publication. URL: \url{http://researchcomputing.ufl.edu}.
Part of this work was performed using the high-performance computers at IRyA-UNAM. We acknowledge the investment over the years from CONACyT and UNAM, as well as the work from the IT staff of this institute. 
This paper makes use of the following ALMA data: ADS/JAO.ALMA\#2017.1.01355.L and ADS/JAO.ALMA\#2013.1.01365.S. ALMA is a partnership of ESO (representing its member states), NSF (USA) and NINS (Japan), together with NRC (Canada), MOST and ASIAA (Taiwan), and KASI (Republic of Korea), in cooperation with the Republic of Chile. The Joint ALMA Observatory is operated by ESO, AUI/NRAO and NAOJ.
The National Radio Astronomy Observatory is a facility of the National Science Foundation operated under cooperative agreement by Associated Universities, Inc.

\end{acknowledgements}

\clearpage
\appendix

\section{Self-calibration \& \bsens comparison}
\label{appendix:selfcalcompare}
We show comparisons between the self-calibrated and un-self-calibrated data
as in Figure \ref{fig:comparison1} for the rest of the target fields.
These are distributed as an online-only supplemental figure set.

We show comparisons between the \bsens and \cleanest data for each field in
Figure \ref{fig:bsenscompG351_B6} and the corresponding online-only figure set.

\section{Self-calibration details}
\label{sec:selfcaldetails}
\refreport{
The details of how each individual field was self-calibrated is included in the header of the released file.  In the \texttt{HISTORY} keywords of the released FITS files, there are entries that look like: \texttt{HISTORY 1: \{`solint': `30s', `gaintype': `T', `calmode': `p', `combine': `scan', `solnorm': False\}
}.  These encode the relevant parameters used in the CASA command \texttt{gaincal}, where the \texttt{1:} in
this example indicates that this was the first iteration of self-calibration.
We also give a table overview of the used parameters in Table \ref{tab:selfcaldetails}.
}

\begin{table*}[htp]
\caption{Selfcal Details}
\resizebox{\textwidth}{!}{
\begin{tabular}{llllll}
\label{tab:selfcaldetails}
Field & Band & $N_{iter}$ & Gaintypes & Cal. Modes & Solution Intervals \\
\hline
G008.67 & B3 & 5 & T,T,T,T,T & p,p,p,p,p & inf,1200s,600s,300s,200s \\
G008.67 & B6 & 5 & T,T,T,T,T & p,p,p,p,p & inf,1200s,600s,300s,200s \\
G010.62 & B3 & 9 & T,T,T,T,T,T,T,T,T & p,p,p,p,p,p,p,ap,p & inf,40s,25s,10s,10s,10s,inf,inf,inf \\
G010.62 & B6 & 5 & T,T,T,T,T & p,p,p,p,p & inf,40s,25s,10s,inf \\
G012.80 & B3 & 7 & T,T,T,T,T,T,T & p,p,p,p,p,p,a & inf,1200s,300s,300s,inf,inf,inf \\
G012.80 & B6 & 6 & G,G,G,G,G,G & p,p,p,p,p,ap & inf,inf,1200s,600s,inf,inf \\
G327.29 & B3 & 2 & G,T & p,p & inf,60s \\
G327.29 & B6 & 5 & G,G,G,G,G & p,p,p,p,p & inf,60s,20s,10s,5s \\
G328.25 & B3 & 4 & T,T,T,T & p,p,p,p & inf,inf,inf,inf \\
G328.25 & B6 & 4 & T,T,T,T & p,p,p,p & inf,300s,90s,60s \\
G333.60 & B3 & 6 & T,T,T,T,T,T & p,p,p,p,p,a & inf,15s,5s,int,inf,inf \\
G333.60 & B6 & 6 & T,T,T,T,T,T & p,p,p,p,p,a & inf,15s,5s,int,inf,inf \\
G337.92 & B3 & 4 & T,T,T,T & p,p,p,p & inf,300s,60s,30s \\
G337.92 & B6 & 4 & T,T,T,T & p,p,p,p & inf,300s,60s,30s \\
G338.93 & B3 & 3 & T,T,T & p,p,p & inf,inf,60s \\
G338.93 & B6 & 6 & G,G,G,G,G,G & p,p,p,p,p,p & inf,60s,30s,20s,10s,5s \\
G351.77 & B3 & 4 & T,T,T,T & p,p,p,p & inf,90s,60s,30s \\
G351.77 & B6 & 4 & T,T,T,T & p,p,p,p & inf,150s,60s,30s \\
G353.41 & B3 & 6 & T,T,T,T,T,T & p,p,p,p,p,p & inf,inf,inf,inf,inf,inf \\
G353.41 & B6 & 6 & T,T,T,T,G,G & p,p,p,p,p,p & inf,inf,inf,inf,inf,inf \\
W43-MM1 & B3 & 4 & T,T,T,T & p,p,p,p & inf,inf,300s,int \\
W43-MM1 & B6 & 4 & T,T,T,T & p,p,p,p & inf,inf,inf,inf \\
W43-MM2 & B3 & 4 & T,T,T,T & p,p,p,p & inf,inf,inf,inf \\
W43-MM2 & B6 & 5 & G,G,G,G,G & p,p,p,p,p & inf,1200s,600s,300s,int \\
W43-MM3 & B3 & 5 & G,T,T,T,T & p,p,p,p,p & inf,inf,200s,int,inf \\
W43-MM3 & B6 & 5 & G,G,G,G,G & p,p,p,p,p & inf,1200s,600s,300s,int \\
W51-E & B3 & 7 & G,T,T,T,T,T,T & p,p,p,p,p,p,p & inf,inf,inf,inf,int,int,inf \\
W51-E & B6 & 7 & T,T,T,T,T,T,T & p,p,p,p,p,p,p & inf,inf,inf,inf,inf,int,int \\
W51-IRS2 & B3 & 4 & T,T,T,T & p,p,p,p & inf,inf,inf,inf \\
W51-IRS2 & B6 & 9 & T,T,T,T,T,T,T,T,T & p,p,p,p,p,p,p,p,a & 60s,60s,60s,60s,60s,60s,60s,inf,inf \\
\hline
\end{tabular}
}\par The comma-separated lists give the parameters, in order, for each iteration of self-calibration.

\end{table*}

\section{Data handling}
\label{sec:datahandling}
We briefly describe some of the challenges we encountered handling the ALMA-IMF data set and solutions we reached, as these problems and solutions may be used to guide resource planning for future programs.
While the raw data products were relatively modest ($\sim40$ TB), the data set exploded to $\sim200$ TB after intermediate data products were created.
Initially, the large size of individual data sets ($\sim$5-20 TB per band, per field) prevented us from performing data reduction in a centralized manner, and first-pass quality assessment and reduction work was performed independently on different machines by individual researchers.
Members of the data reduction team used the common pipeline to self-calibrate and image the data, and they uploaded the selected imaging and calibration parameters to the ALMA-IMF github repository.
This process was effective, but rather slow.

In 2019, we gained access to substantial resources on the Hipergator supercomputer at the University of Florida, including enough storage to process all of the ALMA-IMF data.
At this point, we re-processed all of the measurement sets using the same machine and using the team-developed imaging parameters.
We were then able to perform both image and visibility quality assessment more uniformly.
Analysis of the full $uv$ data or cube data were not practical prior to this centralization effort.
The fully-processed visibility data, after needed calibration and splitting, ranged from $\sim0.5$ to $\sim4$ TB per science goal, where a science goal encompassed all data for a single band for a single field (including 7m, 12m, and TP data).
The visibility data total to 41 TB fully unpacked.
The imaging data, including the cubes, were much larger, while the continuum data products total to $<100$ GB.

To distribute data among the team, we used the Globus data distribution service, which allows controlled access to the data on the supercomputer system.
The ALMA data reduction pipeline weblogs and other images were hosted on the same machine via a web hosting service running an Apache web server.

The data processing for the continuum data alone generally took from several hours for the smallest fields to several days for the largest.
The supercomputer system allowed us to parallelize imaging across different fields and bands, so the full continuum data sets can be imaged in $<1$ week.
We iterated many times internally to produce the final products, each time performing additional quality assessment tasks.

The data analysis and visualization work was done with a variety of tools,
including the CASA viewer \citep{McMullin2007}, CARTA \citep{Comrie2021}, ds9
\citep{Joye2003}, GILDAS-CLASS (\url{https://www.iram.fr/IRAMFR/GILDAS/}), glue
\citep{Robitaille2017}, and Jupyter  notebooks \citep{Kluyver2016jupyter}.
Images shown in the paper were mostly produced with python analysis scripts and jupyter notebooks stored in the ALMA-IMF github repository, though some were produced with GILDAS-CLASS.
The python scripts used numpy \citep{vanderWalt2011,Harris2020}, scipy \citep{Virtanen2020}, astropy \citep{AstropyCollaboration2013,AstropyCollaboration2018}, spectral-cube \citep{Ginsburg2019b}, radio-beam \citep{Koch2018}, and CASA-6 (\url{https://casa.nrao.edu/casadocs/casa-5.6.0/introduction/casa6-installation-and-usage}).
The plots were made with matplotlib and tools built on matplotlib \citep{Hunter2007}.
\section{Central Frequencies}
\label{sec:centralfreq}
We report the central frequencies computed for each of the observed bands, for each field, given a set of assumed spectral indices $\alpha$ in Table \ref{tab:centralfreqs}.

\begin{table*}
\caption{Central Frequencies}
\begin{tabular}{lrrrrrrrrrr}
\label{tab:centralfreqs}
\\ 
\hline 
& \multicolumn{10}{c}{\bsens} \\
& \multicolumn{5}{c}{B3} & \multicolumn{5}{c}{B6} \\
Field & 0 & 2 & 3 & 3.5 & 4 & 0 & 2 & 3 & 3.5 & 4\\
\hline\\
G333.60      &     99.680 &    100.301 &    100.594 &    100.735 &    100.873 &    228.444 &    228.773 &    228.930 &    229.006 &    229.081\\
G012.80      &     99.655 &    100.275 &    100.568 &    100.710 &    100.848 &    228.379 &    228.708 &    228.865 &    228.941 &    229.016\\
G010.62      &     99.661 &    100.282 &    100.574 &    100.716 &    100.854 &    228.408 &    228.738 &    228.894 &    228.971 &    229.046\\
G353.41      &     99.672 &    100.293 &    100.586 &    100.727 &    100.866 &    228.437 &    228.767 &    228.924 &    229.000 &    229.075\\
G351.77      &     99.669 &    100.289 &    100.582 &    100.723 &    100.862 &    228.419 &    228.749 &    228.905 &    228.982 &    229.057\\
W51-E        &     99.651 &    100.272 &    100.565 &    100.706 &    100.845 &    228.357 &    228.686 &    228.843 &    228.919 &    228.994\\
W43-MM1      &     99.632 &    100.252 &    100.545 &    100.686 &    100.824 &    228.866 &    229.234 &    229.409 &    229.494 &    229.577\\
G328.25      &     99.675 &    100.295 &    100.588 &    100.730 &    100.868 &    228.431 &    228.761 &    228.918 &    228.995 &    229.070\\
G338.93      &     99.677 &    100.298 &    100.591 &    100.732 &    100.871 &    228.456 &    228.786 &    228.943 &    229.019 &    229.094\\
G327.29      &     99.678 &    100.298 &    100.591 &    100.733 &    100.871 &    228.433 &    228.763 &    228.920 &    228.997 &    229.072\\
G008.67      &     99.650 &    100.270 &    100.563 &    100.704 &    100.843 &    228.388 &    228.718 &    228.875 &    228.951 &    229.026\\
G337.92      &     99.673 &    100.294 &    100.587 &    100.729 &    100.867 &    228.439 &    228.769 &    228.926 &    229.002 &    229.077\\
W51-IRS2     &     99.650 &    100.270 &    100.563 &    100.704 &    100.842 &    228.368 &    228.698 &    228.855 &    228.931 &    229.006\\
W43-MM3      &     99.628 &    100.248 &    100.541 &    100.682 &    100.820 &    228.339 &    228.669 &    228.826 &    228.902 &    228.977\\
W43-MM2      &     99.628 &    100.248 &    100.540 &    100.682 &    100.820 &    228.338 &    228.668 &    228.825 &    228.901 &    228.976\\
\hline
\hline
& \multicolumn{10}{c}{\cleanest} \\
& \multicolumn{5}{c}{B3} & \multicolumn{5}{c}{B6} \\
Field & 0 & 2 & 3 & 3.5 & 4 & 0 & 2 & 3 & 3.5 & 4\\
\hline\\
G333.60      &     99.717 &    100.328 &    100.617 &    100.756 &    100.892 &    228.468 &    228.816 &    228.981 &    229.062 &    229.141\\
G012.80      &     99.635 &    100.249 &    100.540 &    100.680 &    100.817 &    228.497 &    228.839 &    229.001 &    229.080 &    229.158\\
G010.62      &     99.682 &    100.283 &    100.567 &    100.704 &    100.837 &    228.717 &    229.035 &    229.185 &    229.258 &    229.330\\
G353.41      &    100.662 &    101.188 &    101.431 &    101.547 &    101.660 &    228.901 &    229.212 &    229.359 &    229.431 &    229.500\\
G351.77      &     99.165 &     99.786 &    100.083 &    100.228 &    100.370 &    227.337 &    227.718 &    227.901 &    227.991 &    228.079\\
W51-E        &    100.583 &    101.085 &    101.315 &    101.426 &    101.533 &    228.346 &    228.682 &    228.841 &    228.918 &    228.995\\
W43-MM1      &     98.679 &     99.307 &     99.611 &     99.759 &     99.906 &    229.060 &    229.424 &    229.596 &    229.680 &    229.762\\
G328.25      &    100.595 &    101.133 &    101.381 &    101.500 &    101.615 &    226.890 &    227.288 &    227.480 &    227.575 &    227.668\\
G338.93      &     99.701 &    100.394 &    100.723 &    100.882 &    101.037 &    228.682 &    229.001 &    229.153 &    229.226 &    229.298\\
G327.29      &    100.822 &    101.389 &    101.651 &    101.776 &    101.898 &    229.023 &    229.308 &    229.442 &    229.507 &    229.571\\
G008.67      &     99.491 &    100.098 &    100.386 &    100.525 &    100.661 &    228.113 &    228.475 &    228.648 &    228.732 &    228.815\\
G337.92      &    100.837 &    101.361 &    101.602 &    101.717 &    101.828 &    226.849 &    227.229 &    227.413 &    227.503 &    227.592\\
W51-IRS2     &    100.295 &    100.869 &    101.135 &    101.263 &    101.387 &    227.922 &    228.278 &    228.448 &    228.530 &    228.612\\
W43-MM3      &     99.908 &    100.500 &    100.777 &    100.911 &    101.041 &    228.347 &    228.689 &    228.852 &    228.931 &    229.008\\
W43-MM2      &     99.994 &    100.597 &    100.881 &    101.017 &    101.150 &    226.903 &    227.307 &    227.502 &    227.597 &    227.692\\
\hline 
\hline 
\end{tabular}
\par All frequencies given in GHz.  Headings give the spectral index $\alpha$.
\end{table*}

\section{Data releases}
\label{sec:datareleases}
There are several internal data releases.
We publicly release two, and we describe the differences here.
The February 2021 data release was used for the core catalog.
The June 2021 data also include the re-calibration performed by the ALMA observatory in QA3.

The data products included in the release are the CASA \texttt{tclean}-produced
multi-term multi-frequency synthesis products \citep[as described in Section
\ref{sec:imaging};][]{Rau2011}.

For the February 2021 release, we include only the robust=0 files, but we
include four different stages: the dirty images, created prior to
self-calibration (suffix \texttt{dirty\_preselfcal}), the pre-self-calibrated
images using the final, post-self-calibration model as a startmodel (suffix
\texttt{preselfcal\_finalmodel}, the pre-self-calibrated images cleaned with
\texttt{tclean} (suffix \texttt{preselfcal}, and the final self-calibrated
images (suffix \texttt{selfcal$n$\_finaliter}).  Only the latter of these, the
final iteration of self-calibration, should be used for further analysis, but
the other can be important tools for validation of the data.
We include the same set of files for both the \cleanest and \bsens data.

The June 2021 release have overall properties quite similar to the February
data.  There are no systematic or significant changes between the continuum
data from the pre- and post-QA3 imaging, though our internal QA process
did catch some additional re-calibration steps that were needed prior to final
acceptance of the data products by ALMA.

\section{W43-MM1 B6 data}
\label{sec:w43mm1b6}
Observations of W43-MM1 in Band 6 were carried out in Cycle 2 between July 2014
and June 2015 (project \#2013.1.01365.S).
A first continuum map and core extraction were presented by \cite{Motte2018} in
an article that helped motivate the ALMA-IMF Large Program.  The spatial and
spectral setup presented here are similar to that of this Cycle 2 pilot
project, with the exception of the largest spectral window, which was centered
on 233.450 GHz instead of 232.450 GHz.

The W43-MM1 B6 data shown here have been re-reduced using the ALMA-IMF data
pipeline. There are some minor differences compared to the process described
in Sect.~\ref{sec:data}.  First, no \texttt{cont.dat} produced by the
\texttt{find\_continuum} procedure was available.  Therefore, the continuum
selection for the \cleanest map has been done manually, guided by that of
the nearby and evolutionary similar W43-MM2 region.  This continuum
selection was been based on a single EB and directly applied to the whole 12m
data.  For the cleaning and self-calibration steps, optimum parameters
determined by the ALMA-IMF team have been applied. The resulting \cleanest
image shows a slight improvement in the RMS, about 30 \% lower, compared to the
continuum map presented by \cite{Motte2018}. There is also a significant
reduction of sidelobes around the central region.
Further analysis of these data, including a comparison between the two continuum images, will be described in a paper in preparation by Nony et al. 

\section{\bsens without CO and N$_2$H+}
\label{appendix:bsens_noco}
As noted in Section \ref{sec:bsens}, the \bsens images of some fields exhibited
extended emission correlated with a single bright line, either CO or N$_2$H+.
We therefore have produced a third variant of continuum image in addition to
the \cleanest and \bsens images that we call \bsens\texttt{-nobright}.  These
images use all of the available bandwidth, but exclude the CO (in band 6) and
the N$_2$H+ (in band 3) windows entirely.  
The resulting bandwidth is less than the \bsens but greater than the \cleanest
data.  These images were otherwise produced in the same manner as the \bsens
data as described in Section \ref{sec:bsens}.

\section{Supplemental figure sets}
\label{appendix:suppfigures}
\begin{figure*}
    \centering
    \includegraphics[width=0.22\textwidth]{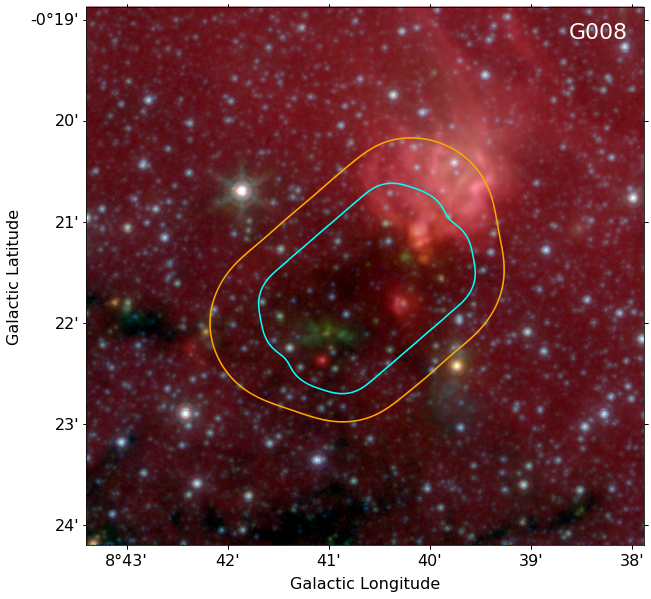}
    \includegraphics[width=0.22\textwidth]{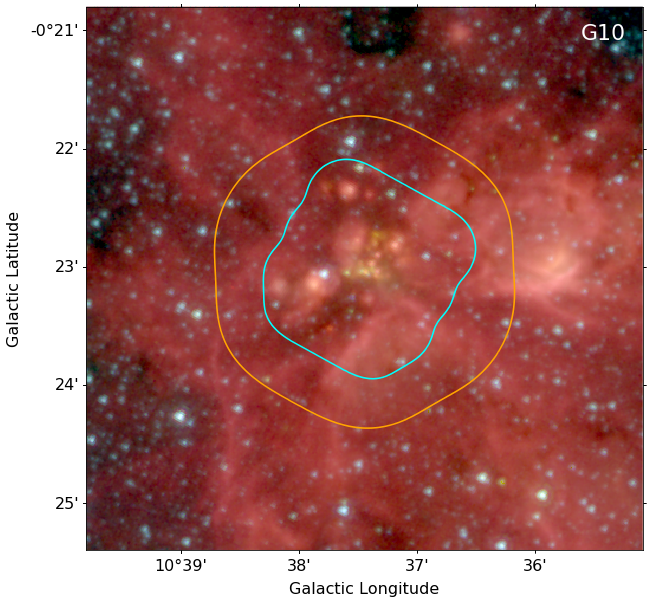}
    \includegraphics[width=0.22\textwidth]{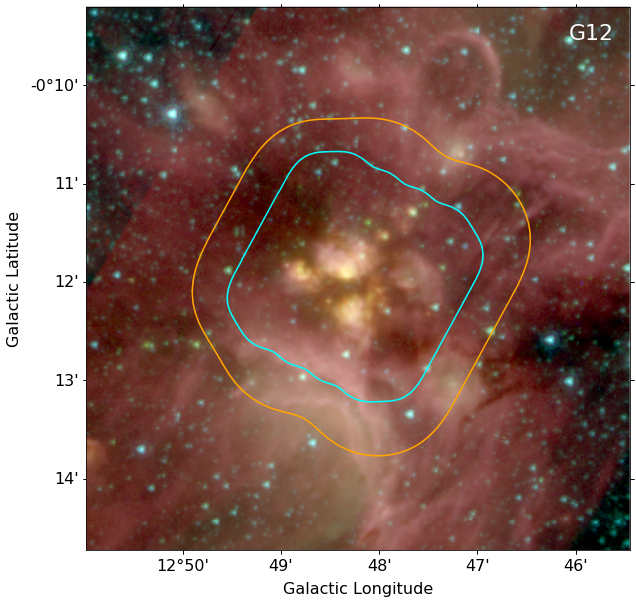}
    \includegraphics[width=0.22\textwidth]{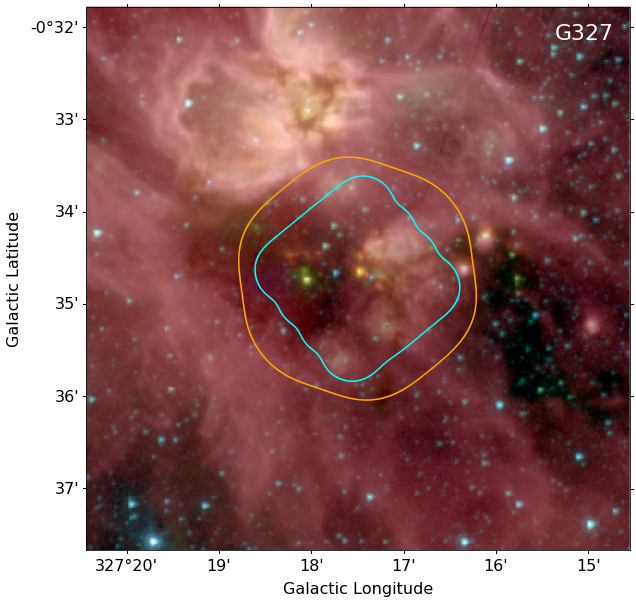}
    \includegraphics[width=0.22\textwidth]{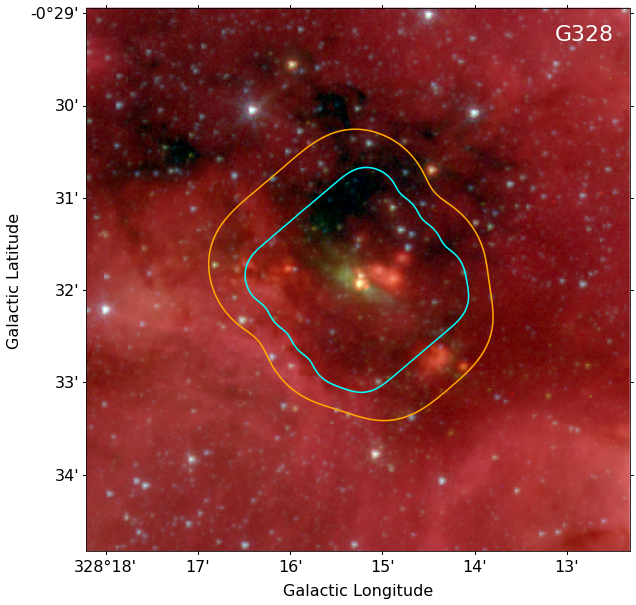}
    \includegraphics[width=0.22\textwidth]{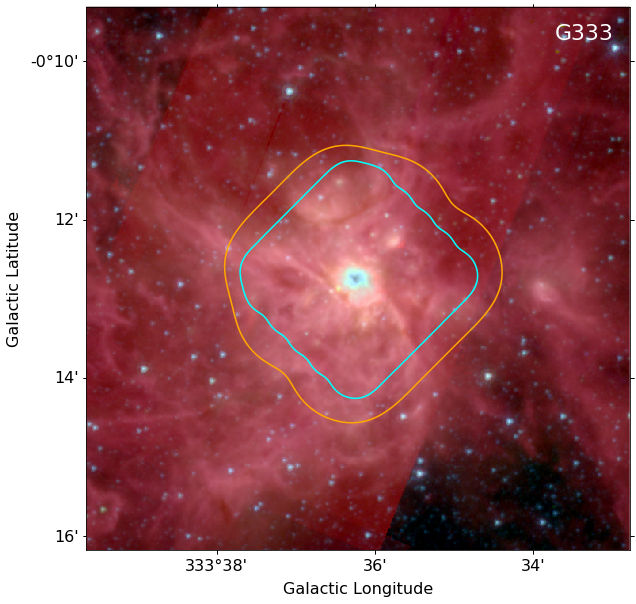}
    \includegraphics[width=0.22\textwidth]{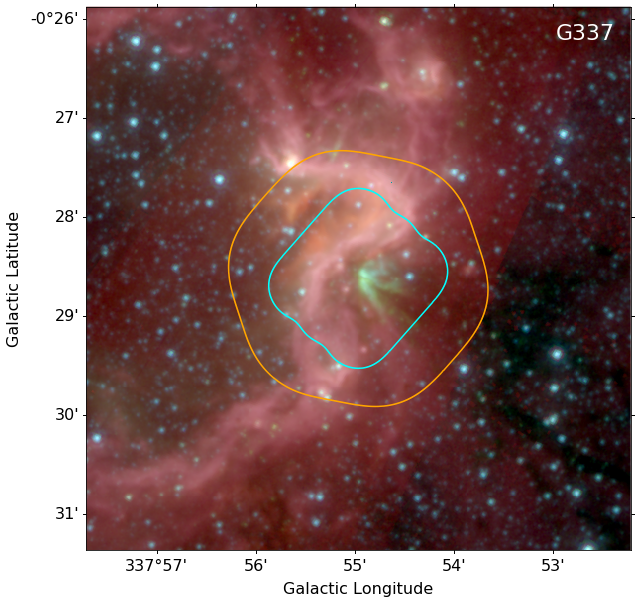}
    \includegraphics[width=0.22\textwidth]{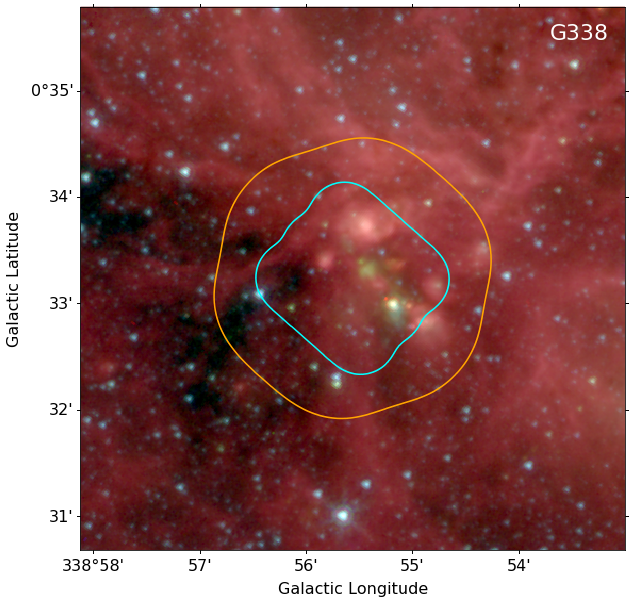}
    \includegraphics[width=0.22\textwidth]{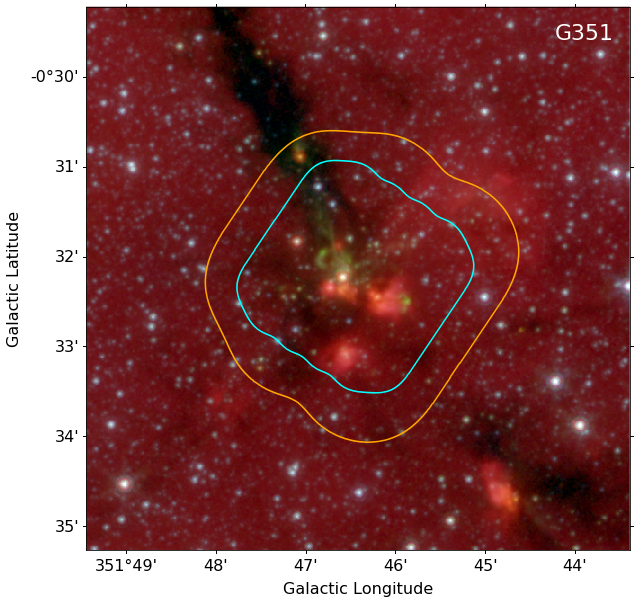}
    \includegraphics[width=0.22\textwidth]{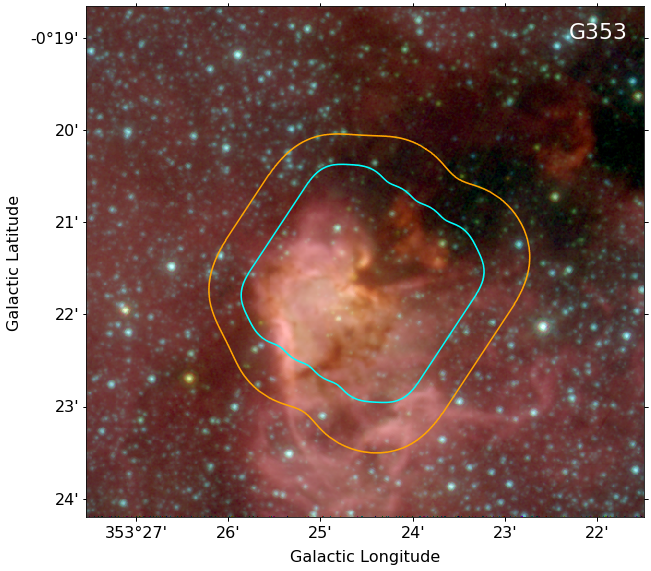}
    \includegraphics[width=0.22\textwidth]{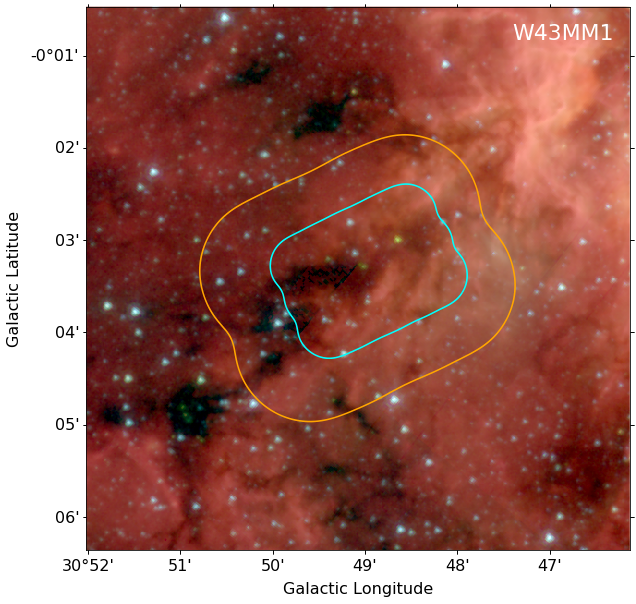}
    \includegraphics[width=0.22\textwidth]{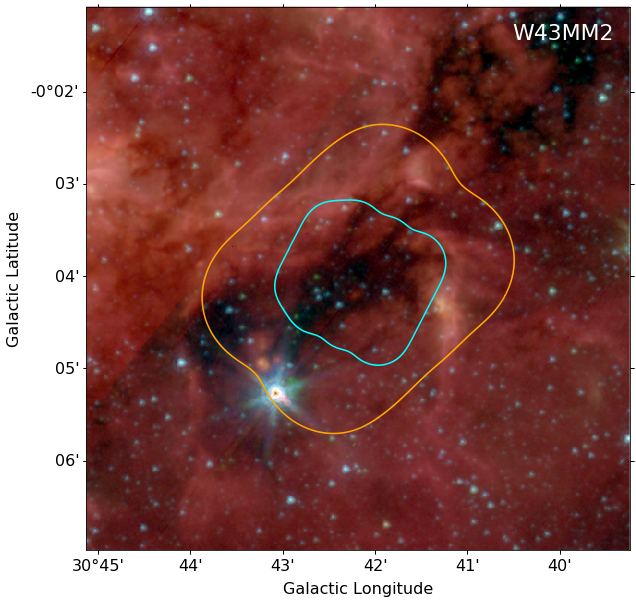}
    \includegraphics[width=0.22\textwidth]{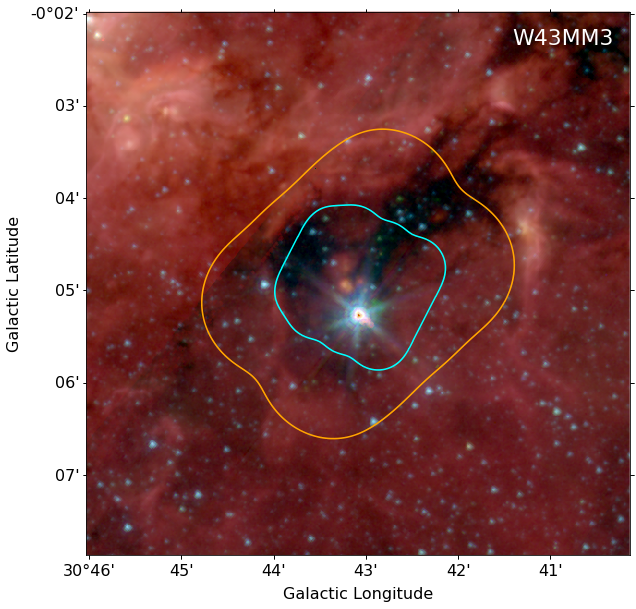}
    \includegraphics[width=0.22\textwidth]{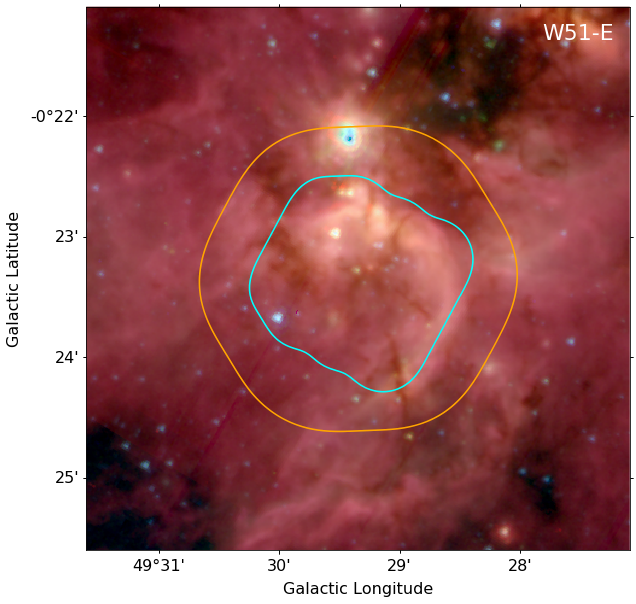}
    \includegraphics[width=0.22\textwidth]{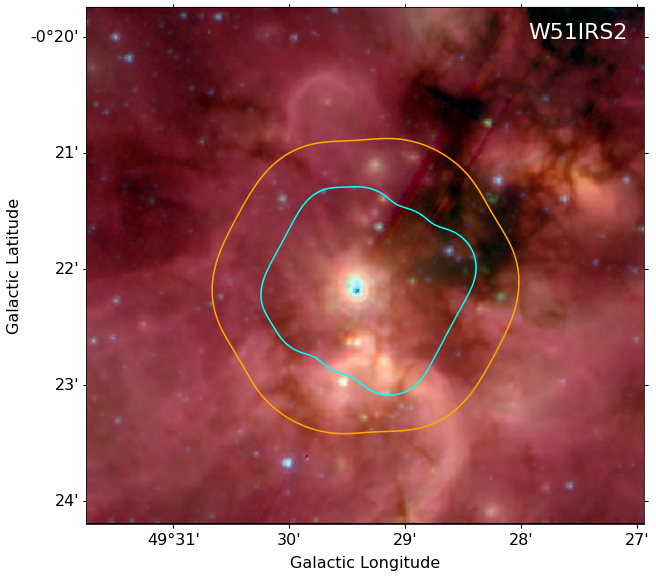}
    \caption{Overview figure showing the mosaic fields-of-view on Spitzer GLIMPSE 3-color images
    \citep{Benjamin2003,Churchwell2009}.  In order, they are: G008, G010, G012, G327, G328,
    G333, G337, G338, G351, G353, W43-MM1, W43-MM2, W43-MM3, W51-E, and W51-IRS2.
    The Band~3 FOV is shown in orange and Band~6 in cyan.}
    \label{fig:overview}
\end{figure*}

\begin{figure*}
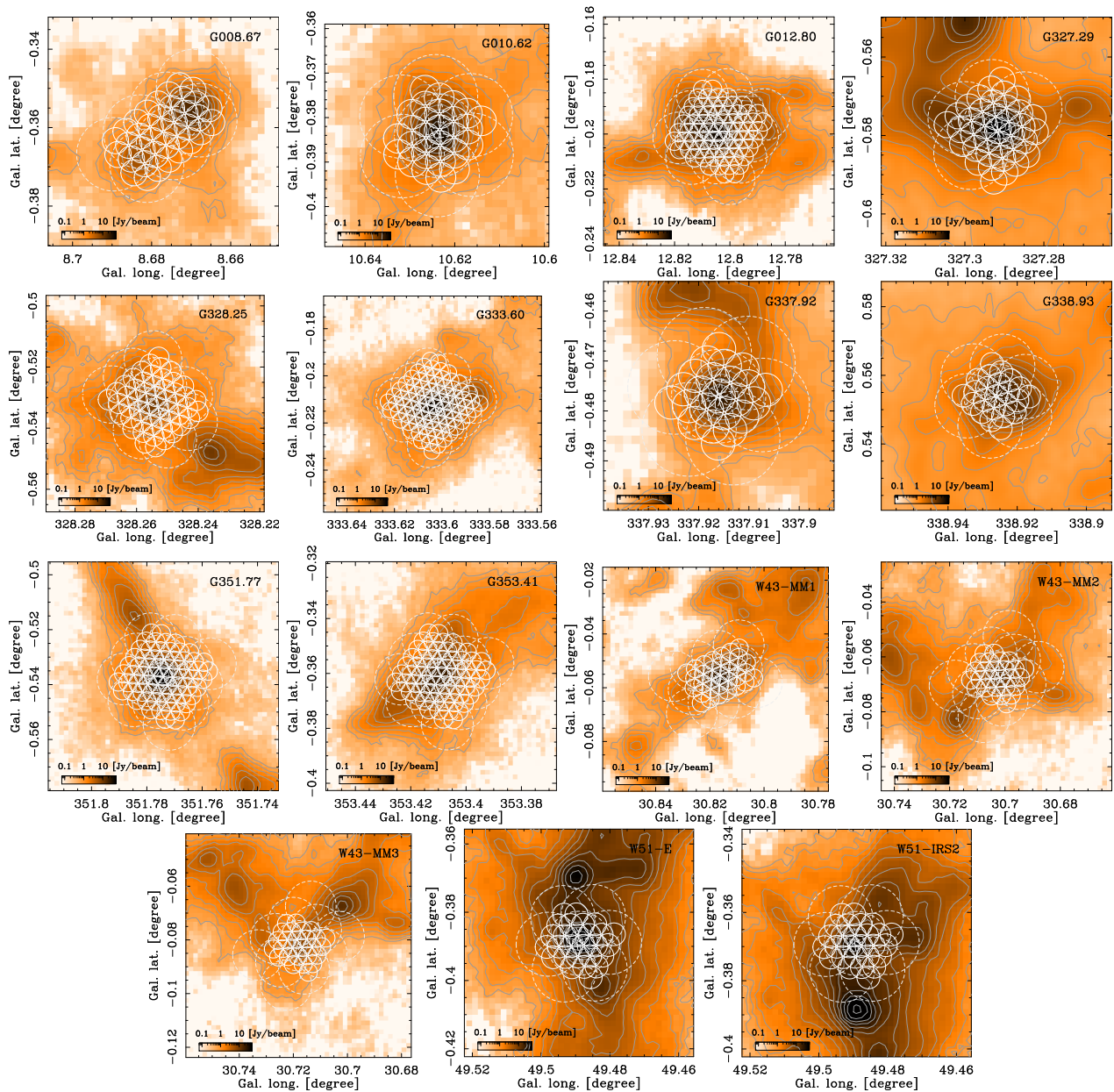

    \centering
    \includegraphics[width=0.22\textwidth]{f67}
    \includegraphics[width=0.22\textwidth]{f68}
    \includegraphics[width=0.22\textwidth]{f69}
    \includegraphics[width=0.22\textwidth]{f70}
    \includegraphics[width=0.22\textwidth]{f71}
    \includegraphics[width=0.22\textwidth]{f72}
    \includegraphics[width=0.22\textwidth]{f73}
    \includegraphics[width=0.22\textwidth]{f74}
    \includegraphics[width=0.22\textwidth]{f75}
    \includegraphics[width=0.22\textwidth]{f76}
    \includegraphics[width=0.22\textwidth]{f77}
    \includegraphics[width=0.22\textwidth]{f78}
    \includegraphics[width=0.22\textwidth]{f79}
    \includegraphics[width=0.22\textwidth]{f80}
    \includegraphics[width=0.22\textwidth]{f81}
    \caption{Overview plot showing individual pointings overlaid on ATLASGAL 
    continuum emission maps.   In order, they are: G008, G010, G012, G327, G328,
    G333, G337, G338, G351, G353, W43-MM1, W43-MM2, W43-MM3, W51-E, and W51-IRS2.
    }
    \label{fig:overview_pointings}
\end{figure*}

\begin{figure*}
    \centering
    \includegraphics[width=0.22\textwidth]{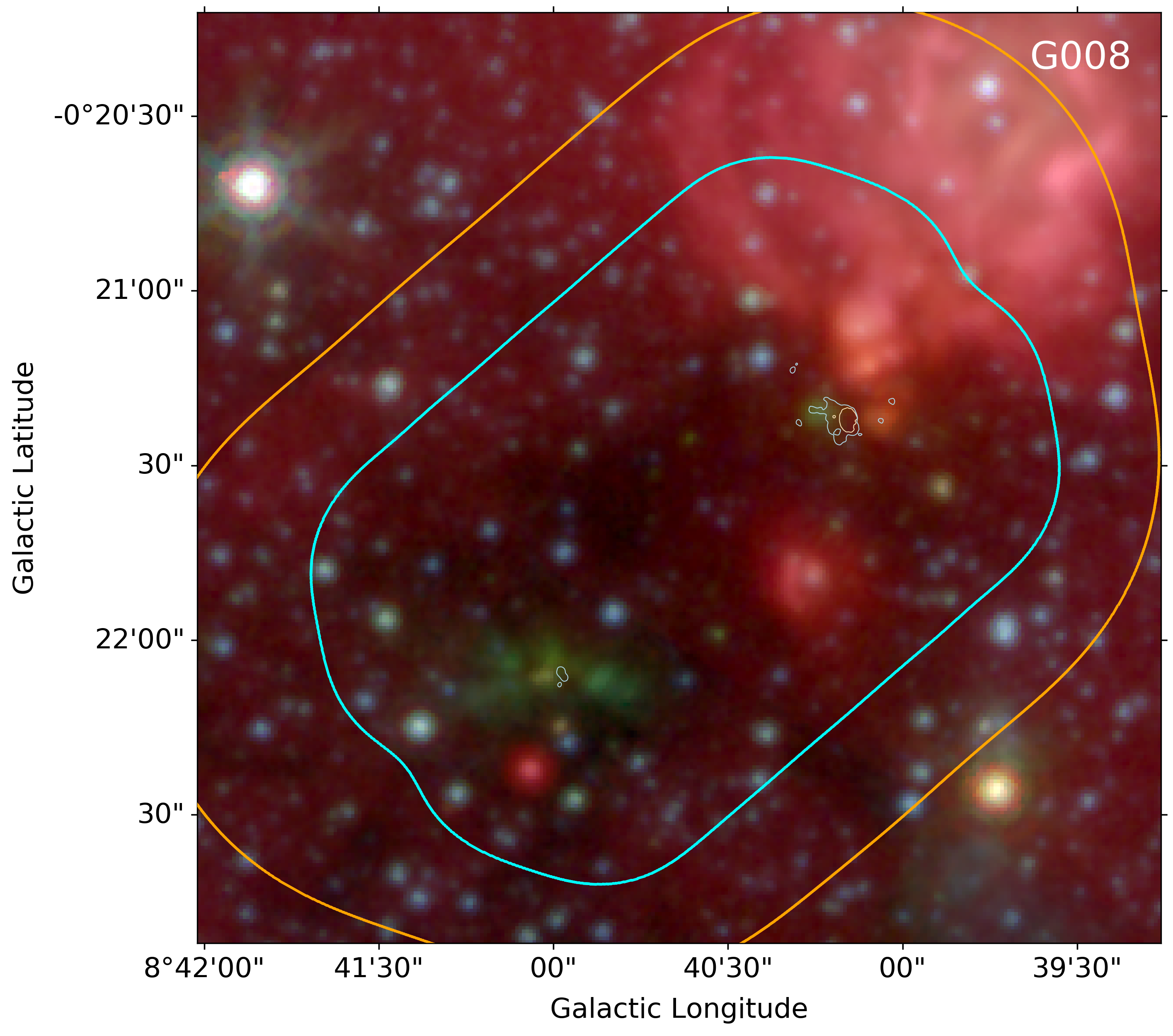}
    \includegraphics[width=0.22\textwidth]{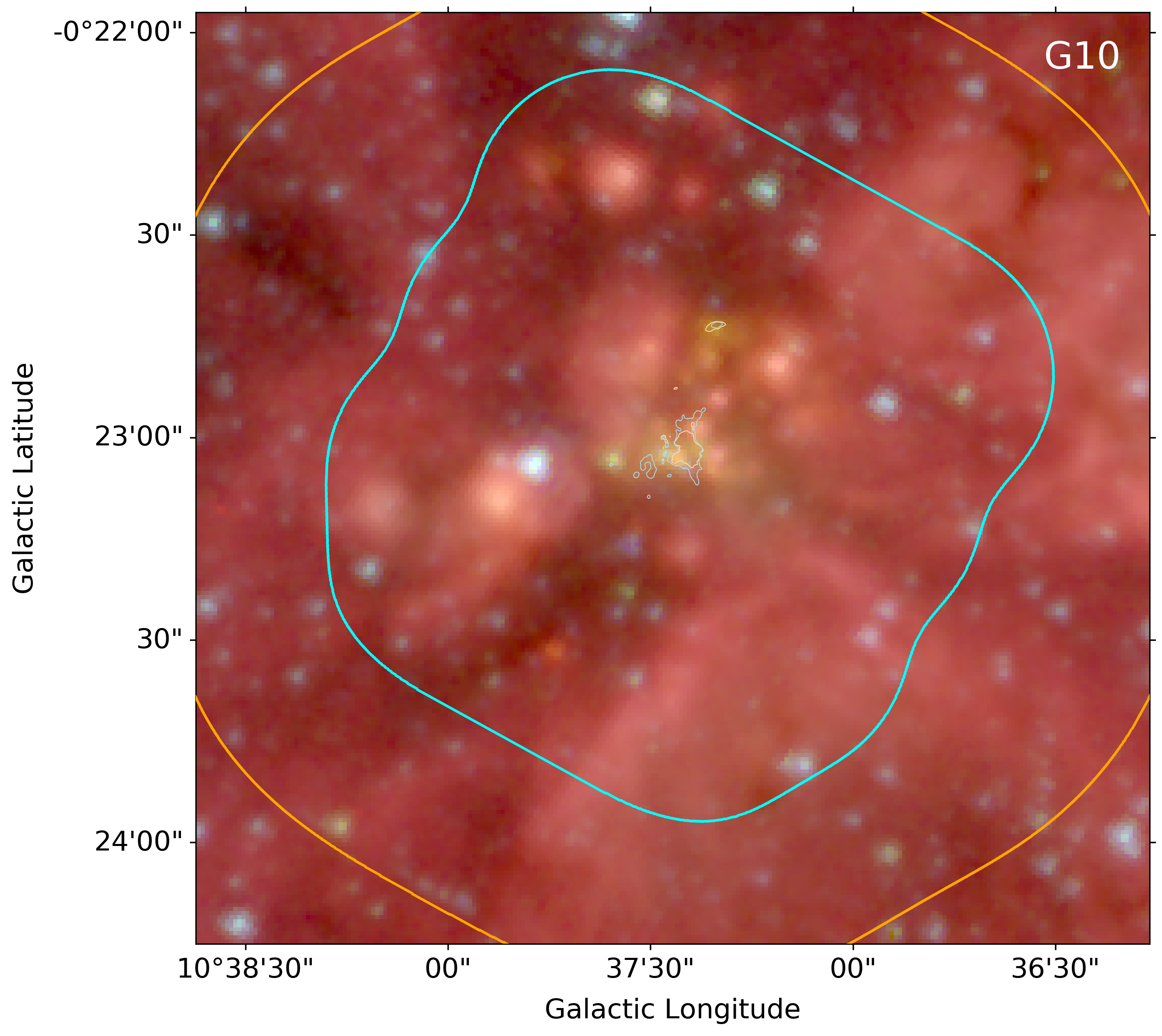}
    \includegraphics[width=0.22\textwidth]{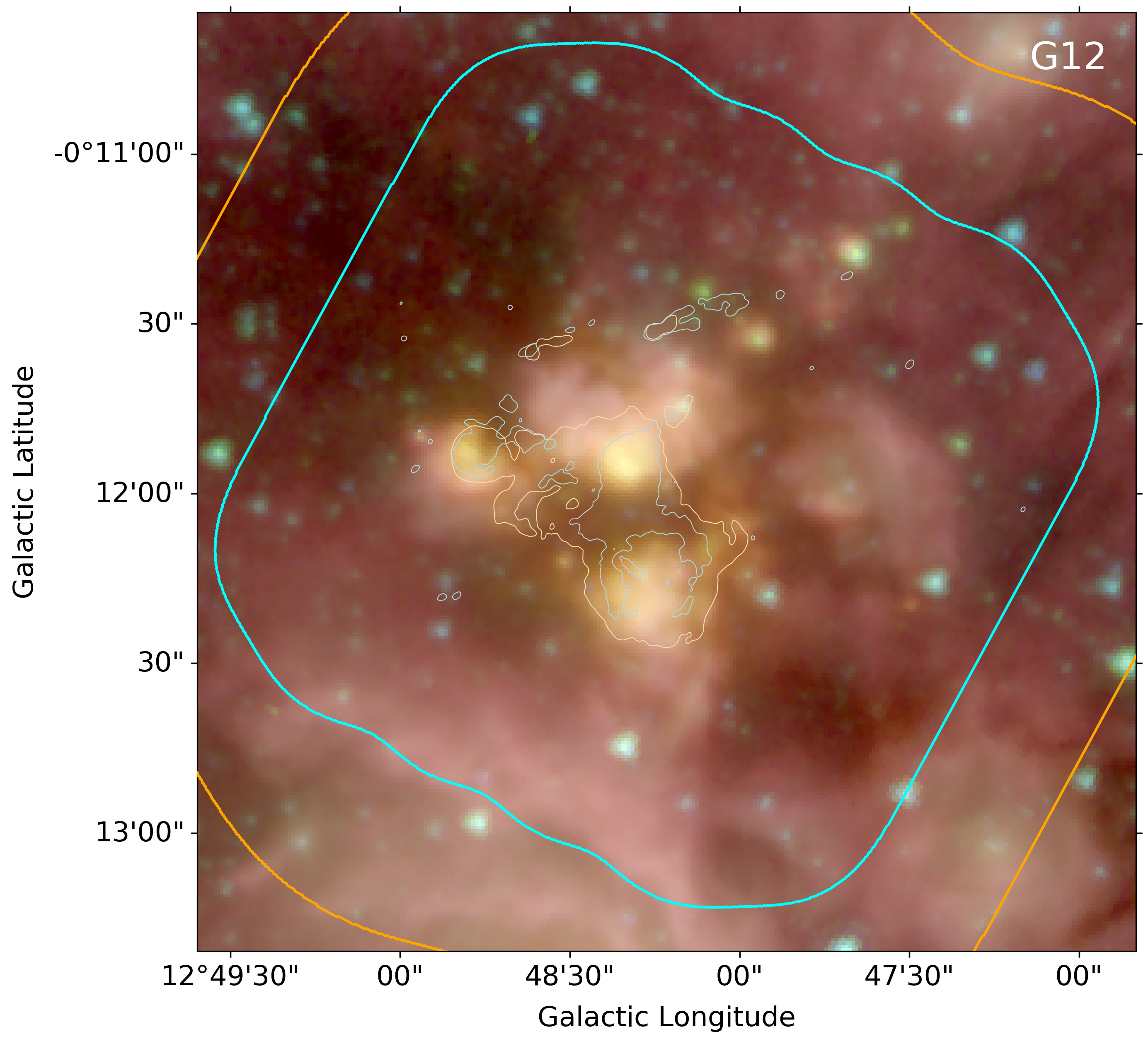}
    \includegraphics[width=0.22\textwidth]{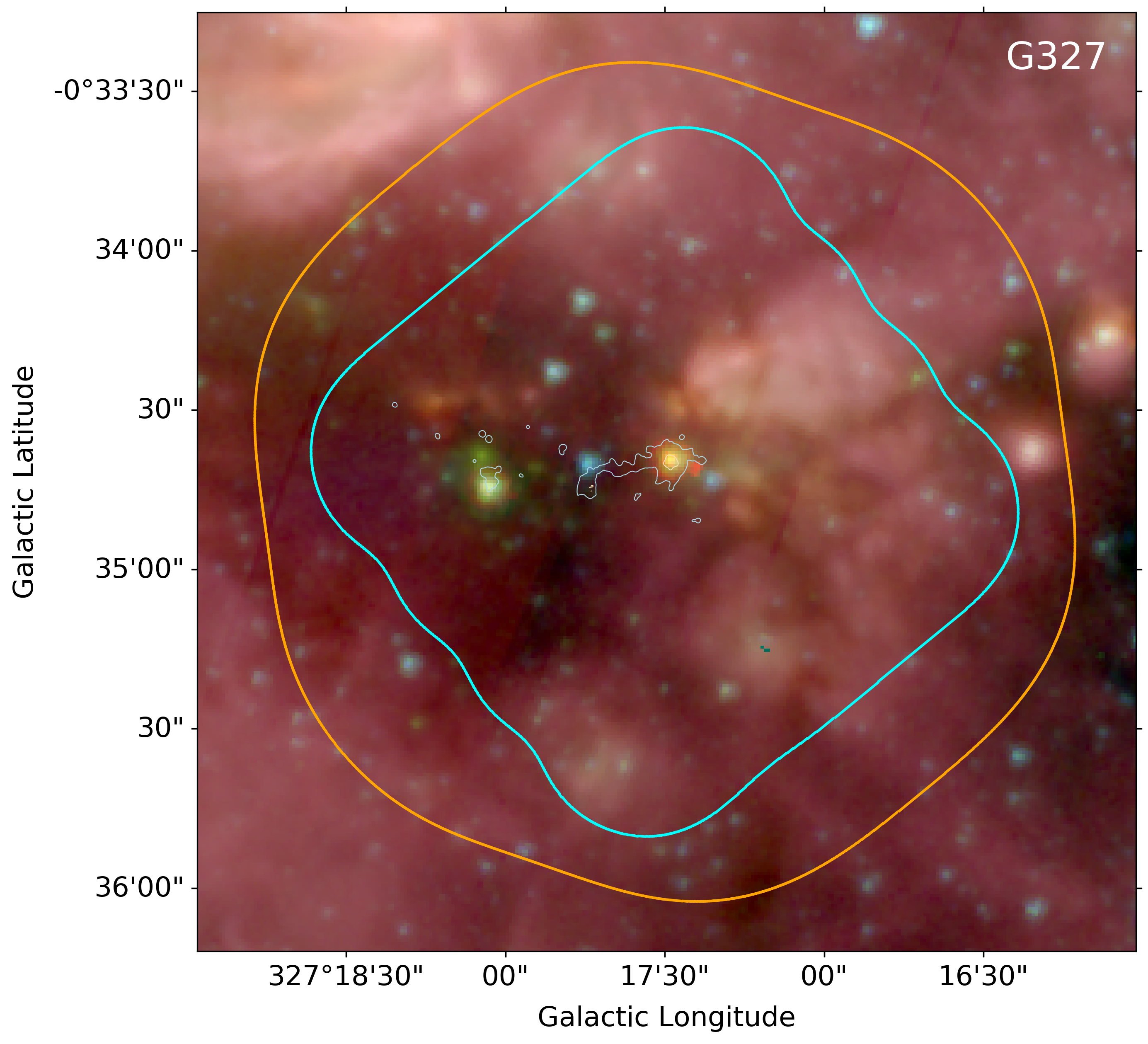}
    \includegraphics[width=0.22\textwidth]{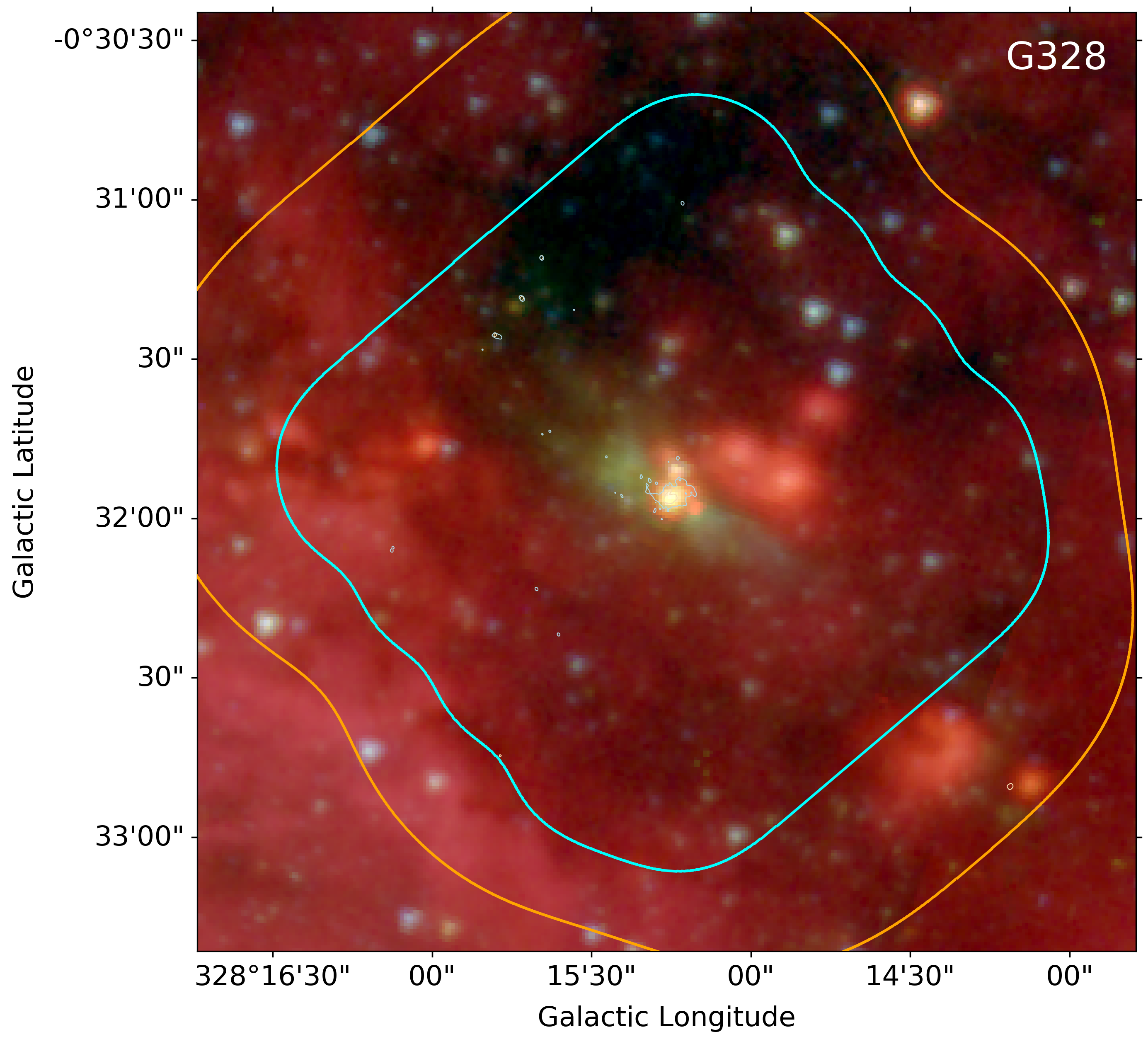}
    \includegraphics[width=0.22\textwidth]{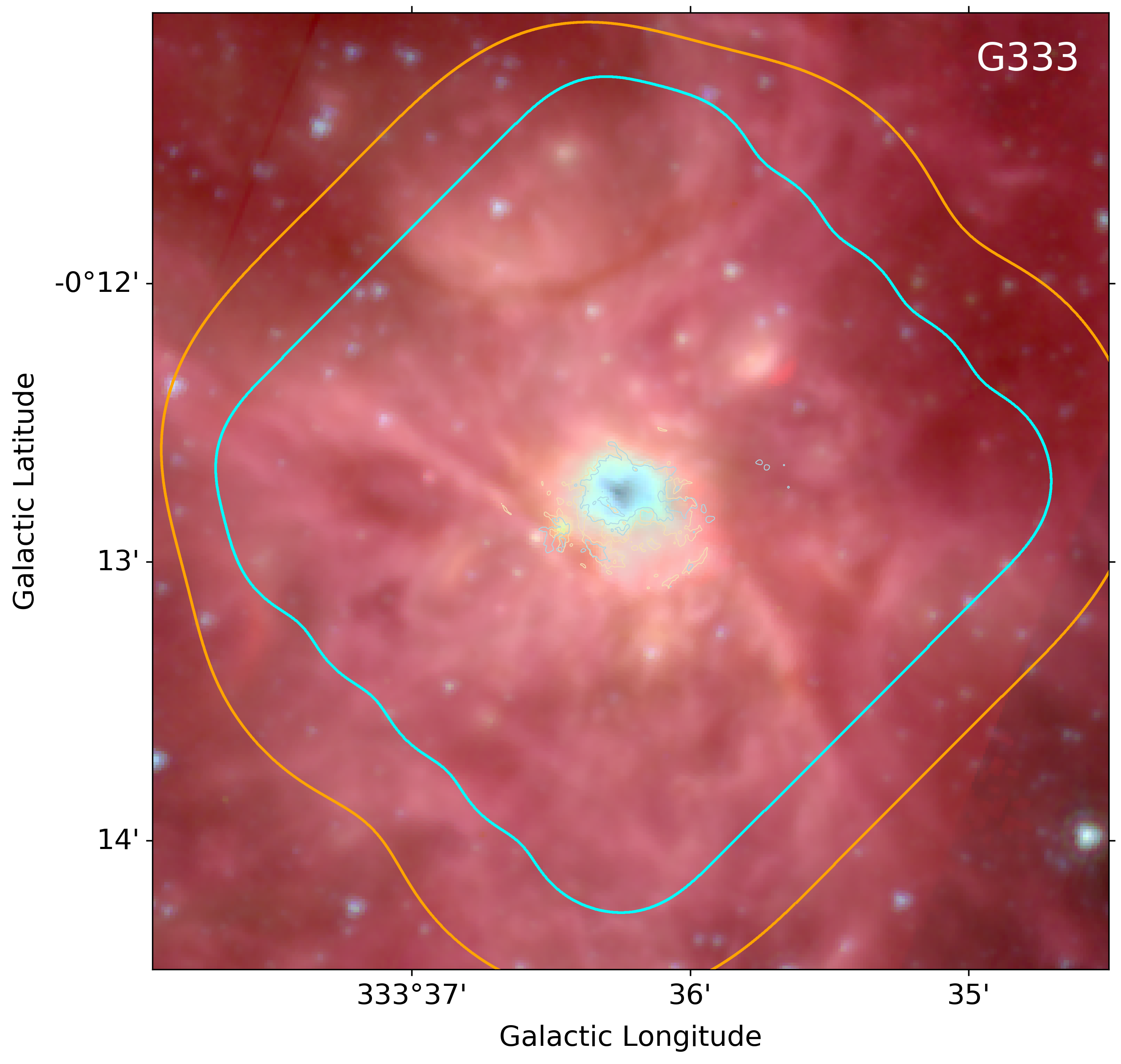}
    \includegraphics[width=0.22\textwidth]{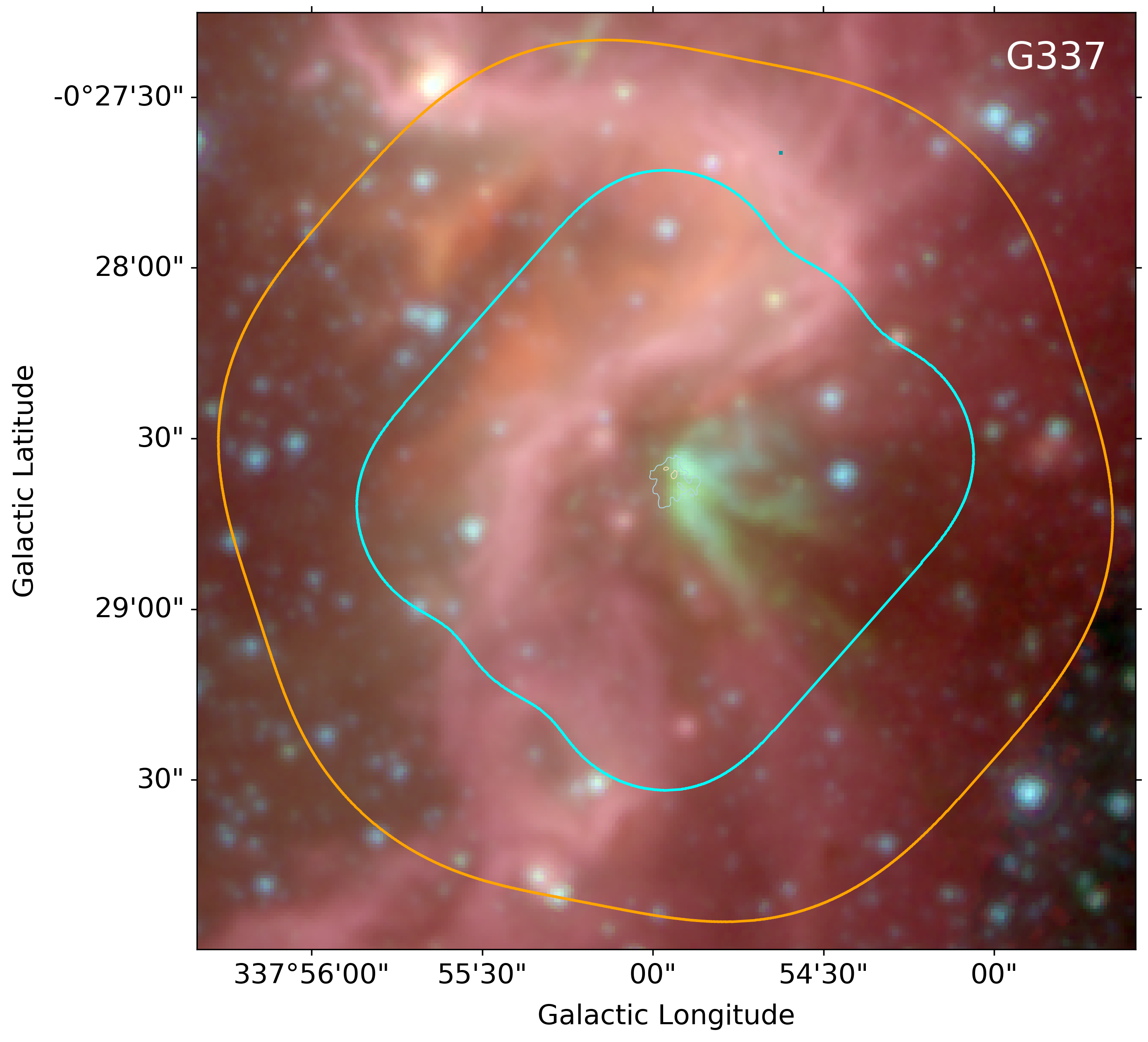}
    \includegraphics[width=0.22\textwidth]{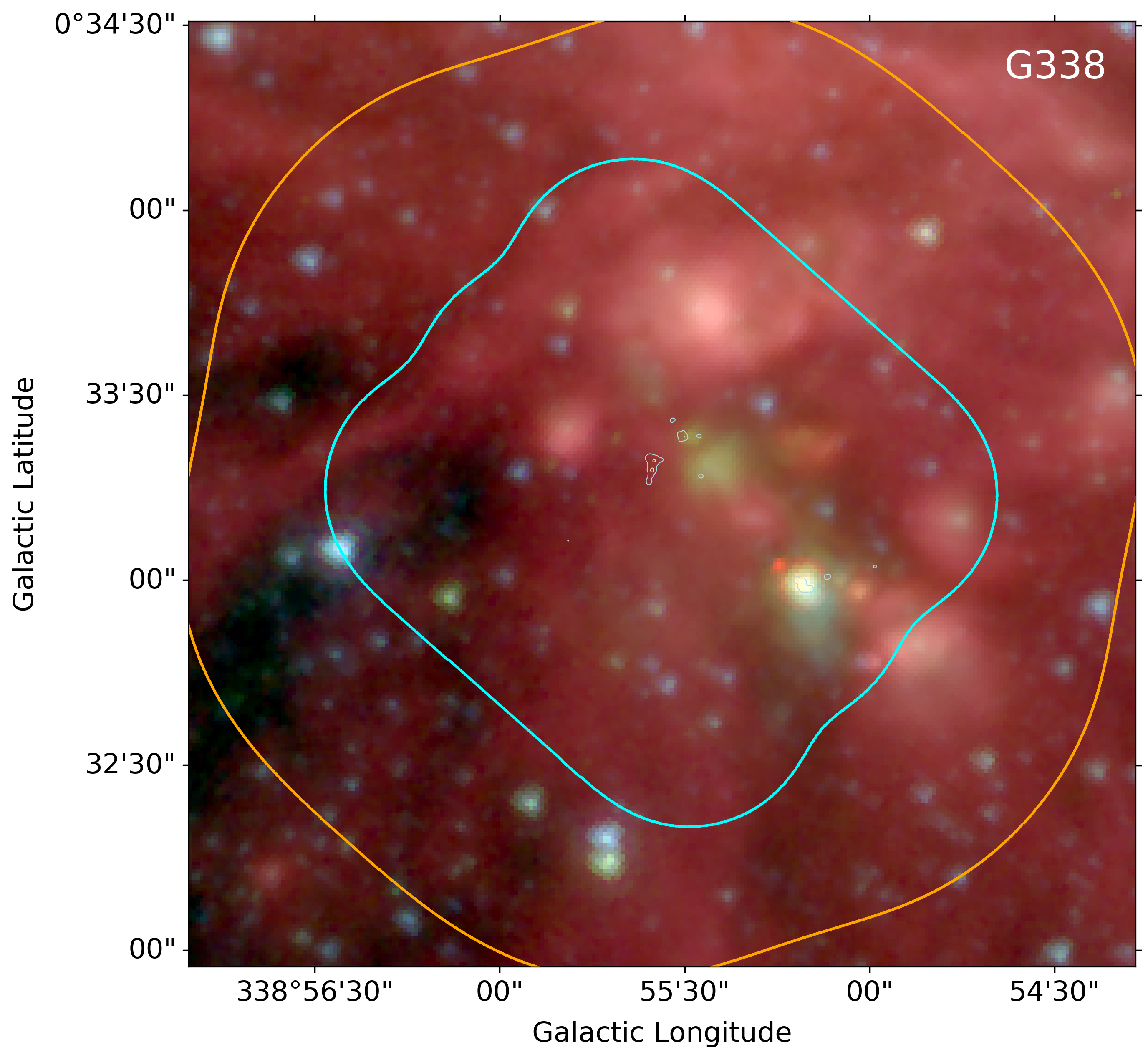}
    \includegraphics[width=0.22\textwidth]{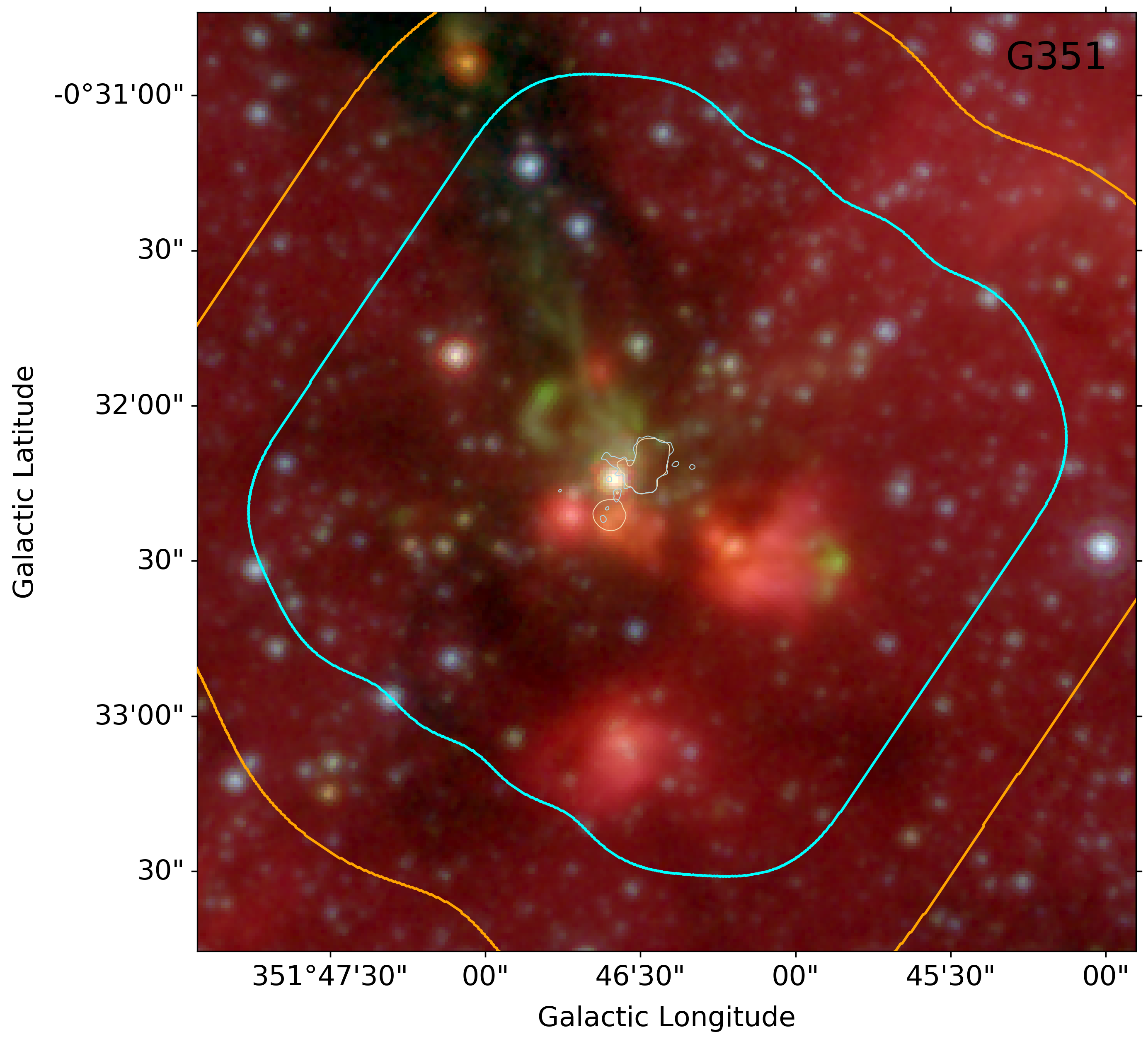}
    \includegraphics[width=0.22\textwidth]{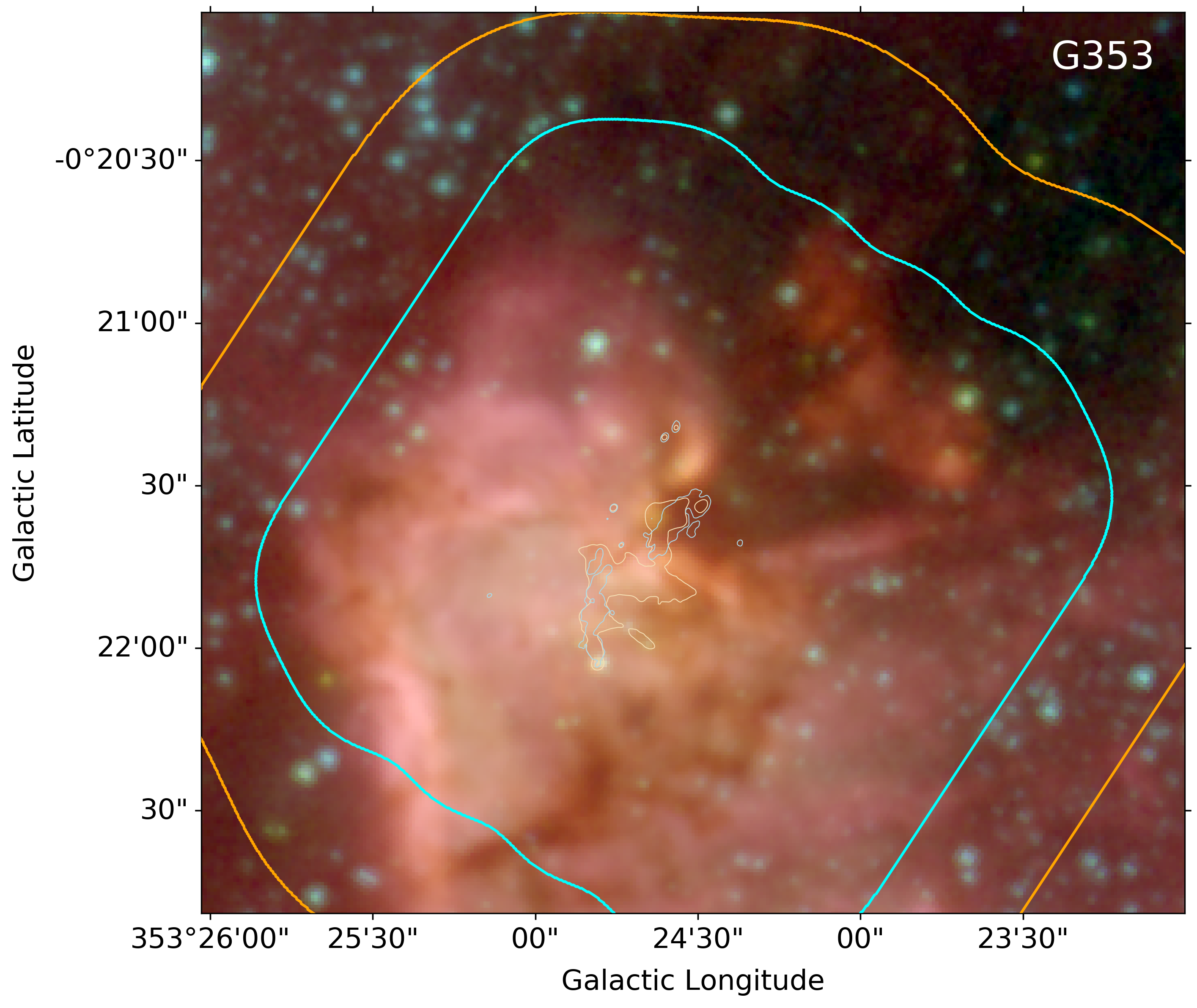}
    \includegraphics[width=0.22\textwidth]{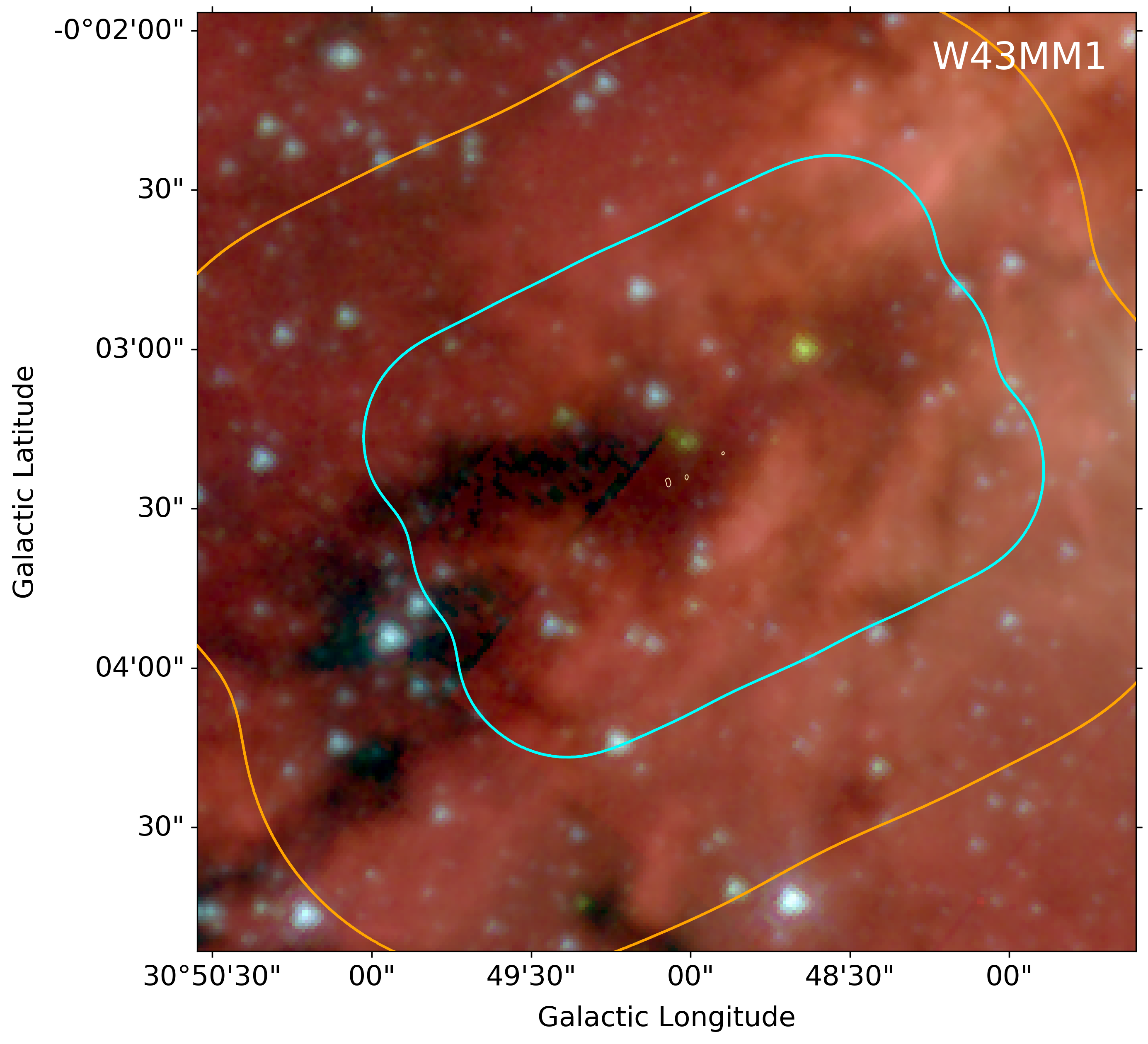}
    \includegraphics[width=0.22\textwidth]{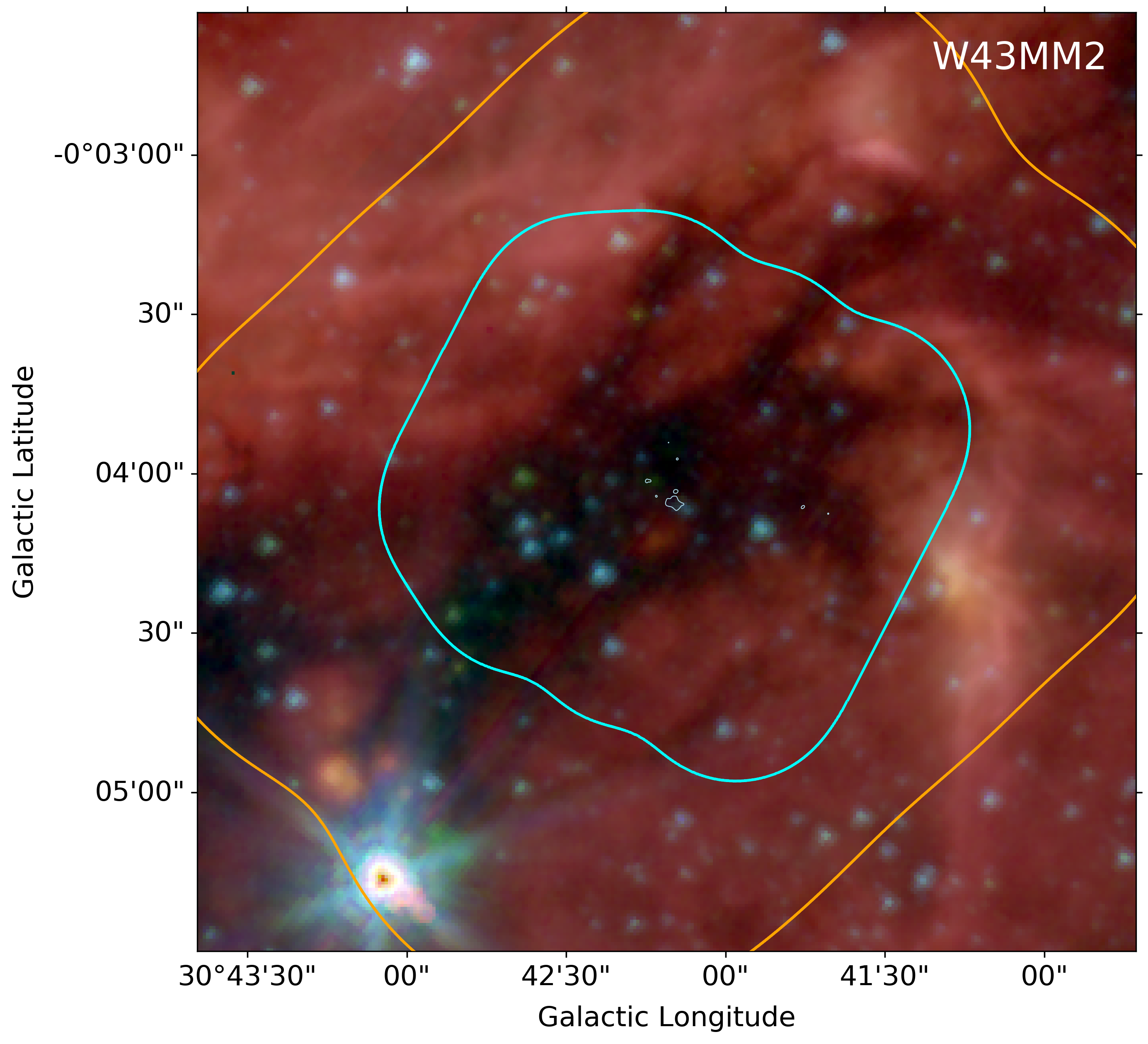}
    \includegraphics[width=0.22\textwidth]{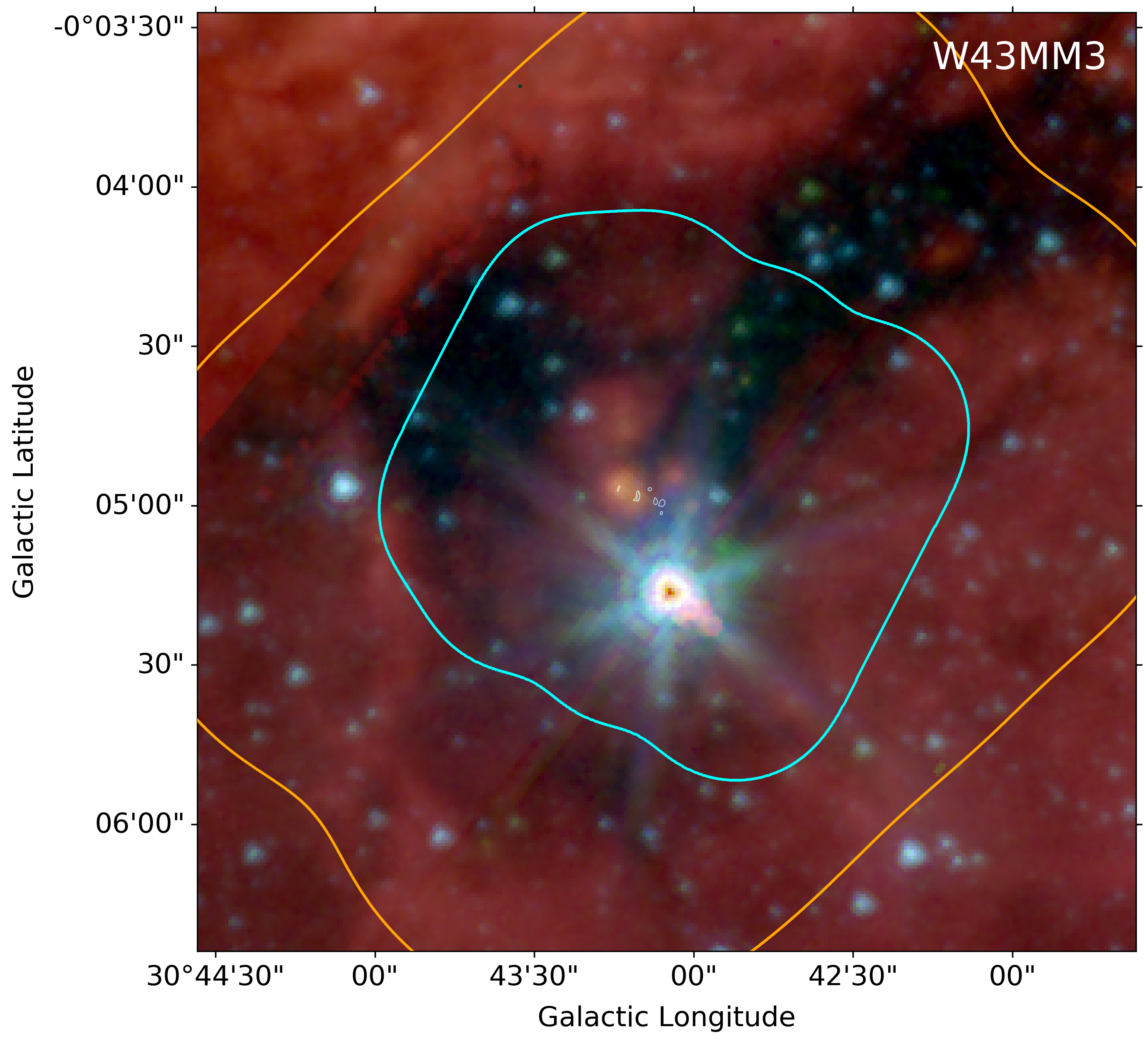}
    \includegraphics[width=0.22\textwidth]{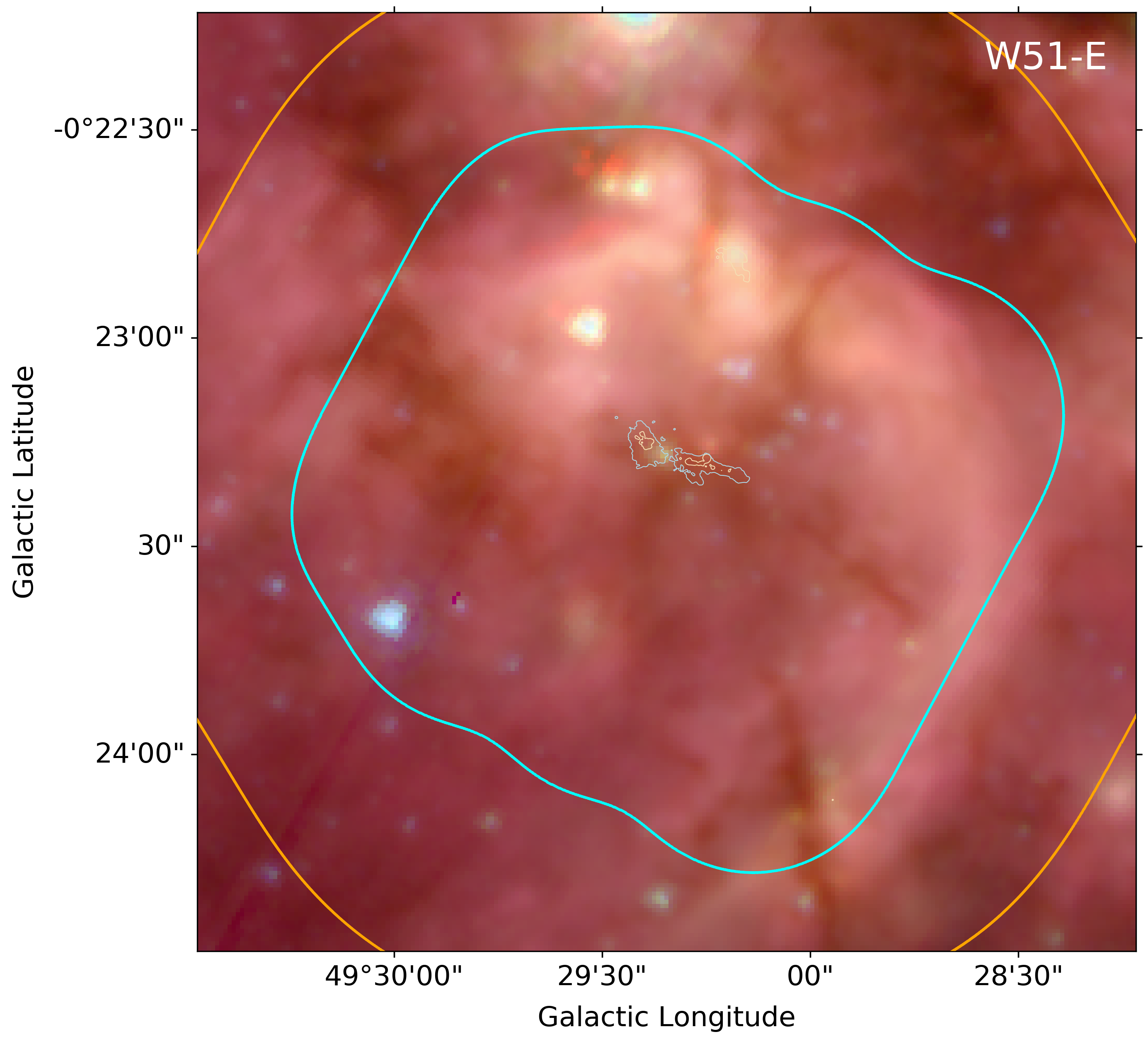}
    \includegraphics[width=0.22\textwidth]{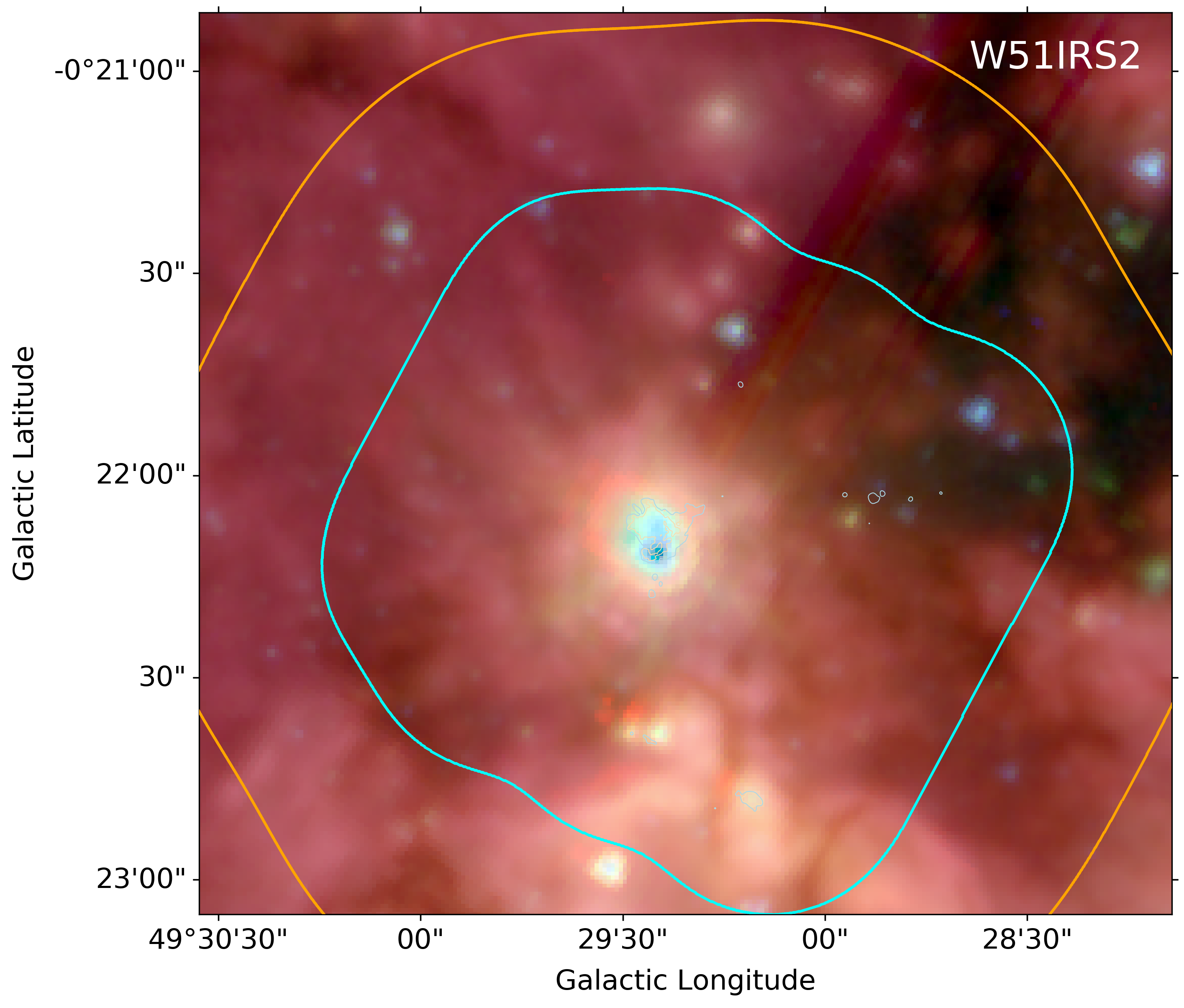}
    \caption{Overview figure showing the mosaic fields-of-view on Spitzer GLIMPSE 3-color images
    \citep{Benjamin2003,Churchwell2009}.  In order, they are: G008, G010, G012, G327, G328,
    G333, G337, G338, G351, G353, W43-MM1, W43-MM2, W43-MM3, W51-E, and W51-IRS2.
    The B3 FOV is shown in orange and B6 in cyan.  Contours are overlaid highlighting the brightest
    detected sources.}
    \label{fig:overview_contour}
\end{figure*}

We distribute several supplemental figure sets reproducing Figures
\ref{fig:g327b6_uvhistogram}, \ref{fig:noise_vs_robust}, and
\ref{fig:noisehistogram} for each field.  As noted in the main text, these are
distributed as the PDF files \texttt{combined\_uvhistograms.pdf},
\texttt{combined\_noise\_and\_beams\_vs\_robust.pdf}, and
\texttt{combined\_flux\_histograms.pdf}, respectively.
We show additional figures to highlight where the ALMA-IMF pointings are in the
context of Spitzer data (Fig. \ref{fig:overview}; Fig.
\ref{fig:overview_contour} shows a single contour from the ALMA-IMF data
overlaid) and ATLASGAL data (Fig. \ref{fig:overview_pointings}).

The fields not shown in the main text from the overview figure (\ref{fig:overview_multicolor_B3}) are also included in this Appendix.

\begin{figure*}
    \centering
    \includegraphics[width=0.48\textwidth]{f26}
    \includegraphics[width=0.48\textwidth]{f27}
    \includegraphics[width=0.48\textwidth]{f28}
    \includegraphics[width=0.48\textwidth]{f29}
    \includegraphics[width=0.48\textwidth]{f30}
    \includegraphics[width=0.48\textwidth]{f31}
    \caption{
    Continued from Fig. \ref{fig:overview_multicolor_B3}.
    }
\end{figure*}
\begin{figure*}\ContinuedFloat
    \includegraphics[width=0.48\textwidth]{f32}
    \includegraphics[width=0.48\textwidth]{f33}
    \includegraphics[width=0.48\textwidth]{f34}
    \includegraphics[width=0.48\textwidth]{f35}
    \includegraphics[width=0.48\textwidth]{f36}
    \caption{Continued from Fig. \ref{fig:overview_multicolor_B3}.}
\end{figure*}

\begin{figure*}
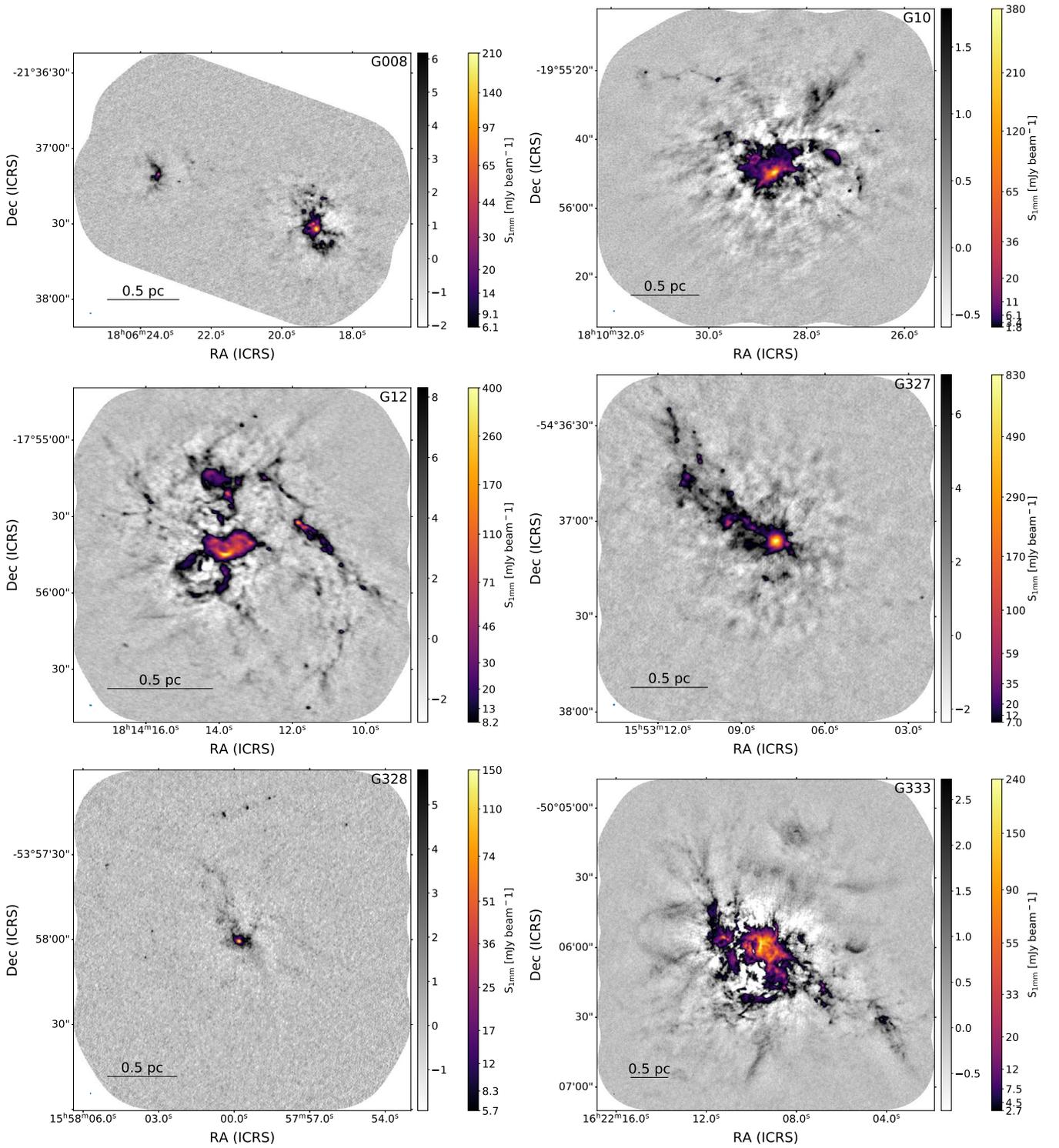

    \centering
    \includegraphics[width=0.48\textwidth]{f37}
    \includegraphics[width=0.48\textwidth]{f38}
    \includegraphics[width=0.48\textwidth]{f39}
    \includegraphics[width=0.48\textwidth]{f40}
    \includegraphics[width=0.48\textwidth]{f41}
    \includegraphics[width=0.48\textwidth]{f42}
    \caption{Overview plot showing B6
    continuum emission maps.  See Fig. \ref{fig:overview_multicolor_B3}.
    }
    \label{fig:overview_multicolor_B6}
\end{figure*}
\begin{figure*}\ContinuedFloat
    \includegraphics[width=0.48\textwidth]{f43}
    \includegraphics[width=0.48\textwidth]{f44}
    \includegraphics[width=0.48\textwidth]{f45}
    \includegraphics[width=0.48\textwidth]{f46}
    \includegraphics[width=0.48\textwidth]{f47}
    \includegraphics[width=0.48\textwidth]{f48}
    \caption{Continued from previous page.}
\end{figure*}
\begin{figure*}\ContinuedFloat
    \includegraphics[width=0.48\textwidth]{f49}
    \includegraphics[width=0.48\textwidth]{f50}
    \includegraphics[width=0.48\textwidth]{f51}
    \caption{Continued from previous page.}
\end{figure*}

\subsection{Overflow figures}
There was insufficient space in the body of the text for several figures that further describe the data.
We include these figures here.

\begin{figure*}
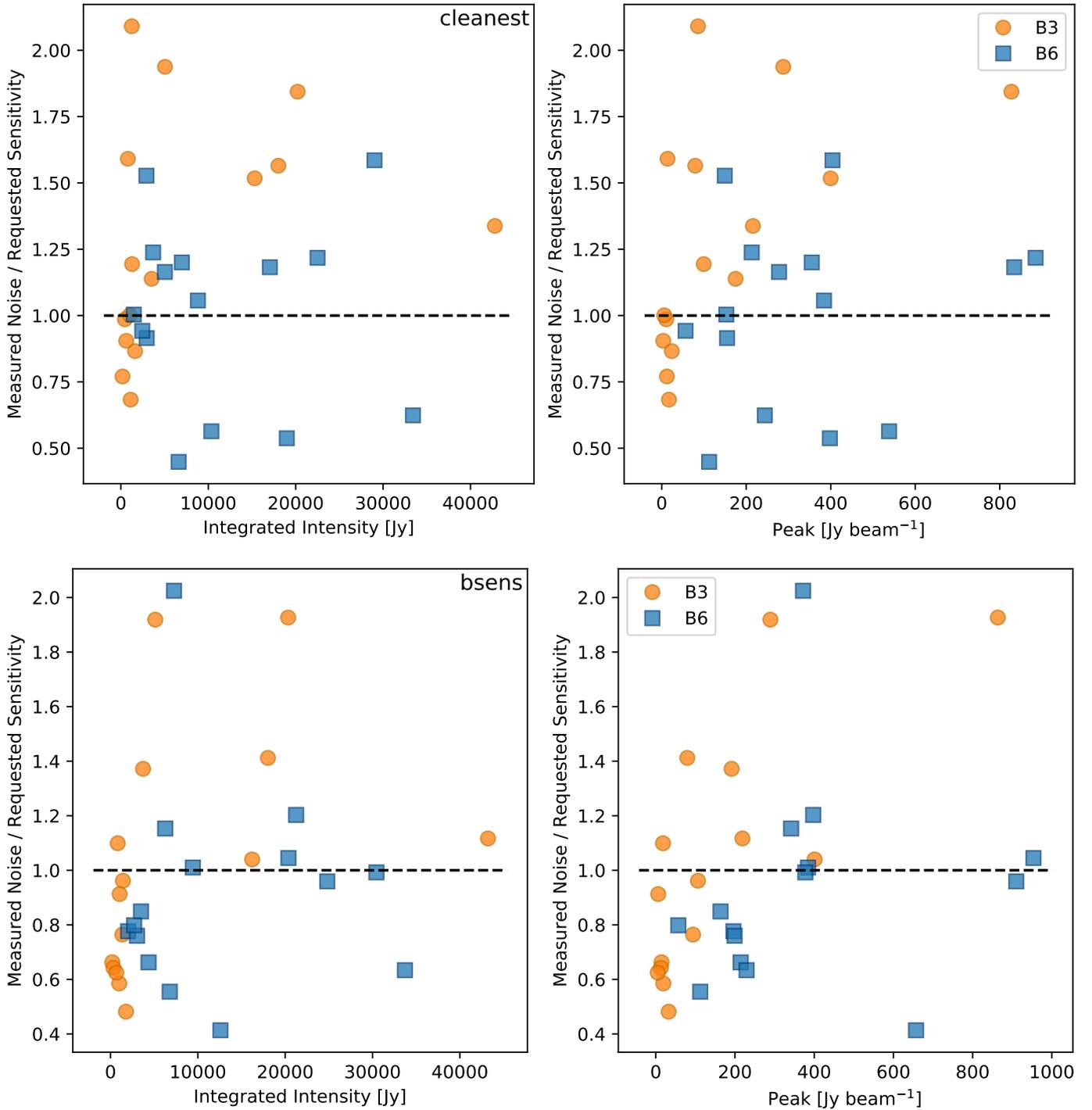

    \includegraphics[width=\textwidth]{f97}
    \includegraphics[width=\textwidth]{f98}
    \caption{Excess noise in the images compared to the thermal noise level that was requested
    in the proposal.
    The top panels show \texttt{\cleanest} continuum data, the bottom show
    \texttt{\bsens}.
    The Y-axis shows the measured noise level in an image with the requested beam (i.e.,
    $\sigma_{MAD} (\Omega_{syn} / \Omega_{req})^{1/2}$; see \S \ref{sec:noisegoals})  divided by
    the requested noise level; higher values indicate higher measured noise.
    The horizontal dashed
    line indicates where the measured noise exactly matches the requested noise.
    The X-axis gives the total flux in the image (left panels) and peak
    intensity (right panels).  We compare the noise to the total and peak
    flux to search for correlations that may explain the excess noise as
    deriving from some form of dynamic range limit, but we do not observe
    any clear correlation.
    The excess noise is caused by calibration errors, unresolved
    structure, and other possibilities discussed in Section \ref{sec:calerrors} 
    }
    \label{fig:noise_excess}
\end{figure*}

\begin{figure*}
\includegraphics[width=\textwidth]{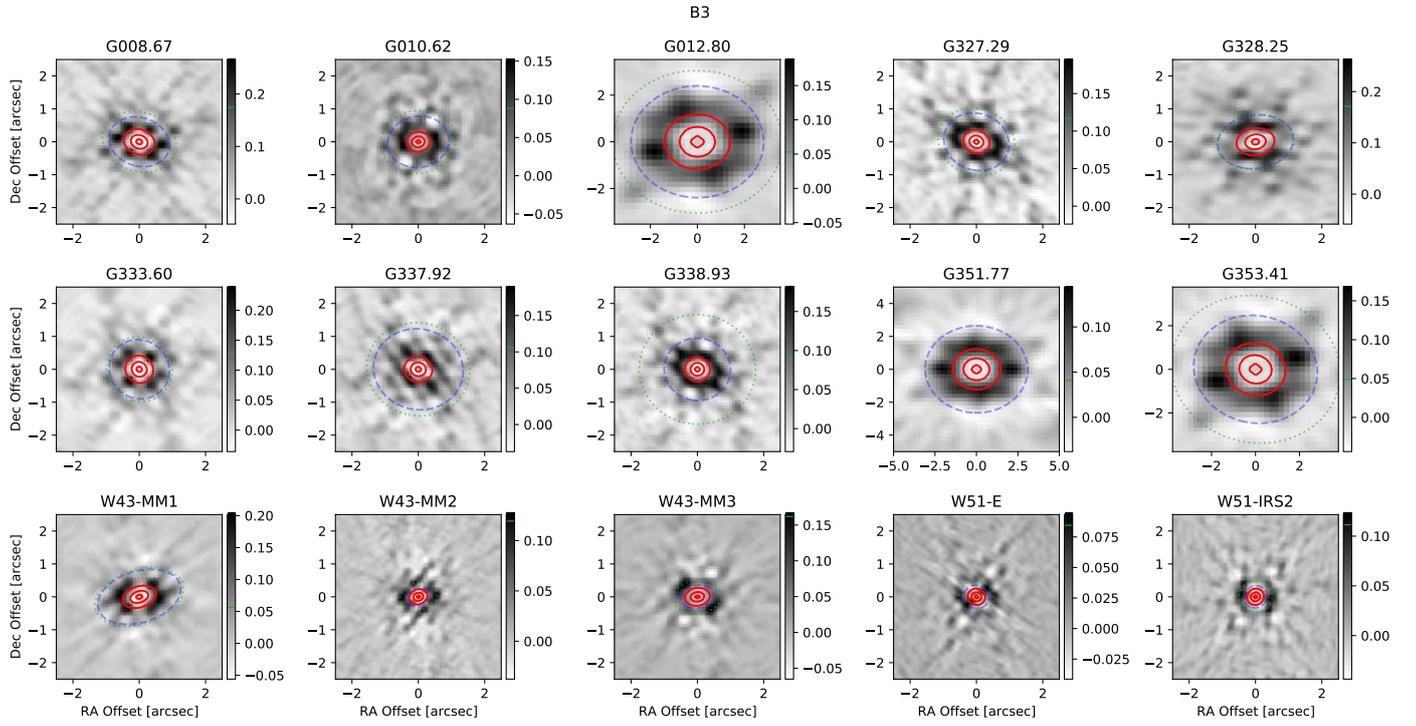}
\caption{
Band 3 PSF figures.  These PSF figures show the difference between the dirty
PSF and the synthesized beam in grayscale with the synthesized beam contoured
on at levels (0.1, 0.5, 0.9) times the peak in red.  The colorscale is on a
normalized scale, where the peak response of the PSF is defined to be unity;
the peak is not shown on the scale because the central Gaussian representing
the synthesized beam has been subtracted.  The blue dashed ellipse shows the
location of the first minimum in the elliptically-averaged, squared PSF
response profile, while green dotted shows the location of the first positive
peak (the first sidelobe) in the PSF. The green dotted curve identifies a peak
location directly identified from the 2D PSF, not from its elliptical profile.
The green dotted curve is not shown if it is beyond the displayed area. }
\label{fig:b3psfs}
\end{figure*}

\begin{figure*}
\includegraphics[width=\textwidth]{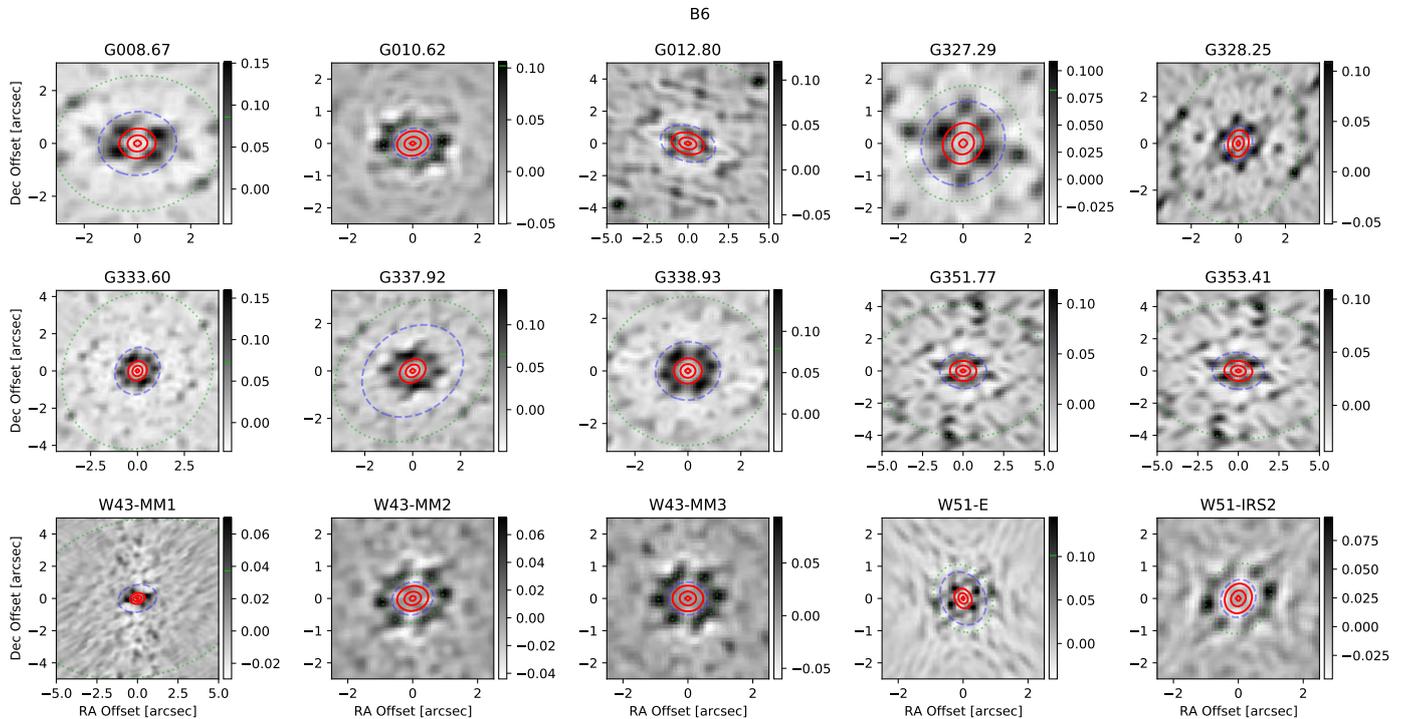}
\caption{
Band 6 PSF figures.
See Figure \ref{fig:b3psfs}.}
\label{fig:b6psfs}
\end{figure*}

\begin{figure*}
    \includegraphics[width=\textwidth]{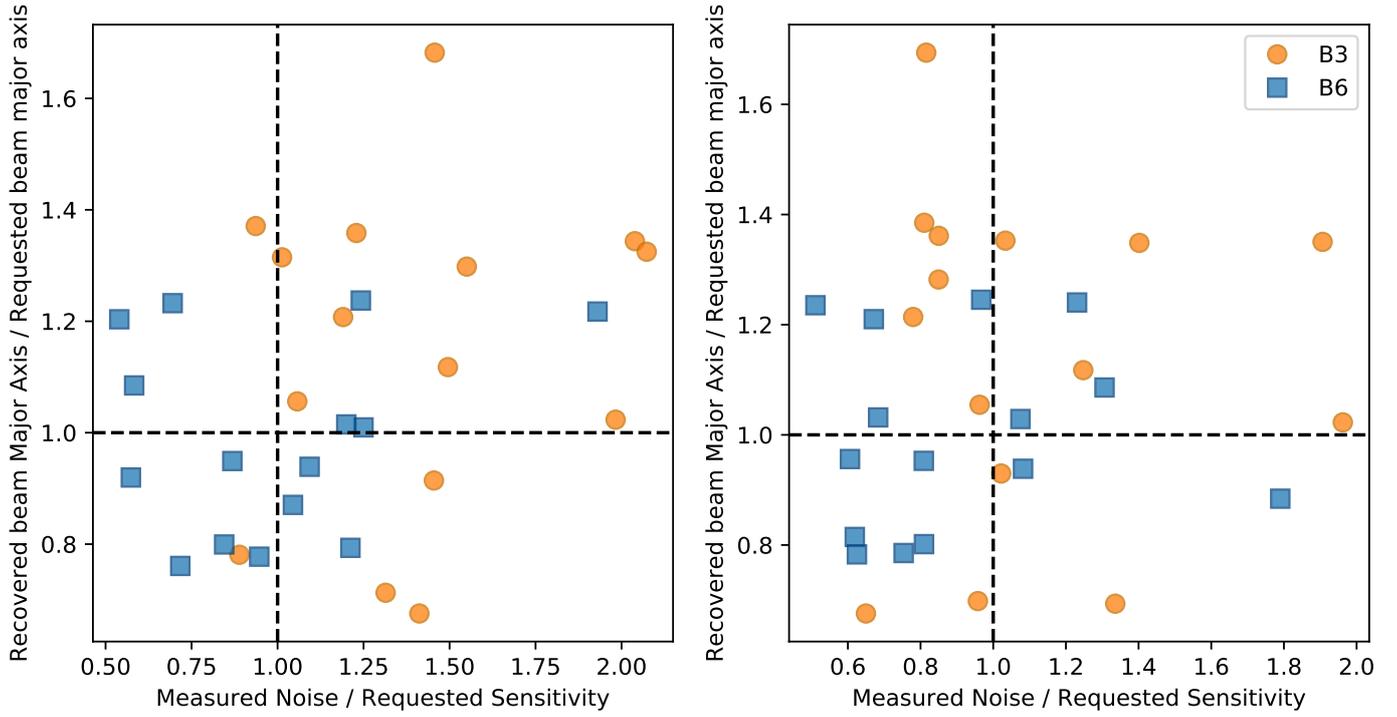}
    \caption{Comparison of the retrieved to requested beam sizes plotted
    against the ratio of the measured noise to the requested.
    (left) \texttt{\cleanest} continuum (right) \texttt{\bsens} continuum.
    The bottom-left quadrant shows where the beam is smaller (better) and the
    noise is lower (better) than requested.  The bottom-right quadrant
    shows noise that is higher than requested but a beam smaller than requested;
    in this quadrant, smoothing the data brings them closer to the target
    noise.  The upper-right quadrant contains those regions whose beam sizes
    are too large and which have excess noise; see \ref{sec:calerrors}.}
    \label{fig:beamsize}
\end{figure*}

\begin{figure*}[htp]
    \includegraphics[width=0.5\textwidth]{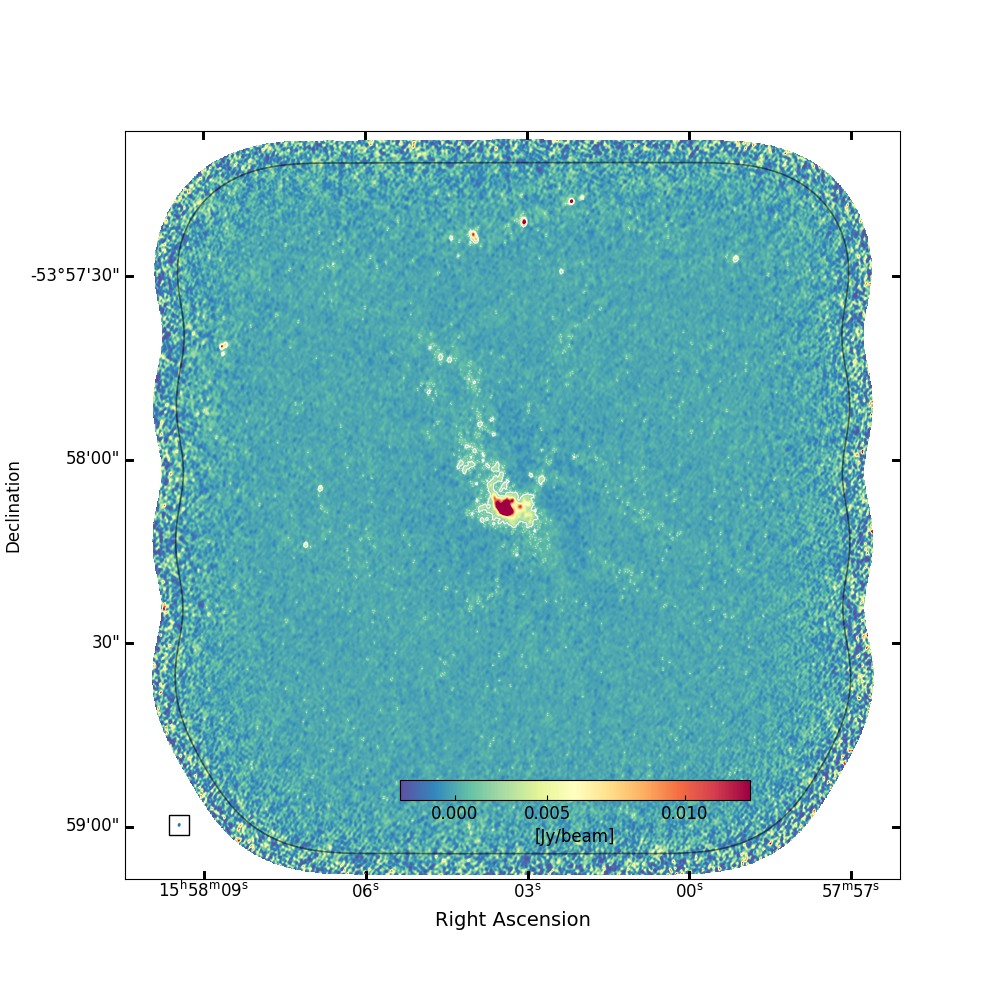}
    \includegraphics[width=0.5\textwidth]{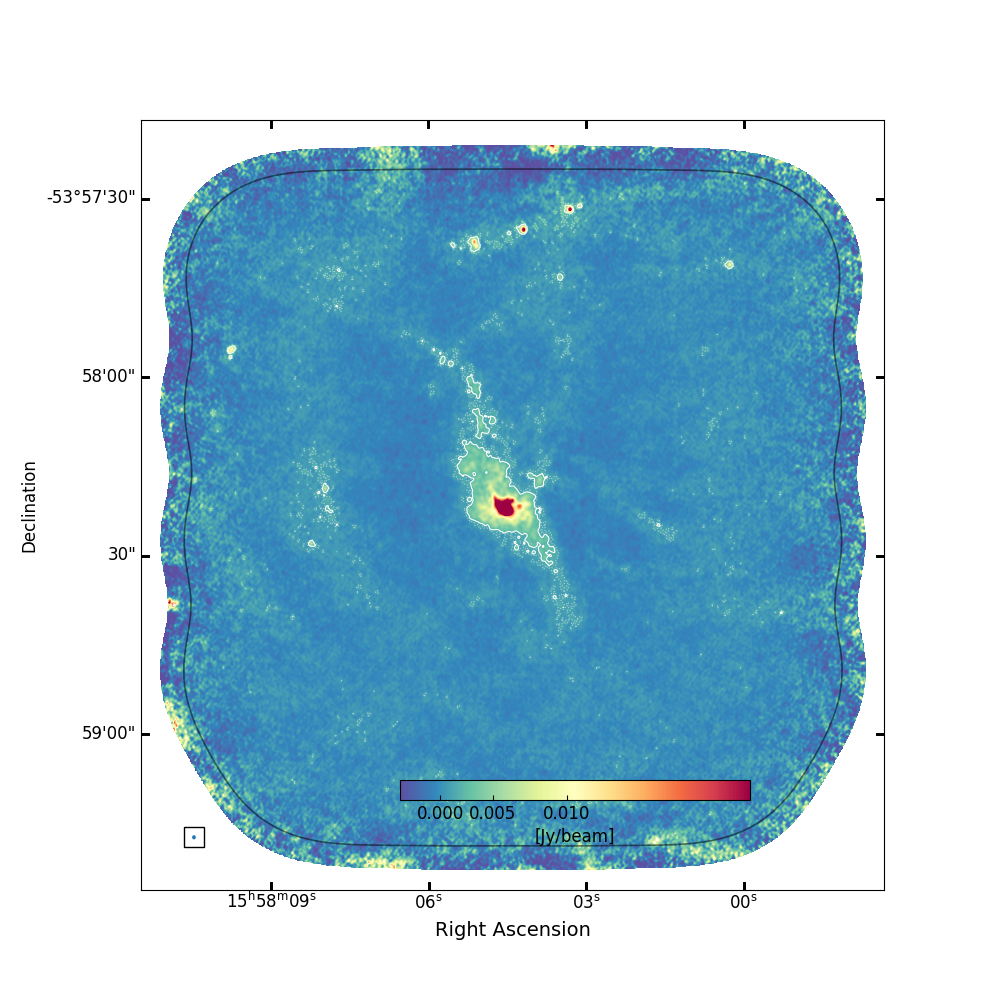}
    \caption{\textbf{Left:} ALMA 12m only continuum mosaic of the G328.25 clump
    in Band~3. White contours show 3 and 5 times the \textit{rms} noise
    measured in an emission free region prior to the primary beam correction.
    Black contours show where the primary beam sensitivity drops to 20\%. The
    synthesized beam is shown in the lower left corner, and the color bar is
    displayed in the bottom of the image. \textbf{Right:} ALMA 7m and 12m array
    combined continuum image in B3. Contours are the same as on the left panel,
    where the corresponding noise is measured on the combined map. The
    synthesized beam is shown in the lower left corner, and the color bar is
    displayed in the bottom of the image.}
    \label{fig:G328_7m12m}
\end{figure*}

From \S \ref{sec:noisegoals}, Figure \ref{fig:noise_excess} shows the excess noise compare to that requested.
From \S \ref{sec:psfprops}, Figures \ref{fig:b3psfs} and \ref{fig:b6psfs} show images of the PSFs in B3 and B6, respectively.  Figure \ref{fig:beamsize} compares the achieved to the requested beam sizes.
From \S \ref{sec:7m12m}, Figure \ref{fig:G328_7m12m} shows the effect of jointly imaging 7m + 12m data for one field.

\begin{figure*}
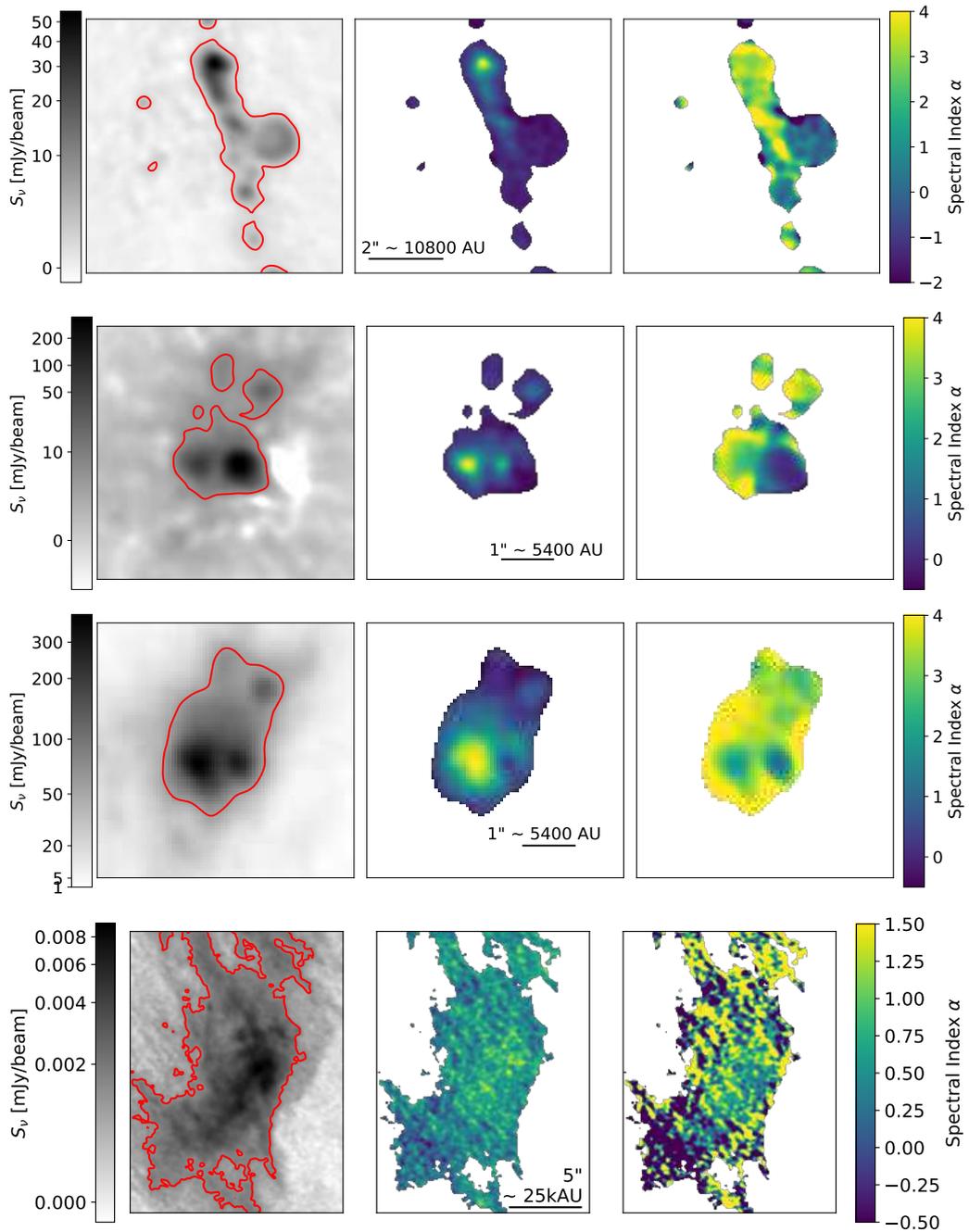

    \centering
    \includegraphics[width=0.75\textwidth]{f106}
    \includegraphics[width=0.75\textwidth]{f107}
    \includegraphics[width=0.75\textwidth]{f108}
    \includegraphics[width=0.75\textwidth]{f109}
    \caption{W51-E B3 Spectral Index maps distinguish HII regions from dust-dominated objects.
    In each panel, the left image shows the \texttt{tt0} (Taylor term 0), which is our approximation
    of the continuum level and is used to create a mask.  The middle panel shows the \texttt{tt1} image.
    The rightmost panel shows $\alpha=\frac{\texttt{tt1}}{\texttt{tt0}}$, the spectral index, and we have
    truncated the display to a plausible physical range $-2 < \alpha < 4$; values beyond this range most likely represent measurement
    errors.
    (top) W51 e1/e8.  The circular object to the right is an ultracompact HII region.
    (top-middle) W51 e2 B3.  This source splits into e2e, the dust-dominated ($\alpha\sim4$) source to the left,
    and e2w, the hypercompact HII region with $\alpha\sim0-1$.  W51 e2w shows signs of a changing
    spectral index across the band, as it appears to be the source of the symmetric ringing errors that
    span the image.
    (bottom-middle) W51 e2 B6.  The HII region e2w remains relatively flat, though somewhat more positively sloped than a pure free-free source; it contains at least some dust.  W51 e2e has a slightly shallower slope than at B3, indicating that it is optically thick ($\alpha=2$).
    (bottom) W51 IRS1 / Main, the extended HII region that dominates the overall image.
    There is no clear detection in the \texttt{tt1} term, suggesting $\alpha\sim0$, which is
    expected for an optically thin HII region.
    See also Fig. 2k in \paperone.
    }
    \label{fig:w51spindx}
\end{figure*}

\begin{figure*}
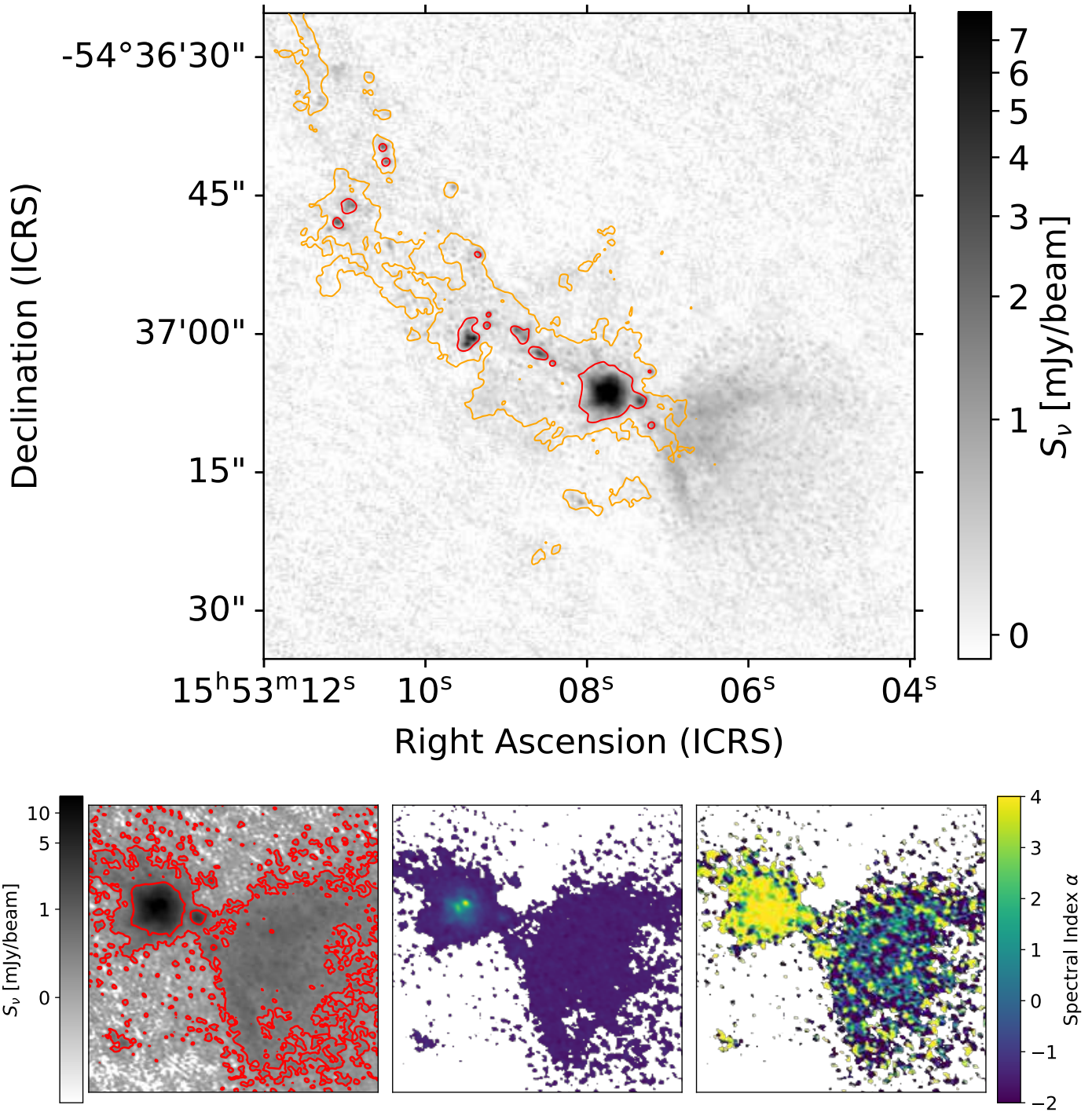

    \centering
    \includegraphics[width=\textwidth]{f110}
    \includegraphics[width=\textwidth]{f111}
    \caption{(top) The G327.29 B3 (\cleanest) image is shown in greyscale with contours
    of the B6 data overlaid (orange: 3 mJy/beam, red: 20 mJy/beam).  While the compact
    sources match well between the two bands, the HII region to the southwest appears only
    in B3. (lower panels) As in Figure \ref{fig:w51spindx}, these panels show Taylor terms 0 (left) and 1 (middle)
    and the spectral index $\alpha$ (right) of a cutout around the brightest core.  The dust-dominated, compact source has a steep $\alpha\sim4$,
    while the extended, free-free-dominated HII region has a flat index $\alpha\sim0$.
    See also Fig. 2f of \paperone.
    }
    \label{fig:g327contour_freefree}
\end{figure*}

\begin{figure*}
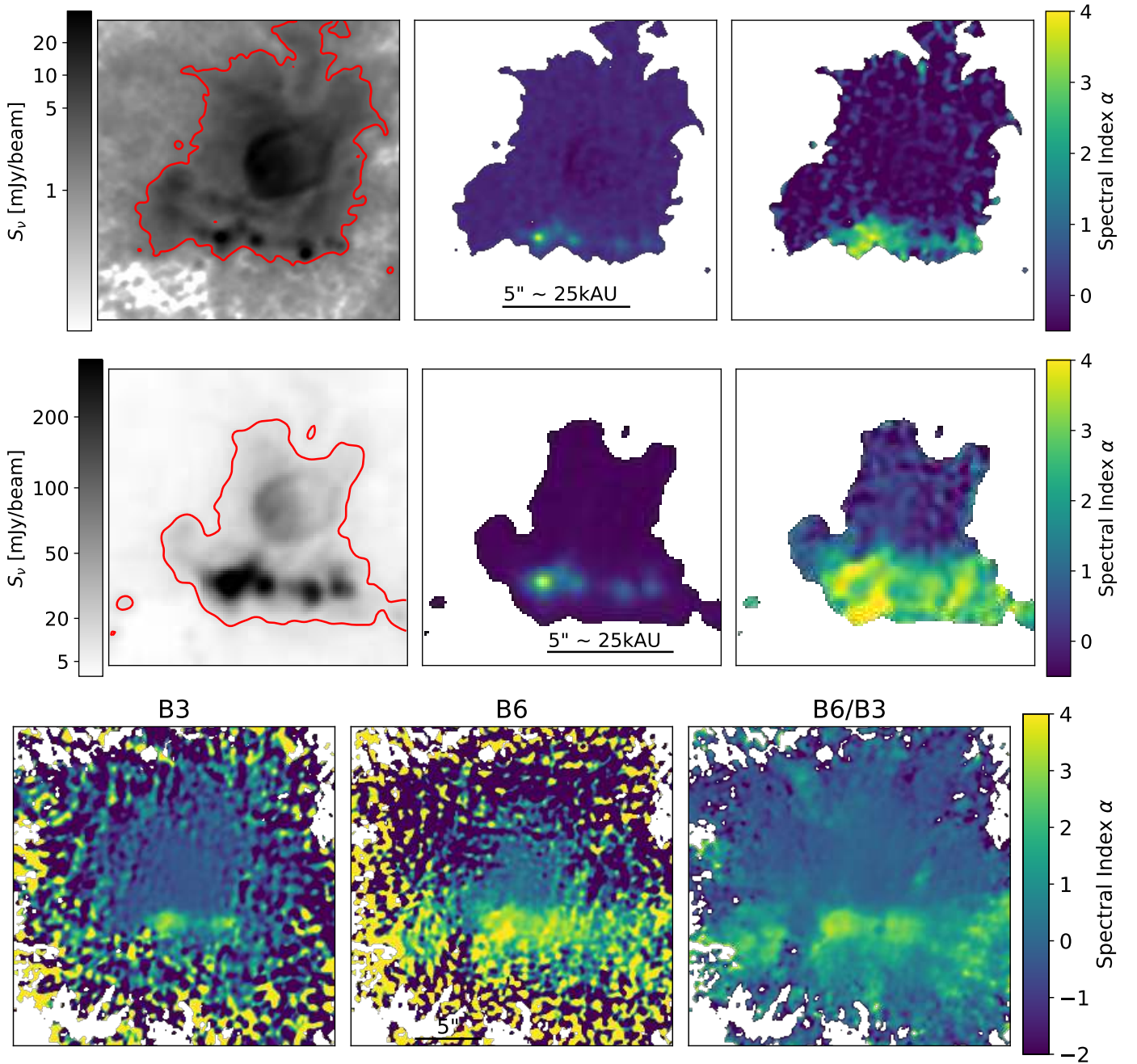

    \centering
    \includegraphics[width=\textwidth]{f114}
    \includegraphics[width=\textwidth]{f115}
    \includegraphics[width=\textwidth]{f116}
    \caption{Spectral index images of W51-IRS2 in band 3 (top), band 6 (middle),
    and calculated from B3/B6 (bottom).
    As in Figures \ref{fig:w51spindx} and \ref{fig:g327contour_freefree},
    for the top two rows,
    the left panel is the \texttt{tt0} term, the middle panel is \texttt{tt1},
    and the right panel is the derived spectral index $\alpha$.
    The compact sources along the south end of the image are clearly dust-dominated,
    with $\alpha\sim3-4$ in both bands.
    The brightest compact source, W51 north, is evidently optically thick
    at 1 mm but thin at 3 mm, with $\alpha_{1mm}\approx2$ and $\alpha_{3mm}\approx4$,
    as has been previously observed \citep{Ginsburg2017,Goddi2020}.
    See also Fig. 2m of \paperone.
    }
    \label{fig:w51irs2spindx}
\end{figure*}

The figures from Section \ref{sec:spectralindex} are also included in this appendix.
Figure \ref{fig:W51e_zooms} shows inset enlarged images of W51-E.
Figure \ref{fig:w51spindx} shows spectral index images of different parts of W51E.
Figure \ref{fig:g327contour_freefree} shows the G327 region, highlighting the difference between the extended HII region and the compact dust emission.
Figures \ref{fig:W51irs2_zooms} and  \ref{fig:w51irs2spindx} show the zoom-in and spectral index images for W51-IRS2.

\begin{figure*}
    \centering
    \includegraphics[width=\textwidth]{f104}
    \includegraphics[width=\textwidth]{f105}
    \caption{Inset zoom figures of the B3 (top) and B6 (bottom) images of W51-E.
    The insets highlight the regions shown in detail in Figure \ref{fig:w51spindx}.
    The central region, containing W51 e1 through e10, is shown in both insets.
    W51 e1/e8 is the southern part of this region, shown in Fig. \ref{fig:w51spindx}a.
    W51 e2 is the northern part of this region, shown in Fig. \ref{fig:w51spindx}b and c.
    The diffuse HII region W51 Main shown as a zoom panel to the right of the \emph{top} figure showing the B3 image; its spectral
    index map is shown in Fig. \ref{fig:w51spindx}d.
    }
    \label{fig:W51e_zooms}
\end{figure*}

\begin{figure*}
    \centering
    \includegraphics[width=\textwidth]{f112}
    \includegraphics[width=\textwidth]{f113}
    \caption{Inset zoom figures in the B3 (top) and B6 (bottom) images of W51-IRS2.
    The central inset highlights the region shown in detail in Figure \ref{fig:w51irs2spindx}.
    }
    \label{fig:W51irs2_zooms}
\end{figure*}

Finally, from Section \ref{sec:hotcoreoutflow}, the \bsens-vs-\cleanest comparison for G351.77 is shown in Figure \ref{fig:bsenscompG351_B6}.  The remaining fields have the same diagnostic images in Appendix \ref{appendix:selfcalcompare}.
Figure \ref{fig:absorption_bsens_diff} shows the difference between the \bsens and \cleanest fields for two images where spectral absorption is an important effect.

\begin{figure*}
\includegraphics[width=\textwidth]{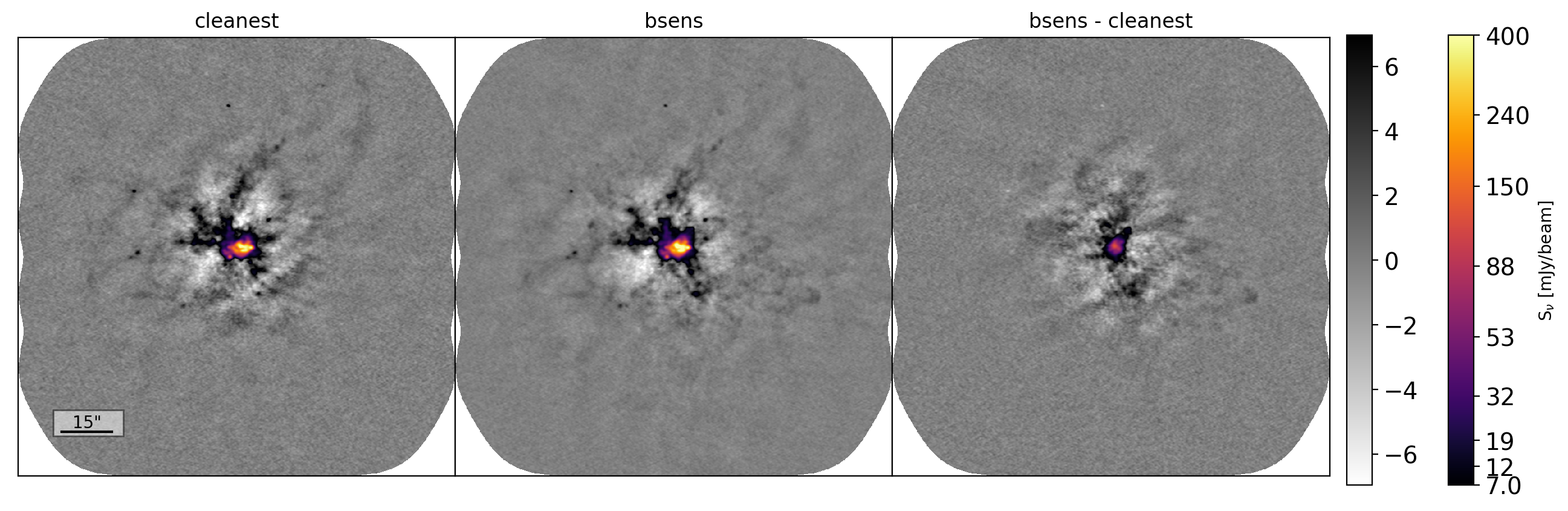}   
\caption{G351.77 B6 \bsens vs \cleanest comparison.  The \bsens (left) and \cleanest (middle)
self-calibrated images are compared in the \bsens-minus-\cleanest image (right).
Additional figures for the remaining 29 mosaic images are shown in Appendix \ref{appendix:selfcalcompare}} 
\label{fig:bsenscompG351_B6}       
\end{figure*}

\begin{figure*}[htp]
    \includegraphics[width=0.48\textwidth]{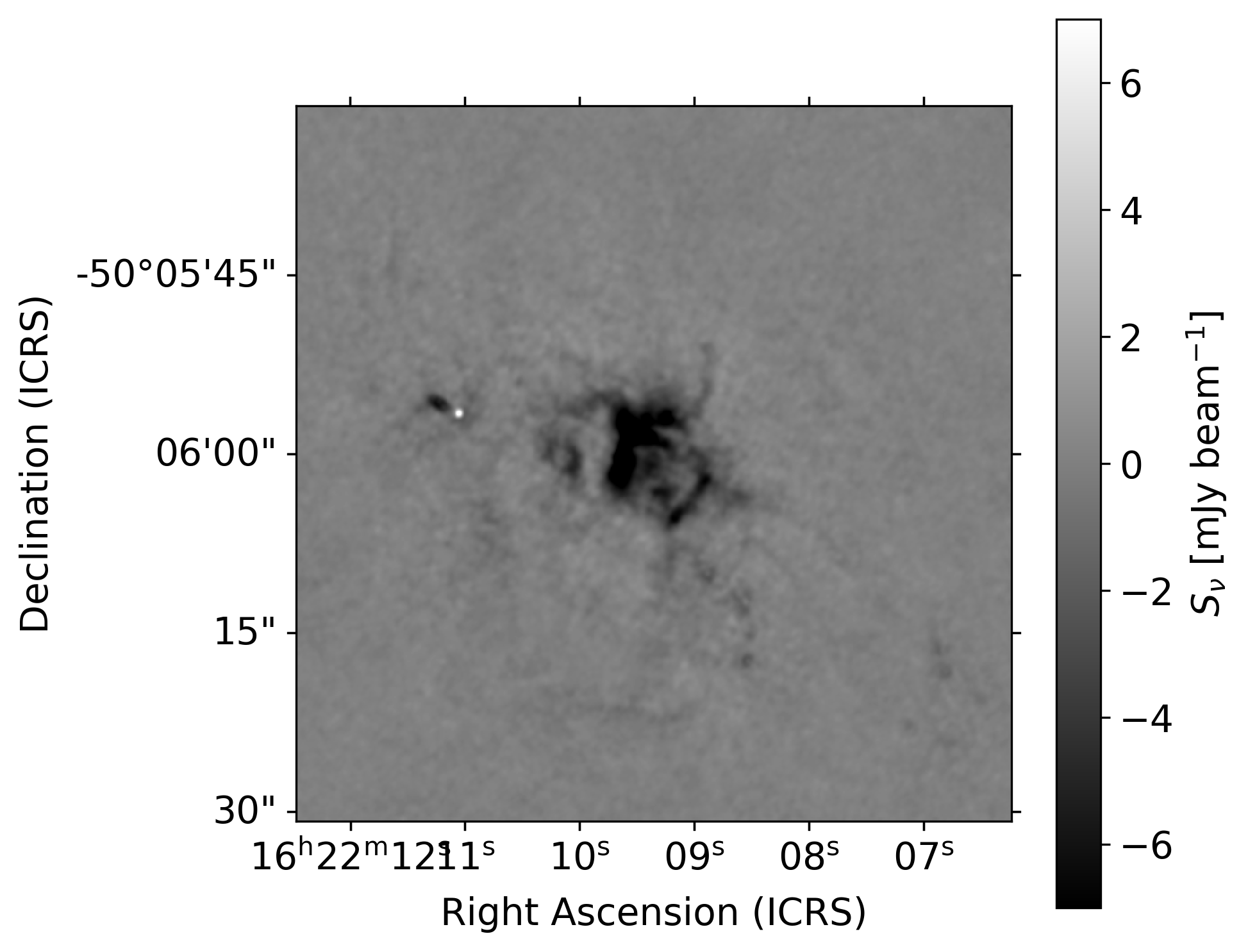}
    \includegraphics[width=0.48\textwidth]{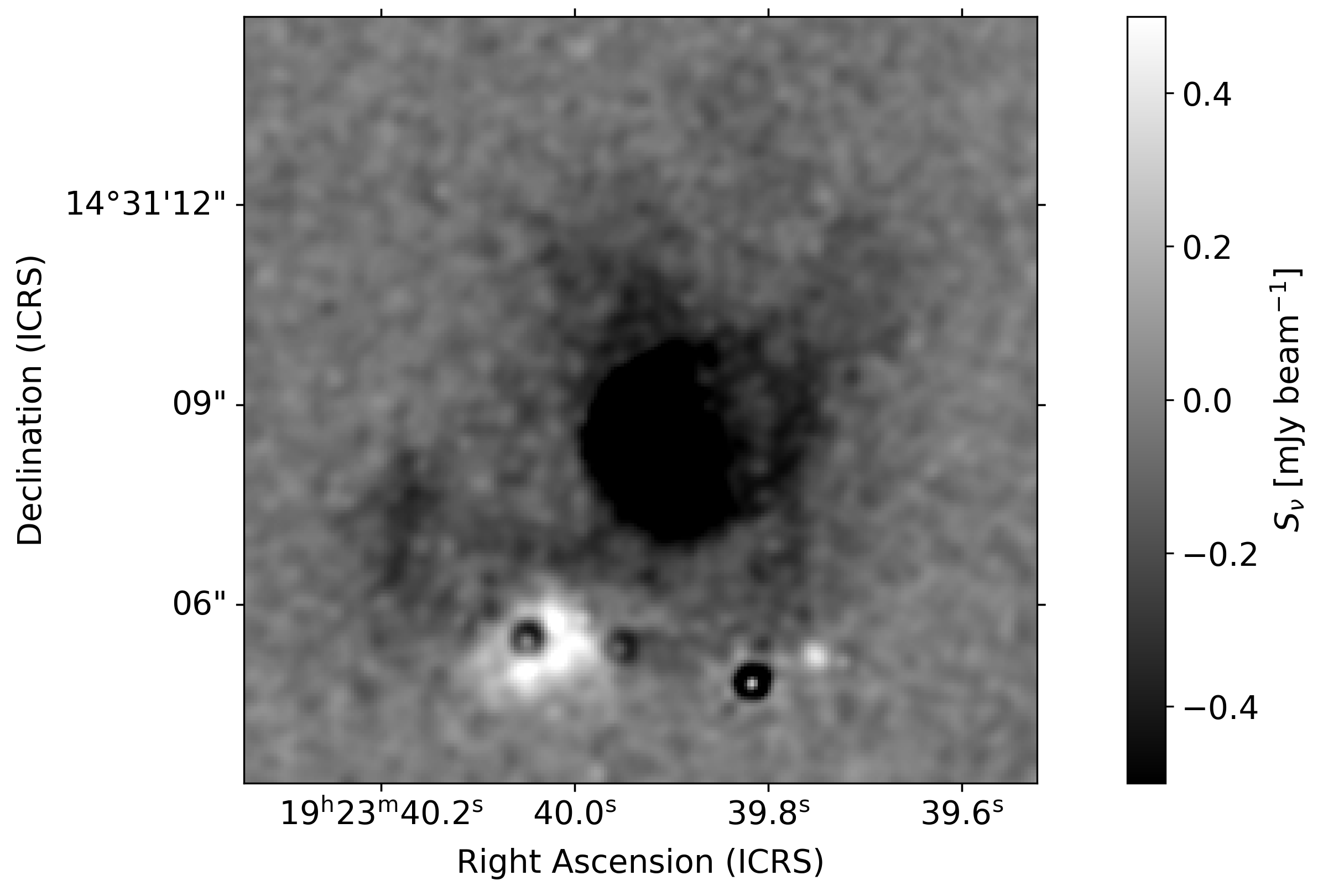}
    \caption{Difference of \bsens minus \cleanest images for the two fields that contain
    strong HII regions against which molecular absorption is observed: G333.60 B6
    (left) and W51-IRS2 B3 (right).  The absorption regions dominate the integrated
    flux of the field because of their large spatial extents,
    but there are also regions where \bsens is greater than \cleanest because
    of excess emission from hot cores.
    }
    \label{fig:absorption_bsens_diff}
\end{figure*}

\section{Observation Table}
We include Table \ref{tab:observations} here.

\clearpage
\onecolumn

\begin{longtable}{cccccccll}
\caption{Summary of observations}

\label{tab:observations}
\\
\hline
Region & Band & D        & Type     & $N_{ptg}$ & FOV$^{(2)}$                                                    & Config. & $t_{int}$ & Obs. Dates \\
       &      &          & $^{(1)}$ &           &  (12m only)                                                    &         &           &            \\
       &      & [kpc]    &          &           & [$\mathrm{{}^{\prime\prime}}\times\mathrm{{}^{\prime\prime}}$] &         &  [hr]     & \\
\hline
 \endfirsthead
 \caption{continued.}\\
 \hline
Region & Band & D        & Type     & $N_{ptg}$ & FOV$^{(2)}$                                                    & Config. & $t_{int}$ & Obs. Dates \\
       &      &          & $^{(1)}$ &           &  (12m only)                                                    &         &           &            \\
       &      & [kpc]    &          &           & [$\mathrm{{}^{\prime\prime}}\times\mathrm{{}^{\prime\prime}}$] &         &  [hr]     & \\
\hline
 \endhead
 \hline

\endfoot

G008.67 & B3 & 3.4 & I  &  7 & 190$\times$125 & TM1;C43-6 & 0.27 & 2018-01-07 \\
        &    &     &    &  7 &                & TM2;C43-2 & 0.12 & 2018-05-08\\
        &    &     &    &  3 &                & 7M        & 1.20 & 2017-11-28,2017-12-14\\

G008.67 & B6 & 3.4 & I  & 37 & 132$\times$87 & TM1;C43-4,2 & 0.78 & 2018-03-31, 2019-03-28 \\
        &    &     &    & 13 &               & 7M  & 3.49 & 2017-12-16, 2018-01-09\\
        &    &     &    &    &               &     &      & 2018-01-17\\ 
        &    &     &    &    &               &     &      & 2018-03-13,2018-03-27\\

G010.62 & B3 & 4.95 & E & 7 & $150\times160$ &  TM1;C43-6 & 1.89 & 2017-12-28$^\dagger$,2017-12-31 \\
        &    &      &   & 7 &                &  TM2;C43-3 & 0.90 & 2018-04-23 \\ 
        &    &      &   & 3 &                &   7M & 3.54 & 2017-12-16, 2017-12-20\\ 
        &    &      &   &   &                &      &      & 2017-12-21$^\dagger$,2017-12-26\\

G010.62 & B6 & 4.95 & E & 27 & $98\times90$ & TM1;C43-4,3 & 1.71 & 2018-03-27$^\dagger$, 2018-08-31$^\dagger$ \\
        &    &      &   & 27 &              & TM2;C43-1   & 0.67 & 2018-07-07\\
        &    &      &   & 10 &              & 7M          & 10.93 & 2018-03-29, 2018-03-31$^\dagger$\\
        &    &      &   &    &              &             &       & 2018-04-09, 2018-04-13\\
        &    &      &   &    &              &             &       & 2018-04-15$^\dagger$, 2018-05-05\\
        &    &      &   &    &              &             &       & 2018-05-06, 2018-05-07\\
        &    &      &   &    &              &             &       & 2018-05-15$^\dagger$, 2018-05-16\\

G012.80 & B3 & 2.4 & E &  13 & $190\times180$ & TM1;C43-4 & 0.10 & 2018-03-17 \\
        &    &     &   &  13 &                & TM2;C43-1 & 0.10 & 2018-07-12 \\
        &    &     &   &   5 &                & 7M  & 0.53 & 2017-12-07, 2018-03-17 \\

G012.80 & B6 & 2.4 & E &  67  & $132\times132$ & TM1;C43-2 & 0.27 &  2018-05-07\\
        &    &     &   &  27  &                & 7M  & 1.85 & 2017-10-19, 2017-12-03\\

G327.29 & B3 & 2.5 & Y &   7 & $160\times152$ & TM1;C43-6 & 0.22 & 2017-12-30 \\
        &    &     &   &   7 &                & TM2;C43-2 & 0.10 & 2018-05-09 \\
        &    &     &   &   3 &                &  7M & 1.05 & 2017-10-17, 2017-10-29\\

G327.29 & B6 & 2.5 & Y & 39 & $105\times109$ & TM1; C43-4,3& 0.67 & 2018-03-31, 2018-04-29\\
        &    &     &   & 14 &                &  7M         & 1.88 & 2017-11-09, 2017-11-18\\

G328.25 & B3 & 2.5 & Y & 10 & $160\times180$ & TM1;C43-5 & 0.31 & 2018-01-10 \\
        &    &     &   & 10 &   & TM2;C43-1 & 0.14 & 2018-07-01 \\
        &    &     &   &  3 &     & 7M & 1.50 & 2017-11-14, 2017-11-12 \\
        
G328.25 & B6 & 2.5 & Y & 52 &  $120\times120$ &  TM1;C43-4 & 0.43 & 2018-03-31 \\
        &    &     &   & 17 &  &  7M & 3.80 & 2017-11-22, 2017-12-12 \\
        &    &     &   &  & & & & 2017-12-13,             2017-12-17 \\

G333.60 & B3 & 4.2 & E & 14 & $190\times180$ & TM1;C43-5 & 0.86 & 2018-01-09 \\
        &    &     &   & 14 &                & TM2;C43-2 & 0.50 & 2018-05-08 \\
        &    &     &   &  7 &                & 7M  &  3.88 & 2017-11-19, 2017-11-28$^\dagger$ \\
        &    &     &   &    &                &   &   &  2017-12-05, 2017-12-07 \\

G333.60 & B6 & 4.2 & E & 85 & $143\times143$ & TM1;C43-4 & 1.78 & 2018-03-13, 2018-03-22\\
        &    &     &   & 85 &                & TM2;C43-1 & 0.71 & 2018-06-09  \\
        &    &     &   & 33 &                & 7M & 6.56&  2017-12-28, 2018-01-01\\
        &    &     &   &    &                &  & & 2018-01-06,  2018-01-09$^\dagger$ \\
        &    &     &   &    &                &  & & 2018-03-06, 2018-03-07$^\ddagger$\\

G337.92 & B3 & 2.7 &  Y$^{(1)}$  &  7 & $160\times152$  & TM1;C43-5 & 0.42 & 2018-01-01 \\
        &    &     &             &  7 &                 & TM2;C43-2 & 0.27 & 2018-05-08 \\
        &    &     &             &  3 &                 & 7M & 1.92 & 2017-11-14, 2017-11-16$^\dagger$  \\ 
        
G337.92 & B6 &  2.7   &  Y$^{(1)}$   & 27 & $92\times86$ & TM1;C43-4 & 0.46 & 2018-03-22\\
        &    &        &              & 27 &              & TM2;C43-1 & 0.18 & 2018-07-01  \\
        &    &        &              &  7 &              & 7M  & 1.67 & 2017-11-26, 2017-12-09 \\

G338.93 & B3 & 3.9 & Y & 7 & $152\times160$& TM1;C43-5 & 0.50 & 2018-01-01\\
        &    &     &   & 7 &               & TM2;C43-2 & 0.28 & 2018-05-08\\
        &    &     &   & 3 &               & 7M  & 1.92 & 2017-12-14$^\dagger$,2017-12-21\\
G338.93 & B6 & 3.9 & Y & 27 & $86\times92$ & TM1;C43-4 & 0.48 & 2018-03-13  \\
        &    &     &   & 27 &              & TM2;C43-1 & 0.18 & 2018-06-28\\
        &    &     &   &  7 &              & 7M  & 1.67 & 2017-12-17,2017-12-23\\

G351.77 & B3 & 2 & I &  14 & $190\times180$  & TM1;C43-4 & 0.1  &  2018-03-17 \\
        &    &   &   &  14 &                 & TM2;C43-1 & 0.1  & 2017-07-12\\
        &    &   &   &   5 &                 & 7M  & 0.43 & 2017-11-14\\
       
G351.77 & B6 & 2 & I &  67 & $132\times132$ & TM1;C43-3 & 0.25 & 2018-05-02 \\ 
        &    &   &   &  27 &                & 7M  & 1.38 & 2018-01-04, 2018-03-12\\
G353.41 & B3 & 2.0 &  I$^{(1)}$  & 14 & $190\times180$ & TM1;C43-4 & 0.10 & 2018-03-17\\
        &    &     &             & 14 &                & TM2;C43-1 & 0.10 & 2018-07-12\\
        &    &     &             &  5 &                & 7M  & 0.43 & 2017-11-14$^\dagger$\\

G353.41 & B6 & 2.0 &  I$^{(1)}$  & 67 &$131\times131$ & TM1;C43-3 & 0.25 & 2018-05-02\\
        &    &     &             & 27 &               &  7M & 1.38 & 2018-01-17,2018-03-25\\

W43-MM1 & B3 & 5.5 & Y & 11 & $190\times150$ & TM1;C43-6 & 2.66  & 2017-12-24$^\ddagger$\\
        &    &     &   & 11 &                & TM2;C43-3 & 2.00 &  2018-08-20$^\dagger$, 2019-04-03\\
        &    &     &   &    &                &           &      & 2019-04-10\\
        &    &     &   &  5 &                & 7M  & 7.54 &  2018-01-09,2018-01-13\\
        &    &     &   &    &                &   & & 2018-01-14, 2018-01-16\\
        &    &     &   &    &                &   & & 2018-01-20,2018-01-24$^\dagger$\\
        &    &     &   &    &                &   & & 2018-03-06\\
        
W43-MM1 & B6\footnote{The W43-MM1 B6 data come from program 2013.1.01365.S \citep{Motte2018}} 
             & 5.5 & Y & 33 & $117\times53$ & TM1;C34-6, C34-5 & 3.0 & 2014-09-02,2015-06-04\\
        &    &     &   &    &              &   &  &             2015-06-05 \\
        &    &     &   & 33 &                & TM2;C34-1 & 1.28  & 2015-04-05$^\dagger$ \\
        &    &     &   & 11 &              &  7M & 1.17 &  2014-06-06,2014-06-07$^\dagger$ \\
        &    &     &   &    &              &   &  & 2014-06-11$^\ddagger$, 2014-06-12\\
        &    &     &   &    &              &   &  & 2015-04-05, 2015-04-28 \\
        &    &     &   &    &              &   &  & 2015-04-29\\
W43-MM2 & B3 & 5.5 & Y & 11 & $190\times150$ & TM1;C43-6,C43-7 & 3.01 & 2017-12-07, 2017-12-12\\
        &    &     &   &    &                &     &       & 2017-12-14\\ 
        &    &     &   & 11 &                & TM2;C43-3 & 1.32  & 2018-04-26$^\dagger$ \\
        &    &     &   &  3 &                &  7M & 5.73  & 2017-11-28,2017-12-14\\
        &    &     &   &    &                &     &       & 2017-12-17$^\dagger$,2017-12-19\\
        &    &     &   &    &                &     &       & 2017-12-21,2017-12-26\\
W43-MM2 & B6 & 5.5 & Y & 27 & $90\times98$ & TM1;C43-4 & 3.34 & 2018-09-06,2018-09-07\\
        &    &     &   &    &              &           &      & 2018-11-29,2018-12-11\\
        &    &     &   & 27 &              & TM2;C43-1 & 0.67 & 2018-06-29\\
        &    &     &   & 10 &              &  7M & 6.4 &  2017-10-26,2017-11-10\\
        &    &     &   &    &              &     & &  2017-12-24,2018-01-20\\
        &    &     &   &    &              &     & &  2018-03-25,2018-03-27\\
        &    &     &   &    &              &     & &  2018-04-01,2018-04-06\\

W43-MM3 & B3 & 5.5 & I & 11 & $190\times150$ & TM1;C43-6 & 3.02 & 2017-12-17$^\dagger$,2017-12-21\\ 
        &    &     &   & 11 &                & TM2;C43-3 & 1.38 & 2018-04-23,2018-04-26\\
        &    &     &   &  3 &                & 7M  & 6.06 & 2017-12-31 $^\dagger$,2018-01-01\\
        &    &     &   &    &                & & & 2018-01-04$^\dagger$, 2018-01-07\\
        &    &     &   &    &                & & & 2018-01-14, 2018-01-16\\

W43-MM3 & B6 & 5.5 & I & 27 & $100\times90$ & TM1;C43-4 & 1.42 & 2018-03-23$^\dagger$\\ 
        &    &     &   & 27 &               & TM2;C43-1 & 0.58 & 2018-06-27\\
        &    &     &   & 10 &               &  7M & 6.42 & 2017-10-17,  2017-10-23\\
        &    &     &   &    &               &     &      & 2017-10-28, 2017-11-05\\
        &    &     &   &    &               &     &      & 2017-12-24, 2018-01-18\\
        &    &     &   &    &               &     &      & 2018-03-22,  2018-03-23\\

W51-E   & B3 & 5.4  & I & 7 & $150\times160$ & TM1;C43-6,C43-7 & 3.58 & 2017-11-30, 2017-12-03\\
        &    &      &   &   &                &     & &2017-12-12\\
        &    &      &   &   &                &     &  & 2017-12-14, 2017-12-26$\dagger$\\
        &    &      &   & 7 &                & TM2;C43-3 & 0.86 &2018-04-23\\
        &    &      &   & 3 &                & 7M  & 4.09 & 2017-11-14, 2017-11-16\\
        &    &      &   &   &                &     & & 2017-11-20, 2017-12-05\\
        &    &      &   &   &                &     & &2017-12-07\\

W51-E   & B6 & 5.4  & I & 27 & $100\times90$ & TM1;C43-5 & 1.76 & 2018-01-20, 2018-01-21 \\
        &    &      &   & 27 &               & TM2;C43-1 & 0.71 &2018-06-27\\
        &    &      &   & 10 &               &  7M & 7.83  & 2017-11-24, 2017-12-01\\
        &    &      &   &    &               &     &  & 2018-03-10, 2018-04-10$^\dagger$\\
        &    &      &   &    &               &     &  &2018-04-11, 2018-04-16\\
        &    &      &   &    &               &     &  & 2018-04-19, 2018-05-02$^\dagger$\\

W51-IRS2 & B3 & 5.4 & E & 7 & $160\times150$ & TM1;C43-6,C43-7 & 2.09 & 2017-12-21$^\dagger$, 2019-08-31  \\
         &    &     &   & 7 &                & TM2;C43-3 & 0.86 &2018-04-26\\
         &    &     &   & 3 &                &  7M & 4.09 & 2017-11-21, 2017-11-23\\
         &    &     &   &   &                &     &   & 2017-12-07, 2017-12-14\\
         &    &     &   &   &                &     &   & 2017-12-24\\

W51-IRS2 & B6 & 5.4 & E & 27 & $92\times98$ & TM1;C43-4 & 1.76 & 2018-09-07$^\dagger$\\
         &    &     &   & 27 &              & TM2;C43-1 & 0.71 &2018-06-29\\
         &    &     &   & 10 &              &  7M &  8.03 & 2018-05-04, 2018-05-05$^\dagger$ \\
         &    &     &   &    &              &  &  & 2018-05-06, 2018-05-07\\
         &    &     &   &    &              &  &  & 2018-05-08, 2018-05-09$^\dagger$\\
         &    &     &   &    &              &  &  & 2018-05-11, 2018-05-12\\

\hline
\multicolumn{9}{p\textwidth}{
$^{(1)}$: 
Types are `young' (Y), `intermediate' (I), and `evolved' (E); the evolutionary
status of each region is defined in \paperone based on global 1mm/3mm
spectral index (Section \ref{sec:spectralindex}), free-free intensity, and
H$_{41\alpha}$ morphology.}\\
\multicolumn{9}{p\textwidth}{
$^{(2)}$: The field of view listed is the box size encompassing the ALMA footprint;
the actual field of view is a subset of this rectangular region.
}\\
\multicolumn{9}{p\textwidth}{
$^\dagger$: 2 EBs executed on the same day.
}\\
\multicolumn{9}{p\textwidth}{
$^\ddagger$: 3 EBs executed on the same day.
}\\

\end{longtable}

\clearpage
\twocolumn


\begin{thebibliography}{37}
\expandafter\ifx\csname natexlab\endcsname\relax\def\natexlab#1{#1}\fi

\bibitem[{{Alves} {et~al.}(2007){Alves}, {Lombardi}, \& {Lada}}]{Alves2007}
{Alves}, J., {Lombardi}, M., \& {Lada}, C.~J. 2007, \aap, 462, L17

\bibitem[{{Astropy Collaboration} {et~al.}(2018){Astropy Collaboration},
  {Price-Whelan}, {Sip{\H{o}}cz}, {G{\"u}nther}, {Lim}, {Crawford}, {Conseil},
  {Shupe}, {Craig}, {Dencheva}, {Ginsburg}, {VanderPlas}, {Bradley},
  {P{\'e}rez-Su{\'a}rez}, {de Val-Borro}, {Aldcroft}, {Cruz}, {Robitaille},
  {Tollerud}, {Ardelean}, {Babej}, {Bach}, {Bachetti}, {Bakanov}, {Bamford},
  {Barentsen}, {Barmby}, {Baumbach}, {Berry}, {Biscani}, {Boquien}, {Bostroem},
  {Bouma}, {Brammer}, {Bray}, {Breytenbach}, {Buddelmeijer}, {Burke},
  {Calderone}, {Cano Rodr{\'\i}guez}, {Cara}, {Cardoso}, {Cheedella}, {Copin},
  {Corrales}, {Crichton}, {D'Avella}, {Deil}, {Depagne}, {Dietrich}, {Donath},
  {Droettboom}, {Earl}, {Erben}, {Fabbro}, {Ferreira}, {Finethy}, {Fox},
  {Garrison}, {Gibbons}, {Goldstein}, {Gommers}, {Greco}, {Greenfield},
  {Groener}, {Grollier}, {Hagen}, {Hirst}, {Homeier}, {Horton}, {Hosseinzadeh},
  {Hu}, {Hunkeler}, {Ivezi{\'c}}, {Jain}, {Jenness}, {Kanarek}, {Kendrew},
  {Kern}, {Kerzendorf}, {Khvalko}, {King}, {Kirkby}, {Kulkarni}, {Kumar},
  {Lee}, {Lenz}, {Littlefair}, {Ma}, {Macleod}, {Mastropietro}, {McCully},
  {Montagnac}, {Morris}, {Mueller}, {Mumford}, {Muna}, {Murphy}, {Nelson},
  {Nguyen}, {Ninan}, {N{\"o}the}, {Ogaz}, {Oh}, {Parejko}, {Parley}, {Pascual},
  {Patil}, {Patil}, {Plunkett}, {Prochaska}, {Rastogi}, {Reddy Janga},
  {Sabater}, {Sakurikar}, {Seifert}, {Sherbert}, {Sherwood-Taylor}, {Shih},
  {Sick}, {Silbiger}, {Singanamalla}, {Singer}, {Sladen}, {Sooley},
  {Sornarajah}, {Streicher}, {Teuben}, {Thomas}, {Tremblay}, {Turner},
  {Terr{\'o}n}, {van Kerkwijk}, {de la Vega}, {Watkins}, {Weaver}, {Whitmore},
  {Woillez}, {Zabalza}, \& {Astropy Contributors}}]{AstropyCollaboration2018}
{Astropy Collaboration}, {Price-Whelan}, A.~M., {Sip{\H{o}}cz}, B.~M., {et~al.}
  2018, \aj, 156, 123

\bibitem[{{Astropy Collaboration} {et~al.}(2013){Astropy Collaboration},
  {Robitaille}, {Tollerud}, {Greenfield}, {Droettboom}, {Bray}, {Aldcroft},
  {Davis}, {Ginsburg}, {Price-Whelan}, {Kerzendorf}, {Conley}, {Crighton},
  {Barbary}, {Muna}, {Ferguson}, {Grollier}, {Parikh}, {Nair}, {Unther},
  {Deil}, {Woillez}, {Conseil}, {Kramer}, {Turner}, {Singer}, {Fox}, {Weaver},
  {Zabalza}, {Edwards}, {Azalee Bostroem}, {Burke}, {Casey}, {Crawford},
  {Dencheva}, {Ely}, {Jenness}, {Labrie}, {Lim}, {Pierfederici}, {Pontzen},
  {Ptak}, {Refsdal}, {Servillat}, \& {Streicher}}]{AstropyCollaboration2013}
{Astropy Collaboration}, {Robitaille}, T.~P., {Tollerud}, E.~J., {et~al.} 2013,
  \aap, 558, A33

\bibitem[{{Benjamin} {et~al.}(2003){Benjamin}, {Churchwell}, {Babler}, {Bania},
  {Clemens}, {Cohen}, {Dickey}, {Indebetouw}, {Jackson}, {Kobulnicky},
  {Lazarian}, {Marston}, {Mathis}, {Meade}, {Seager}, {Stolovy}, {Watson},
  {Whitney}, {Wolff}, \& {Wolfire}}]{Benjamin2003}
{Benjamin}, R.~A., {Churchwell}, E., {Babler}, B.~L., {et~al.} 2003, \pasp,
  115, 953

\bibitem[{{Beuther} \& {Schilke}(2004)}]{Beuther2004}
{Beuther}, H. \& {Schilke}, P. 2004, Science, 303, 1167

\bibitem[{{Brogan} {et~al.}(2018){Brogan}, {Hunter}, \&
  {Fomalont}}]{Brogan2018}
{Brogan}, C.~L., {Hunter}, T.~R., \& {Fomalont}, E.~B. 2018, arXiv e-prints,
  arXiv:1805.05266

\bibitem[{{Churchwell} {et~al.}(2009){Churchwell}, {Babler}, {Meade},
  {Whitney}, {Benjamin}, {Indebetouw}, {Cyganowski}, {Robitaille}, {Povich},
  {Watson}, \& {Bracker}}]{Churchwell2009}
{Churchwell}, E., {Babler}, B.~L., {Meade}, M.~R., {et~al.} 2009, \pasp, 121,
  213

\bibitem[{{Comrie} {et~al.}(2021){Comrie}, {Wang}, {Hsu}, {Moraghan}, {Harris},
  {Pang}, {Pi{\'n}ska}, {Chiang}, {Simmonds}, {Chang}, {Jan}, \&
  {Lin}}]{Comrie2021}
{Comrie}, A., {Wang}, K.-S., {Hsu}, S.-C., {et~al.} 2021, {CARTA: Cube Analysis
  and Rendering Tool for Astronomy}

\bibitem[{{Csengeri} {et~al.}(2018){Csengeri}, {Bontemps}, {Wyrowski},
  {Belloche}, {Menten}, {Leurini}, {Beuther}, {Bronfman}, {Commer{\c{c}}on},
  {Chapillon}, {Longmore}, {Palau}, {Tan}, \& {Urquhart}}]{Csengeri2018}
{Csengeri}, T., {Bontemps}, S., {Wyrowski}, F., {et~al.} 2018, \aap, 617, A89

\bibitem[{{Csengeri} {et~al.}(2014){Csengeri}, {Urquhart}, {Schuller}, {Motte},
  {Bontemps}, {Wyrowski}, {Menten}, {Bronfman}, {Beuther}, {Henning}, {Testi},
  {Zavagno}, \& {Walmsley}}]{Csengeri2014}
{Csengeri}, T., {Urquhart}, J.~S., {Schuller}, F., {et~al.} 2014, \aap, 565,
  A75

\bibitem[{{Enoch} {et~al.}(2008){Enoch}, {Evans}, {Sargent}, {Glenn},
  {Rosolowsky}, \& {Myers}}]{Enoch2008}
{Enoch}, M.~L., {Evans}, Neal~J., I., {Sargent}, A.~I., {et~al.} 2008, \apj,
  684, 1240

\bibitem[{{Ginsburg} {et~al.}(2018){Ginsburg}, {Bally}, {Barnes}, {Bastian},
  {Battersby}, {Beuther}, {Brogan}, {Contreras}, {Corby}, {Darling}, {De Pree},
  {Galv{\'a}n-Madrid}, {Garay}, {Henshaw}, {Hunter}, {Kruijssen}, {Longmore},
  {Lu}, {Meng}, {Mills}, {Ott}, {Pineda}, {S{\'a}nchez-Monge}, {Schilke},
  {Schmiedeke}, {Walker}, \& {Wilner}}]{Ginsburg2018}
{Ginsburg}, A., {Bally}, J., {Barnes}, A., {et~al.} 2018, \apj, 853, 171

\bibitem[{{Ginsburg} {et~al.}(2017){Ginsburg}, {Goddi}, {Kruijssen}, {Bally},
  {Smith}, {Galv{\'a}n-Madrid}, {Mills}, {Wang}, {Dale}, {Darling},
  {Rosolowsky}, {Loughnane}, {Testi}, \& {Bastian}}]{Ginsburg2017}
{Ginsburg}, A., {Goddi}, C., {Kruijssen}, J.~M.~D., {et~al.} 2017, \apj, 842,
  92

\bibitem[{{Ginsburg} {et~al.}(2019{\natexlab{a}}){Ginsburg}, {Koch},
  {Robitaille}, {Beaumont}, {Adamginsburg}, {Sip{\H{o}}cz}, {ZuHone}, {Patra},
  {Jones}, {Lim}, {Stern}, {Rosolowsky}, {Earl}, {De Val-Borro}, {Jrobbfed},
  {Shuokong}, {Kepley}, {Sokolov}, {Badger}, {Maret}, {Garrido}, {Booker}, \&
  {Tollerud}}]{Ginsburg2019b}
{Ginsburg}, A., {Koch}, E., {Robitaille}, T., {et~al.} 2019{\natexlab{a}},
  {radio-astro-tools/spectral-cube: Release v0.4.5}

\bibitem[{{Ginsburg} {et~al.}(2019{\natexlab{b}}){Ginsburg}, {Sip{\H{o}}cz},
  {Brasseur}, {Cowperthwaite}, {Craig}, {Deil}, {Guillochon}, {Guzman},
  {Liedtke}, {Lian Lim}, {Lockhart}, {Mommert}, {Morris}, {Norman}, {Parikh},
  {Persson}, {Robitaille}, {Segovia}, {Singer}, {Tollerud}, {de Val-Borro},
  {Valtchanov}, {Woillez}, {Astroquery Collaboration}, \& {a subset of astropy
  Collaboration}}]{Ginsburg2019}
{Ginsburg}, A., {Sip{\H{o}}cz}, B.~M., {Brasseur}, C.~E., {et~al.}
  2019{\natexlab{b}}, \aj, 157, 98

\bibitem[{{Goddi} {et~al.}(2020){Goddi}, {Ginsburg}, {Maud}, {Zhang}, \&
  {Zapata}}]{Goddi2020}
{Goddi}, C., {Ginsburg}, A., {Maud}, L.~T., {Zhang}, Q., \& {Zapata}, L.~A.
  2020, \apj, 905, 25

\bibitem[{{Harris} {et~al.}(2020){Harris}, {Millman}, {van der Walt},
  {Gommers}, {Virtanen}, {Cournapeau}, {Wieser}, {Taylor}, {Berg}, {Smith},
  {Kern}, {Picus}, {Hoyer}, {van Kerkwijk}, {Brett}, {Haldane}, {del R{\'\i}o},
  {Wiebe}, {Peterson}, {G{\'e}rard-Marchant}, {Sheppard}, {Reddy}, {Weckesser},
  {Abbasi}, {Gohlke}, \& {Oliphant}}]{Harris2020}
{Harris}, C.~R., {Millman}, K.~J., {van der Walt}, S.~J., {et~al.} 2020, \nat,
  585, 357

\bibitem[{{Hunter}(2007)}]{Hunter2007}
{Hunter}, J.~D. 2007, Computing in Science and Engineering, 9, 90

\bibitem[{{Joye} \& {Mandel}(2003)}]{Joye2003}
{Joye}, W.~A. \& {Mandel}, E. 2003, in Astronomical Society of the Pacific
  Conference Series, Vol. 295, Astronomical Data Analysis Software and Systems
  XII, ed. H.~E. {Payne}, R.~I. {Jedrzejewski}, \& R.~N. {Hook}, 489

\bibitem[{Kluyver {et~al.}(2016)Kluyver, Ragan-Kelley, P{\'e}rez, Granger,
  Bussonnier, Frederic, Kelley, Hamrick, Grout, Corlay, Ivanov, Avila, Abdalla,
  \& Willing}]{Kluyver2016jupyter}
Kluyver, T., Ragan-Kelley, B., P{\'e}rez, F., {et~al.} 2016, in Positioning and
  Power in Academic Publishing: Players, Agents and Agendas, ed. F.~Loizides \&
  B.~Schmidt, IOS Press, 87 -- 90

\bibitem[{{Koch} {et~al.}(2018){Koch}, {Ginsburg}, {AKL}, {Rosolowsky},
  {Robitaille}, {de Val-Borro}, {Sipocz}, \& {adamginsburg}}]{Koch2018}
{Koch}, E., {Ginsburg}, A., {AKL}, {et~al.} 2018, {Keflavich/Radio\_Beam: V0.0
  Prerelease (First Tag For Zenodo Doi)}

\bibitem[{{K{\"o}nyves} {et~al.}(2015){K{\"o}nyves}, {Andr{\'e}},
  {Men'shchikov}, {Palmeirim}, {Arzoumanian}, {Schneider}, {Roy}, {Didelon},
  {Maury}, {Shimajiri}, {Di Francesco}, {Bontemps}, {Peretto}, {Benedettini},
  {Bernard}, {Elia}, {Griffin}, {Hill}, {Kirk}, {Ladjelate}, {Marsh}, {Martin},
  {Motte}, {Nguy{\^e}n Luong}, {Pezzuto}, {Roussel}, {Rygl}, {Sadavoy},
  {Schisano}, {Spinoglio}, {Ward-Thompson}, \& {White}}]{Konyves2015}
{K{\"o}nyves}, V., {Andr{\'e}}, P., {Men'shchikov}, A., {et~al.} 2015, \aap,
  584, A91

\bibitem[{{Lu} {et~al.}(2020){Lu}, {Cheng}, {Ginsburg}, {Longmore},
  {Kruijssen}, {Battersby}, {Zhang}, \& {Walker}}]{Lu2020}
{Lu}, X., {Cheng}, Y., {Ginsburg}, A., {et~al.} 2020, \apjl, 894, L14

\bibitem[{{McMullin} {et~al.}(2007){McMullin}, {Waters}, {Schiebel}, {Young},
  \& {Golap}}]{McMullin2007}
{McMullin}, J.~P., {Waters}, B., {Schiebel}, D., {Young}, W., \& {Golap}, K.
  2007, in Astronomical Society of the Pacific Conference Series, Vol. 376,
  Astronomical Data Analysis Software and Systems XVI, ed. R.~A. {Shaw},
  F.~{Hill}, \& D.~J. {Bell}, 127

\bibitem[{{Molinari} {et~al.}(2010){Molinari}, {Swinyard}, {Bally}, {Barlow},
  {Bernard}, {Martin}, {Moore}, {Noriega-Crespo}, {Plume}, {Testi}, {Zavagno},
  {Abergel}, {Ali}, {Anderson}, {Andr{\'e}}, {Baluteau}, {Battersby},
  {Beltr{\'a}n}, {Benedettini}, {Billot}, {Blommaert}, {Bontemps}, {Boulanger},
  {Brand}, {Brunt}, {Burton}, {Calzoletti}, {Carey}, {Caselli}, {Cesaroni},
  {Cernicharo}, {Chakrabarti}, {Chrysostomou}, {Cohen}, {Compiegne}, {de
  Bernardis}, {de Gasperis}, {di Giorgio}, {Elia}, {Faustini}, {Flagey},
  {Fukui}, {Fuller}, {Ganga}, {Garcia-Lario}, {Glenn}, {Goldsmith}, {Griffin},
  {Hoare}, {Huang}, {Ikhenaode}, {Joblin}, {Joncas}, {Juvela}, {Kirk},
  {Lagache}, {Li}, {Lim}, {Lord}, {Marengo}, {Marshall}, {Masi}, {Massi},
  {Matsuura}, {Minier}, {Miville-Desch{\^e}nes}, {Montier}, {Morgan}, {Motte},
  {Mottram}, {M{\"u}ller}, {Natoli}, {Neves}, {Olmi}, {Paladini}, {Paradis},
  {Parsons}, {Peretto}, {Pestalozzi}, {Pezzuto}, {Piacentini}, {Piazzo},
  {Polychroni}, {Pomar{\`e}s}, {Popescu}, {Reach}, {Ristorcelli}, {Robitaille},
  {Robitaille}, {Rod{\'o}n}, {Roy}, {Royer}, {Russeil}, {Saraceno}, {Sauvage},
  {Schilke}, {Schisano}, {Schneider}, {Schuller}, {Schulz}, {Sibthorpe},
  {Smith}, {Smith}, {Spinoglio}, {Stamatellos}, {Strafella}, {Stringfellow},
  {Sturm}, {Taylor}, {Thompson}, {Traficante}, {Tuffs}, {Umana}, {Valenziano},
  {Vavrek}, {Veneziani}, {Viti}, {Waelkens}, {Ward-Thompson}, {White},
  {Wilcock}, {Wyrowski}, {Yorke}, \& {Zhang}}]{Molinari2010}
{Molinari}, S., {Swinyard}, B., {Bally}, J., {et~al.} 2010, \aap, 518, L100

\bibitem[{{Motte} {et~al.}(1998){Motte}, {Andre}, \& {Neri}}]{Motte1998}
{Motte}, F., {Andre}, P., \& {Neri}, R. 1998, \aap, 336, 150

\bibitem[{{Motte} {et~al.}(2022){Motte}, {Bontemps}, {Csengeri}, {Pouteau},
  {Louvet}, {Stutz}, {Cunningham}, {L{\'o}pez-Sepulcre}, {Brouillet},
  {Galv{\'a}n-Madrid}, {Ginsburg}, {Maud}, {Men'shchikov}, {Nakamura}, {Nony},
  {Sanhueza}, {{\'A}lvarez-Guti{\'e}rrez}, {Armante}, {Baug}, {Bonfand},
  {Busquet}, {Chapillon}, {D{\'\i}az-Gonz{\'a}lez}, {Fern{\'a}ndez-L{\'o}pez},
  {Guzm{\'a}n}, {Herpin}, {Liu}, {Olguin}, {Towner}, {Bally}, {Battersby},
  {Braine}, {Bronfman}, {Chen}, {Dell'Ova}, {Di Francesco}, {Gonz{\'a}lez},
  {Gusdorf}, {Hennebelle}, {Izumi}, {Joncour}, {Lee}, {Lefloch}, {Lesaffre},
  {Lu}, {Menten}, {Mignon-Risse}, {Molet}, {Moraux}, {Mundy}, {Nguyen Luong},
  {Reyes}, {Reyes Reyes}, {Robitaille}, {Rosolowsky}, {Sandoval-Garrido},
  {Schuller}, {Svoboda}, {Tatematsu}, {Thomasson}, {Walker}, {Wu}, {Whitworth},
  \& {Wyrowski}}]{Motte2021}
{Motte}, F., {Bontemps}, S., {Csengeri}, T., {et~al.} 2022, \aap, 662, A8

\bibitem[{{Motte} {et~al.}(2018){Motte}, {Nony}, {Louvet}, {Marsh}, {Bontemps},
  {Whitworth}, {Men'shchikov}, {Nguyen Luong}, {Csengeri}, {Maury}, {Gusdorf},
  {Chapillon}, {K{\"o}nyves}, {Schilke}, {Duarte-Cabral}, {Didelon}, \&
  {Gaudel}}]{Motte2018}
{Motte}, F., {Nony}, T., {Louvet}, F., {et~al.} 2018, Nature Astronomy, 2, 478

\bibitem[{{Offner} {et~al.}(2014){Offner}, {Clark}, {Hennebelle}, {Bastian},
  {Bate}, {Hopkins}, {Moraux}, \& {Whitworth}}]{Offner2014}
{Offner}, S.~S.~R., {Clark}, P.~C., {Hennebelle}, P., {et~al.} 2014, in
  Protostars and Planets VI, ed. H.~{Beuther}, R.~S. {Klessen}, C.~P.
  {Dullemond}, \& T.~{Henning}, 53

\bibitem[{{Ohashi} {et~al.}(2016){Ohashi}, {Sanhueza}, {Chen}, {Zhang},
  {Busquet}, {Nakamura}, {Palau}, \& {Tatematsu}}]{Ohashi2016}
{Ohashi}, S., {Sanhueza}, P., {Chen}, H.-R.~V., {et~al.} 2016, \apj, 833, 209

\bibitem[{{Rau} \& {Cornwell}(2011)}]{Rau2011}
{Rau}, U. \& {Cornwell}, T.~J. 2011, \aap, 532, A71

\bibitem[{{Robitaille} {et~al.}(2017){Robitaille}, {Beaumont}, {Qian},
  {Borkin}, \& {Goodman}}]{Robitaille2017}
{Robitaille}, T., {Beaumont}, C., {Qian}, P., {Borkin}, M., \& {Goodman}, A.
  2017, {glueviz v0.13.1: multidimensional data exploration}

\bibitem[{{Sanhueza} {et~al.}(2019){Sanhueza}, {Contreras}, {Wu}, {Jackson},
  {Guzm{\'a}n}, {Zhang}, {Li}, {Lu}, {Silva}, {Izumi}, {Liu}, {Miura},
  {Tatematsu}, {Sakai}, {Beuther}, {Garay}, {Ohashi}, {Saito}, {Nakamura},
  {Saigo}, {Veena}, {Nguyen-Luong}, \& {Tafoya}}]{Sanhueza2019}
{Sanhueza}, P., {Contreras}, Y., {Wu}, B., {et~al.} 2019, \apj, 886, 102

\bibitem[{{Schuller} {et~al.}(2009){Schuller}, {Menten}, {Contreras},
  {Wyrowski}, {Schilke}, {Bronfman}, {Henning}, {Walmsley}, {Beuther},
  {Bontemps}, {Cesaroni}, {Deharveng}, {Garay}, {Herpin}, {Lefloch}, {Linz},
  {Mardones}, {Minier}, {Molinari}, {Motte}, {Nyman}, {Reveret}, {Risacher},
  {Russeil}, {Schneider}, {Testi}, {Troost}, {Vasyunina}, {Wienen}, {Zavagno},
  {Kovacs}, {Kreysa}, {Siringo}, \& {Wei{\ss}}}]{Schuller2009}
{Schuller}, F., {Menten}, K.~M., {Contreras}, Y., {et~al.} 2009, \aap, 504, 415

\bibitem[{{van der Walt} {et~al.}(2011){van der Walt}, {Colbert}, \&
  {Varoquaux}}]{vanderWalt2011}
{van der Walt}, S., {Colbert}, S.~C., \& {Varoquaux}, G. 2011, Computing in
  Science and Engineering, 13, 22

\bibitem[{{Virtanen} {et~al.}(2020){Virtanen}, {Gommers}, {Oliphant},
  {Haberland}, {Reddy}, {Cournapeau}, {Burovski}, {Peterson}, {Weckesser},
  {Bright}, {van der Walt}, {Brett}, {Wilson}, {Millman}, {Mayorov}, {Nelson},
  {Jones}, {Kern}, {Larson}, {Carey}, {Polat}, {Feng}, {Moore}, {VanderPlas},
  {Laxalde}, {Perktold}, {Cimrman}, {Henriksen}, {Quintero}, {Harris},
  {Archibald}, {Ribeiro}, {Pedregosa}, {van Mulbregt}, \& {SciPy 1. 0
  Contributors}}]{Virtanen2020}
{Virtanen}, P., {Gommers}, R., {Oliphant}, T.~E., {et~al.} 2020, Nature
  Methods, 17, 261

\bibitem[{{Zhang} {et~al.}(2015){Zhang}, {Wang}, {Lu}, \&
  {Jim{\'e}nez-Serra}}]{Zhang2015}
{Zhang}, Q., {Wang}, K., {Lu}, X., \& {Jim{\'e}nez-Serra}, I. 2015, \apj, 804,
  141

\end{thebibliography}
\end{document}